\documentclass{article}

\usepackage{amssymb}

\usepackage[authoryear]{natbib}
\usepackage{xurl}
\usepackage{graphicx}
\usepackage[export]{adjustbox}
\usepackage{authblk}
\usepackage{booktabs}
\usepackage{dcolumn}
\usepackage{scalerel}
\usepackage{amsmath}
\usepackage{amsthm}
\usepackage{multicol}
\usepackage{dsfont}

\usepackage{caption}
\usepackage{subcaption}
\DeclareCaptionLabelFormat{andtable}{#1~#2  \&  \tablename~\thetable}

\usepackage{float}
\restylefloat{figure}

\newtheorem{proposition}{Proposition}[section]

\begin{document}

\title{Smoothly Adaptively Centered Ridge Estimator}

\author{Edoardo Belli \\
        edoardo.belli@polimi.it}

\affil{MOX - Modeling and Scientific Computing, Department of Mathematics, Politecnico di Milano, Italy}

\date{}

\maketitle

\begin{abstract}
With a focus on linear models with smooth functional covariates, we propose a penalization framework (SACR) based on the nonzero centered ridge, where the center of the penalty is optimally reweighted in a supervised way, starting from the ordinary ridge solution as the initial centerfunction. In particular, we introduce a convex formulation that jointly estimates the model's coefficients and the weight function, with a roughness penalty on the centerfunction and constraints on the weights in order to recover a possibly smooth and/or sparse solution. This allows for a non-iterative and continuous variable selection mechanism, as the weight function can either inflate or deflate the initial center, in order to target the penalty towards a suitable center, with the objective to reduce the unwanted shrinkage on the nonzero coefficients, instead of uniformly shrinking the whole coefficient function. As empirical evidence of the interpretability and predictive power of our method, we provide a simulation study and two real world spectroscopy applications with both classification and regression.
\end{abstract}

\section{Introduction} 
Smooth and highly collinear data generally arises in applications where high frequency acquisition devices are employed, like chemometrics, spectroscopy and electrical engineering. A natural way of modeling these data-generating processes is by means of functional data analysis (FDA) \citep{book_fda,book_nonparafda}, where each covariate can be seen as a smooth function $x \in L^{2}(I)$ that has been evaluated at sequential timesteps, often with missing values and different spacing between the observations. Without loss of generality, we will focus on scalar-valued functions defined on $I=[0,1]$, but this approach can be extended to deal with vector valued functions defined on multidimensional domains. The two main aspects to be considered are the estimation of the underlying functional covariates from the raw data, and the estimation of the predictive model itself. Let $\mathcal{D}=\lbrace (x_i,y_i) \rbrace_{i=1}^{N}$ be the training set with random i.i.d. functions $x_i \in L^{2}(I)$ and responses $y_i \in \mathbb{R}$, we focus on the scalar on function linear model:

\begin{equation*}
\label{eq:linearmodel}
y_{i} = \beta_{0} + \int_{I} x_i(t)\beta(t)dt + \epsilon_{i}
\end{equation*}

\noindent where $\beta_{0} \in \mathbb{R}$ is the intercept, $\beta$ is the coefficient function and $\epsilon_{i} \sim \mathcal{N}(0,\sigma^{2})$ are random i.i.d. errors. In fact, the penalties that we analyze can also be employed in a generalized linear model (GLM) framework, and we will study classification problems with functional logistic regression as well. Regarding the first aspect, consider the functional covariate $x$ as a finite expansion $x(t)= \sum_{j=1}^{J} \xi_{j} \psi_{j}(t)$ with suitable basis functions $\psi_{j}$ and coefficients $\xi_{j}$, a common approach is to recover each functional sample $x_i$ individually, by means of interpolating or smoothing splines, depending on the amount of noise. If the raw data contains a large number of missing observations, this approach fails as some of the functional covariates may have been observed on just a few points over the domain. This issue can be solved by leveraging the information from the whole dataset, estimating the basis and the coefficients of the expansion by means of functional principal components (fPCA) with local smoothing \citep{fda_sparselongdata} or mixed effects \citep{fda_smoothsplnested,fda_pcasparse}. Once the input functions have been recovered, depending on the approach that has been implemented, it is possible to either work directly with the coefficients of the expansion, or to evaluate the estimated functions on the same dense equispaced $p$-dimensional grid. Without loss of generality, in this work we opt for the grid approach and we recover the functions individually, but the methods that we propose are not directly tied to this choice, as long as the discretized functional samples have the same dimensionality. Regarding the aspect of estimating the predictive model, the coefficient function $\beta$ is also expressed as a  basis expansion, with some form of regularization as an identifiability constraint, given that the theoretical functional linear model is ill-posed. A parsimonious approach is to restrict the number of basis functions, by either fixing a known suitable basis like a Fourier basis, or by considering only the first $K$ eigenfunctions of the covariance operator obtained from fPCA \citep{fda_pred}, which for any given $K$ explains most of the variation of the input functions in the $L^{2}$ sense. On the opposite side of the spectrum, another approach is instead to employ a rich enough basis while at the same time including some form of penalization, typically an $L_{2}$ penalty on $\beta$ or its derivatives in order to impose smoothness \citep{fda_splineflm,fda_splineerrors,fda_smoothsplines,fda_rkhs}, but $L_{1}$-based penalties have also been used \citep{fda_L1reg}. Note that the restricted basis and the penalization approaches are not mutually exclusive, and hybrid techniques have been proposed as well \citep{psplinesgenreg,fda_flirti,fda_sparseest}. In our setting we choose to adopt the penalization approach, by using the following simple grid basis with $p$ dense and equispaced knots placed on the $p$ evaluations corresponding to the evaluation grid of the estimated input functions:

\begin{equation*}
  \beta(t) = \sum_{j=1}^{p} \beta_{j} b_{j}(t)  \hspace{1.5cm}
  b_{j}(t) = 
  \begin{cases}
  1 & \hspace{0.1cm} \text{if} \hspace{0.2cm} \frac{j-1}{p} < t \leq \frac{j}{p} \\
  0 & \hspace{0.1cm} \text{otherwise} 
  \end{cases}
\end{equation*}

\noindent which is a common solution that enables us to use any multivariate method for the numerical estimation, allowing for a proper comparison between different approaches, as the initial FDA preprocessing is shared between all the tested methods. The main objective of this work is to propose an adaptive penalization approach that is able to fit smooth and sparse coefficient functions \citep{chapter_fdasparsity}, ideally being able to recover the regions of the domain in which the covariates have no effect on the response, while at the same time allowing for a smooth behaviour if needed. Given the abundance of $p>>N$ applications with different requirements, it is no surprise that the literature on variable selection in linear models has experienced a significant growth in both the statistical and machine learning communities. What appears to be the most successful framework is based on the well known penalized least squares formulation (in the multivariate notation), and in particular the bridge estimator \citep{bridge}:

\begin{equation*}
\label{eq:bridge}
\underset{\scaleto{\beta \in \mathbb{R}^{p}}{6pt}}{\scaleto{min}{7pt}} \:\:\: \sum_{i=1}^{N} \left( y_{i} - x_{i}^{\top} \beta \right)^{2} + \lambda \sum_{j=1}^{p} |\beta_{j}|^{\gamma}
\end{equation*}  

\noindent where $\gamma>0$, and $\lambda>0$ that controls the strength of the penalization (we omit the intercept). It is known that for $\gamma<1$ this yields a non-convex optimization problem, where in particular for $\gamma \rightarrow 0$ the bridge reduces to best subset selection \citep{asymplasso}. Besides the computational issues, subset selection methods are also known to be unstable \citep{heuristicsinstability}, and for this reason we will focus only on penalty-based approaches, although we are aware of the different stepwise algorithms. Moreover, given that the penalties are not scale-invariant, we will assume that the input data has been standardized. When $\gamma \geq 1$ the problem is instead convex but we pay the price of unwanted shrinkage of the coefficients, which introduces bias. For $\gamma=1$ in particular we obtain the lasso \citep{lasso}, which is a convex relaxation of best subset selection, while $\gamma=2$ corresponds to ridge regression \citep{ridge}. Regarding our specific setting, which deals with high dimensional and highly collinear data, it is not clear which  approach to adopt, as the lasso may exclude important variables from the model and produce nonsmooth coefficient functions, the ridge may yield both nonsparse and nonsmooth ones, while the usual FDA roughness penalty may be too smooth and fail to recover any sharp change in the support of the coefficient function $\beta$. A possible solution is to impose hybrid penalties, like in the case of the elastic net \citep{elasticnet} or the smooth lasso \citep{smoothlasso}. Our proposed approach is instead exclusively based on the nonzero centered $L_{2}$ penalty \citep{ridgeprior,ridgefusion,genridgeinvcov,targetedridge}, and it is also inspired by the adaptive ridge estimator and other reweighted bias reduction techniques, as we will discuss in Section 2. Section 3 describes our method in detail, the applications are shown in Section 4, with concluding remarks in Section 5.

\section{Related Work}
As previously introduced, a significant issue that follows from the convex formulations of the bridge estimator ($\gamma \geq 1$) is that in order to perform variable selection, we inevitably end up with unwanted shrinkage of the "true" coefficients. This is even worse for $\gamma>1$, where the amount of shrinkage increases with the magnitude of the coefficient being estimated \citep{asymplasso}. Moreover, it is known that except when the OLS coefficients are exactly zero, the ridge is not able to yield sparse solutions, although the coefficients can get arbitrarly small for larger values of $\lambda$. It follows that when estimating a sparse model with ridge regression, there could be the need to employ some form of manual thresholding, setting to zero the smaller coefficients while at the same time accepting the overshrinkage of the larger ones, with a tradeoff between $\lambda$ and the threshold. In practice, lasso is usually the preferred choice when sparsity is sought after, as it is able to set coefficients to exactly zero by acting as a soft thresholding operator. However, an issue with the lasso is the fact that the optimal $\lambda$ with respect to prediction gives inconsistent results from the point of view of variable selection \citep{varsellassograph}, and for functional data in particular, the irrepresentable condition \citep{lassoirrepr} is likely to be violated, given that the curves often have high autocorrelation and the response may depend only on a subset of the domain. The main motivation behind our work is the idea of reducing bias by coefficient-wise adaptive tuning of the penalization. While the ordinary ridge regression corresponds to a sphere centered at the origin, which shrinks all the coefficients uniformly towards zero, \cite{ridge} already introduced a generalized form of ridge regression, which allowed to shrink each coefficient individually, resulting in an ellipsoid. In the usual penalized least squares formulation, the generalized ridge can be expressed as:

\begin{equation*}
\label{eq:genridge}
\underset{\scaleto{\beta \in \mathbb{R}^{p}}{6pt}}{\scaleto{min}{7pt}} \:\:\: \sum_{i=1}^{N} \left( y_{i} - x_{i}^{\top} \beta \right)^{2} + \sum_{j=1}^{p}  \lambda_{j}\beta_{j}^{2}
\end{equation*} 

\noindent where the parameters $\lambda_{j}>0$ control the amount of shrinkage on the corresponding coefficients $\beta_{j}$. This type of penalty is also known as the adaptive ridge estimator, and regardless of the loss function, it has been shown to be equivalent to the lasso, in the sense that they recover the same solution \citep{chapter_lassoridge} \citep{chapter_outcomeslassoridge}. In principle one would like to optimize with respect to both $\beta \in \mathbb{R}^{p}$ and $\lambda \in \mathbb{R}^{p}$, but globally the problem is nonconvex, and the adaptive ridge estimator uses an EM approach that is guaranteed to converge to a local optimum. Instead of alternating optimization, other methods are based on iterative refinements of an initial solution $\tilde{\beta} \in \mathbb{R}^{p}$, and we will refer to such methods as two-stage or multi-stage approaches. One of the first is the non-negative garrote (NNG) \citep{nngarrote}, which is closely related to the EM adaptive ridge and has the following formulation:

\begin{equation*}
\label{eq:nngarrote}
\underset{\scaleto{c \in \mathbb{R}^{p}}{6pt}}{\scaleto{min}{7pt}} \:\:\: \sum_{i=1}^{N} \left( y_{i} - \sum_{j=1}^{p} c_{j} x_{ij} \tilde{\beta}_{j} \right)^{2} + \lambda \sum_{j=1}^{p} c_{j}
\end{equation*}
\begin{center}
$ s.t. \:\:\:\:\:\:
c_{j}\geq 0 
$
\end{center}

\noindent where $\lambda >0$ and the fitted coefficients are recovered as $\hat{\beta}_{j} = \hat{c}_{j} \tilde{\beta}_{j}$. The original NNG was initialized with the OLS solution $\tilde{\beta}=\beta^{ols}$, but other works experimented with other initial estimators like the ridge, the lasso, and the elastic net for high dimensional scenarios \citep{onthenngarrote}. On a side note, the NNG was also the inspiration for the original lasso paper \citep{book_sls}. A generalization of the NNG (without the sign constraint) is the adaptive lasso \citep{adalasso}, which assumes a known weight vector $w \in \mathbb{R}^{p}$ and  solves:

\begin{equation*}
\label{eq:adalasso}
\underset{\scaleto{\beta \in \mathbb{R}^{p}}{6pt}}{\scaleto{min}{7pt}} \:\:\: \sum_{i=1}^{N} \left( y_{i} - x_{i}^{\top} \beta \right)^{2} + \lambda \sum_{j=1}^{p} w_{j} |\beta_{j}|
\end{equation*} 

\noindent with \( w_{j}= 1/|\beta_{j}^{ols}|^{\gamma} \), $\gamma>0$ and $\lambda >0$ selected by cross-validation. This is also a two-stage approach and the final coefficients $\hat{\beta}_{j}$ can be computed by setting $\tilde{x}_{ij}=x_{ij}/w_{j}$, solving a lasso problem with $\tilde{x}_{i}$ as input data, and finally recovering the coefficients as $\hat{\beta}_{j}=\tilde{\beta}_{j}/w_{j}$, with $\tilde{\beta}$ the solution of the previous lasso problem. As for the NNG, the initial estimator is not restricted to the OLS and the ridge is suggested in case of collinearity. Another reweighted estimator is the broken adaptive ridge (BAR) \citep{bar}, which is a multi-stage approach that starts from a ridge penalized solution $\hat{\beta}^{0}$ and at each iteration refines the previous one $\hat{\beta}^{k} = g ( \hat{\beta}^{k-1} )$, with $\hat{\beta}^{*} = \lim_{k \to \infty} \hat{\beta}^{k}$ and

\begin{equation*}
\label{eq:bar}
g ( \tilde{\beta} ) = \underset{\scaleto{\beta \in \mathbb{R}^{p}}{6pt}}{\scaleto{arg \:\, min}{10pt}} \:\:\: \sum_{i=1}^{N} \left( y_{i} - x_{i}^{\top} \beta \right)^{2} + \lambda \sum_{j=1}^{p} \beta_{j}^{2}/\tilde{\beta}_{j}^{2} 
\end{equation*} 

\noindent The subsequent iterations share the same $\lambda$, which is fixed starting from $k=1$ and not tuned individually at each step. Instead, the initial solution $\hat{\beta}^{0}$ is not necessarily obtained with the same $\lambda$ and could be further tuned, although empirically the BAR estimator was found to be insensitive to the initial value. All the methods that we have discussed share the common idea of using multiplicative weights in order to reduce bias, but in fact this is not the only viable approach. Nonconcave penalties like the SCAD \citep{scad} and the MCP \citep{mcp} are both based on quadratic splines with singularities at the origin, giving rise to nonconvex optimization problems that depend on different parameters and are often regarded as unstable, although more refined optimization algorithms have been proposed \citep{onestepsparse} \citep{nonconvexpenoptalgo}. Our approach is instead based on a convex formulation, but it is worth considering both the (elastic) SCAD and MCP for comparison purposes. Finally, yet another option to reduce bias is the one adopted by the relaxed lasso \citep{relaxedlasso}, which separates the variable selection aspect from the coefficient estimation one by first fitting a standard lasso model, followed by a second lasso but only including the covariates that correspond to the nonzero coefficients, with a relaxation parameter $\phi \in \left( 0,1 \right]$ in order to reduce unwanted shrinkage. Both problems share the same fixed $\lambda$ and therefore this can be done pathwise, unlinke from the adaptive lasso where the initial estimator has already been optimized with respect to $\lambda$, and then $\lambda$ is tuned again for the reweighted problem. In particular, let $\hat{\beta}^{\lambda}$ be the lasso solution for a fixed $\lambda$ and let $\mathcal{A} _{\lambda} = \{ 1\leq j \leq p \: | \: \hat{\beta}^{\lambda}_{j} \neq 0 \}$, the relaxed lasso solution $\hat{\beta}^{\lambda,\phi}$ is obtained by solving:

\begin{equation*}
\label{eq:relaxo}
\underset{\scaleto{\beta \in \mathbb{R}^{p}}{6pt}}{\scaleto{min}{7pt}} \:\:\: \sum_{i=1}^{N} \left( y_{i} - x_{i}^{\top} \{ \beta \mathds{1}_{\mathcal{A} _{\lambda}} \}  \right)^{2} + \phi \lambda \sum_{j=1}^{p} |\beta_{j}|
\end{equation*}  
\begin{equation*}
\{ \beta \mathds{1}_{\mathcal{A} _{\lambda}} \}_{j} = 
\begin{cases}
\beta_{j} & \: \text{if} \:\:\: j \in \mathcal{A} _{\lambda} \\
0 & \: \text{otherwise} 
\end{cases}
\end{equation*}

\noindent Our approach in a way is built on a similar relaxation scheme, but instead of performing variable selection and parameter estimation sequentially, we do it jointly and without directly removing any covariate from the initial model, by employing an adaptive weight function that acts on the center of the penalty. Like the ordinary ridge, our penalty is spherical, but is based on the nonzero centered ridge \citep{ridgeprior}:

\begin{equation*}
\label{eq:nonzerocenteredridge}
\underset{\scaleto{\beta \in \mathbb{R}^{p}}{6pt}}{\scaleto{min}{7pt}} \:\:\: \sum_{i=1}^{N} \left( y_{i} - x_{i}^{\top} \beta \right)^{2} + \lambda \sum_{j=1}^{p} (\beta_{j} - c_{j})^{2}
\end{equation*} 

\noindent where the center of the sphere $c \in \mathbb{R}^{p}$ is provided by the user and $\lambda>0$ is selected by cross-validation. For $\lambda$ and $c$ fixed, let $X \in \mathbb{R}^{N \times p}$ be the design matrix and $Y \in \mathbb{R}^{N}$ the response vector, the solution can be computed in closed form as:

\begin{equation*}
\hat{\beta}^{\lambda,c} = (X^{\top}X +\lambda I)^{-1}(X^{\top}Y +\lambda c)
\end{equation*}

\noindent Moreover, the expected value of this estimator is:

\begin{equation*}
\mathbb{E}_{Y|X} [  \hat{\beta}^{\lambda,c} ] = (X^{\top}X +\lambda I)^{-1}(X^{\top}X \beta +\lambda c) 
\end{equation*}

\noindent and therefore it is unbiased for $c=\beta$, meaning that the true value of the parameter is used as the center of the penalty. Clearly there would not be the need to fit any model if $\beta$ was already known, but this suggests that the bias will be low if the fixed $c$ is a good approximation of the true regression coefficient, and we propose to find $c$ by adaptively reweighting the ridge solution.

\section{Smoothly Adaptively Centered Ridge}
We already discussed some of the similarities between our approach and other known methods, with specific attention to the generalized/adaptive ridge and the nonzero centered ridge. The main downside of the adaptive ridge is that optimizing with respect to the shrinkage parameters yields a nonconvex problem, while for the nonzero centered ridge we need to specify the center of the penalty. Our focus is on the smooth $p>>N$ setting and in particular we will refer to the FDA terminology. We propose a convex formulation that allows for adaptive tuning of the type of shrinkage that is imposed on each region of the domain of the coefficient function $\beta$. Instead of employing a variable shrinkage parameter function like in the adaptive ridge, we uniformly shrink $\beta$ as in the ordinary ridge, while at the same time jointly optimizing the center of the penalty. Let $\tilde{\beta}^{\lambda}$ be the solution of the ordinary ridge for a fixed $\lambda$, we introduce a smooth weight function $w:I \rightarrow \mathbb{R}^{+}$ that acts on $\tilde{\beta}^{\lambda}$ and fit our estimator by solving the following convex problem with linear constraints: 

\begin{equation}
\label{eq:sacr}
\begin{aligned}
\underset{\scaleto{\beta_{0},\beta,w}{6pt}}{\scaleto{min}{7pt}} \:\:\: \sum_{i=1}^{N} \left[ y_{i}  -\beta_{0} - \int_{I}x_{i}(t)\beta(t)dt \right]^{2} + \lambda \phi \int_{I} \big[ \beta(t) -w(t)\tilde{\beta}^{\lambda}(t) \big]^{2}dt \\
+ \: \lambda(1-\phi) \int_{I} \big[ D^{2}w(t)\tilde{\beta}^{\lambda}(t)\big]^{2}dt \\
\end{aligned}
\end{equation} 
\begin{center}
$ s.t. \:\:\:\:\:\:
\begin{cases}
\int_{I} w(t)dt=|I| \\
w(t)\geq 0 
\end{cases}
$
\end{center}

\noindent where $\beta_{0} \in \mathbb{R}$ is the intercept, $\lambda>0$ and $\phi \in \left( 0,1 \right]$ that controls the balance between the two penalty terms. Both $\lambda$ and $\phi$ are selected by cross-validation and the $\lambda$ used in Problem \ref{eq:sacr} is the same as the one used for computing $\tilde{\beta}^{\lambda}$. The first term of the penalty is a nonzero centered ridge that shrinks $\beta$ uniformly towards $w\tilde{\beta}^{\lambda}$, while the second term is a roughness penalty on the center of the previous one. The weight function $w$ can be seen as an adaptive density which either contracts or dilates the initial center $\tilde{\beta}^{\lambda}$, allowing the nonzero centered penalty to selectively shrink $\beta$ towards zero in the regions of the domain that are not correlated with the response, while reducing the unwanted shrinkage in the informative regions. This sparsity inducing behaviour is motivated by the constraints imposed on $w$, which necessarily lead to a tradeoff between inflating and deflating $\tilde{\beta}^{\lambda}$, as proven in Proposition \ref{eq:tradeoff}.

\begin{proposition}
\label{eq:tradeoff}
Let $I \subset \mathbb{R}$ be a closed interval and let $w:I \rightarrow \mathbb{R}^{+}$ be a smooth function such that $\int_{I} w(t)dt=|I|$. Consider the closed subintervals $I_{i} \subset I$ such that $\mu(I_{i}) \neq 0$ and $I_{i} \cap I_{j} = \emptyset$ for $i\neq j$. Partition $I$ in disjoint intervals $I_{<}$, $I_{>}$ and $I_{=}$ such that $I = I_{>} \cup I_{<} \cup I_{=}$ where: 

\begin{equation*}
\begin{aligned}
I_{>} &= \bigcup_{i} I_{i}: \hspace{0.1cm}  w(t)>1  \hspace{0.2cm} \forall t \in I_{i} \\
I_{<}  &= \bigcup_{i} I_{i}: \hspace{0.1cm}  w(t)<1  \hspace{0.2cm} \forall t \in I_{i} \\
I_{=}  &= \bigcup_{i} I_{i}: \hspace{0.1cm}  w(t)=1  \hspace{0.2cm} \forall t \in I_{i}
\end{aligned}
\end{equation*}

then $|I_{>}| \neq 0 \iff |I_{<}| \neq 0$.  

\begin{proof}
From the first mean value theorem for integrals follows that:

\begin{equation*}
\begin{aligned}
\int_{I_{>}} w(t)dt &= w(c_{>})|I_{>}|, \hspace{0.5cm}  c_{>} \in I_{>}^{o}, \hspace{0.5cm} w(c_{>}) >1 \\
\int_{I_{<}} w(t)dt &= w(c_{<})|I_{<}|, \hspace{0.5cm}  c_{<} \in I_{<}^{o}, \hspace{0.5cm} w(c_{<}) <1 \\
\int_{I_{=}} w(t)dt &= w(c_{=})|I_{=}|, \hspace{0.5cm}  c_{=} \in I_{=}^{o}, \hspace{0.5cm} w(c_{=}) =1 \\
\end{aligned}
\end{equation*}

\noindent therefore, from $I = I_{>} \cup I_{<} \cup I_{=}$ results that:

\begin{equation*}
\begin{aligned}
\int_{I} w(t)dt &= \int_{I_{>}} w(t)dt + \int_{I_{<}} w(t)dt + \int_{I_{=}} w(t)dt \\
|I|             &= w(c_{>})|I_{>}| \hspace{0.1cm} + \hspace{0.1cm} w(c_{<})|I_{<}| + |I_{=}| \\
|I|             &= w(c_{>})|I_{>}| \hspace{0.1cm} + \hspace{0.1cm} w(c_{<})|I_{<}| + |I| - |I_{>}| - |I_{<}| \\
0               &= |I_{>}|[w(c_{>})-1] \hspace{0.1cm} + \hspace{0.1cm} |I_{<}|[w(c_{<})-1] \\
\end{aligned}
\end{equation*}

\noindent by construction we know that $[w(c_{>})-1] > 1$ and $[w(c_{<})-1] < 1$, proving that $I_{>}$ and $I_{<}$ are either both null sets or both non-null sets. 

\end{proof}
\end{proposition}

It follows that when the true $\beta$ is provided as $\tilde{\beta}^{\lambda}$, there is no need to either inflate or deflate the center of the penalty, and therefore the optimal $w$ is 1 almost everywhere, leading to an unbiased estimator as it is equal to the nonzero centered ridge with $c=w\tilde{\beta}^{\lambda}=\beta$. In practice there is no guarantee that solving Problem \ref{eq:sacr} with the optimal center will lead to uniform unitary weights, as  $w$ is penalized and jointly estimated with $\beta$ from the data. From the geometrical perspective, the weight function acts as an anisotropic scaling on $\tilde{\beta}^{\lambda}$, which has the effect of adaptively moving the center of the penalty. This has the advantage of nonuniform shrinkage between the coefficients, as in the adaptive ridge with its ellipsoidal penalty, while at the same time keeping the tractable convex formulation of the spherical penalty, since $\lambda$ is a scalar that is selected by cross-validation. Therefore, the adaptive shrinkage of the coefficient function is the result of an adaptive target, and not of an adaptive shrinkage intensity. Our approach can be described as a continuous way of doing variable selection, which is executed jointly with the estimation of the regression coefficients. As the weight function is not included in the model and is never used for prediction, we are not adding further parameters to the model itself, although we are doubling the parameters to be estimated. Regarding the two terms of the penalty, it is worth noting that since the roughness penalty is imposed on the center of the first term, the coefficient function is only indirectly penalized with respect to its roughness, by being pushed towards an adaptively scaled center. Imposing the roughness penalty on the center itself, instead of on the weight function only, ensures the option of retrieving a smooth centerfunction, with $\phi$ controlling the amount of smoothness, and not just a smooth scaling of $\tilde{\beta}^{\lambda}$. The choice of the ridge solution as the initial center is quite natural, as it is stable and almost always not exactly zero. This latter property of the $L_{2}$ penalty is often regarded as a problem or at least an incovenience, but in our case is instead welcomed, as the weight function is multiplicative and would not be able to inflate an initial zero coefficient. It follows that using a sparse coefficient function as the initial center equals to excluding multiple variables from the model, which is not a problem if the initial zero coefficients should indeed be zero, but is also not necessary to produce sparse or at least interpretable solutions, as the variable shrinkage induced by the weight function should already push to zero the coefficients of the unwanted variables. In practice, higher values of $\lambda$ will correspond to higher shrinkage of the coefficient function towards the centerfunction, where the selected $\lambda$ depends on how suitable the centerfunction is. Therefore, for a sparse and adequate centerfunction, the selected $\lambda$ will be high and the fitted coefficient function can get arbitrarly close to zero where needed, without the tradeoff of the ordinary ridge, where we pay the price of unwanted shrinkage on the nonzero coefficients. 

Until now we only considered the context of regression, but in fact our approach can be generalized to the GLM framework as follows:      

\begin{equation*}
\label{eq:gensacr}
\begin{aligned}
\underset{\scaleto{\beta_{0},\beta,w}{6pt}}{\scaleto{min}{7pt}} \:\:\: J(\beta_{0},\beta,x,y) + Pen_{\lambda \phi}(\beta,w) 
\end{aligned}
\end{equation*} 
\begin{center}
$ s.t. \:\:\:\:\:\:
\begin{cases}
\int_{I} w(t)dt=|I| \\
w(t)\geq 0 
\end{cases}
$
\end{center} 

\noindent where $J$ can be any convex loss function, as in the case of functional logistic regression, and it is independent from the weight function.

With respect to the numerical optimization, as the proposed formulation is quadratic with linear equality and inequality constraints, we opted for interior point methods, which are a class of optimization algorithms that are often regarded as state of the art for these types of problems \citep{iposurvey}. In particular, we employ the solver IPOPT \citep{ipopt} which is based on a primal-dual interior point algorithm with filter line search. The worst-case number of iterations is $O(\sqrt{n})$ with $n$ the number of variables, although interior point methods usually converge in a few steps. At each iteration the dominating cost is $O(n^{3})$ for applying Newton's method in order to solve a system of equations, and therefore the overall worst-case computational cost is $O(n^{3.5})$. In our specific scenario we have $n=2p$, as our formulation doubles the amount of parameters to be estimated.

\section{Applications}
In this section we provide some empirical results of the performance of our method (SACR) in the contex of FDA with $p>>N$. In particular, we show a simulation study and two real world applications, one for classification and one for regression. We compare SACR with multiple penalized methods that are known for inducing sparsity and/or smoothness, like the lasso, adaptive lasso, relaxed lasso, NNG, ridge, BAR, elastic net, elastic SCAD, elastic MCP and the roughness penalty. The base implementations for the lasso, ridge and elastic net are the ones from \textit{scikit-learn} \citep{scikit}, while for the adaptive lasso, relaxed lasso, NNG and BAR we implemented our own wrappers based on those. The elastic SCAD and elastic MCP are available in the
CRAN package \textit{ncvreg} \citep{package_ncvreg}, while we also use our own python implementation of the roughness penalized functional linear model. We interface with the solver IPOPT by modeling the optimization problems with Pyomo \citep{book_pyomo}.

\subsection{Simulation Study}
\begin{figure}[H]
  \centering
    \begin{subfigure}[c]{0.32\textwidth}
      \includegraphics[width=\textwidth]{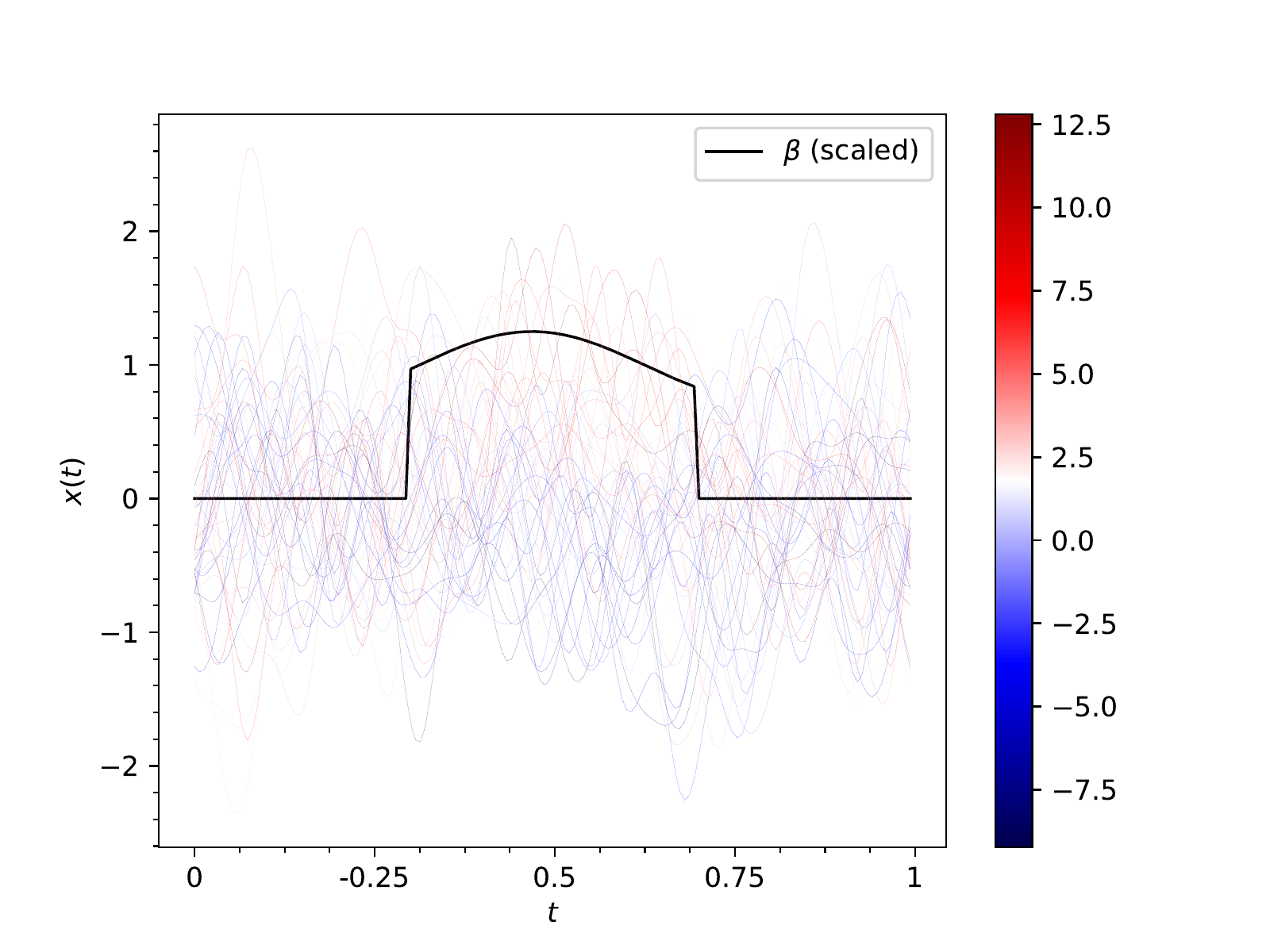}
      \caption{$x_{i}$ - independent}
    \end{subfigure}
    \begin{subfigure}[c]{0.32\textwidth}
      \includegraphics[width=\textwidth]{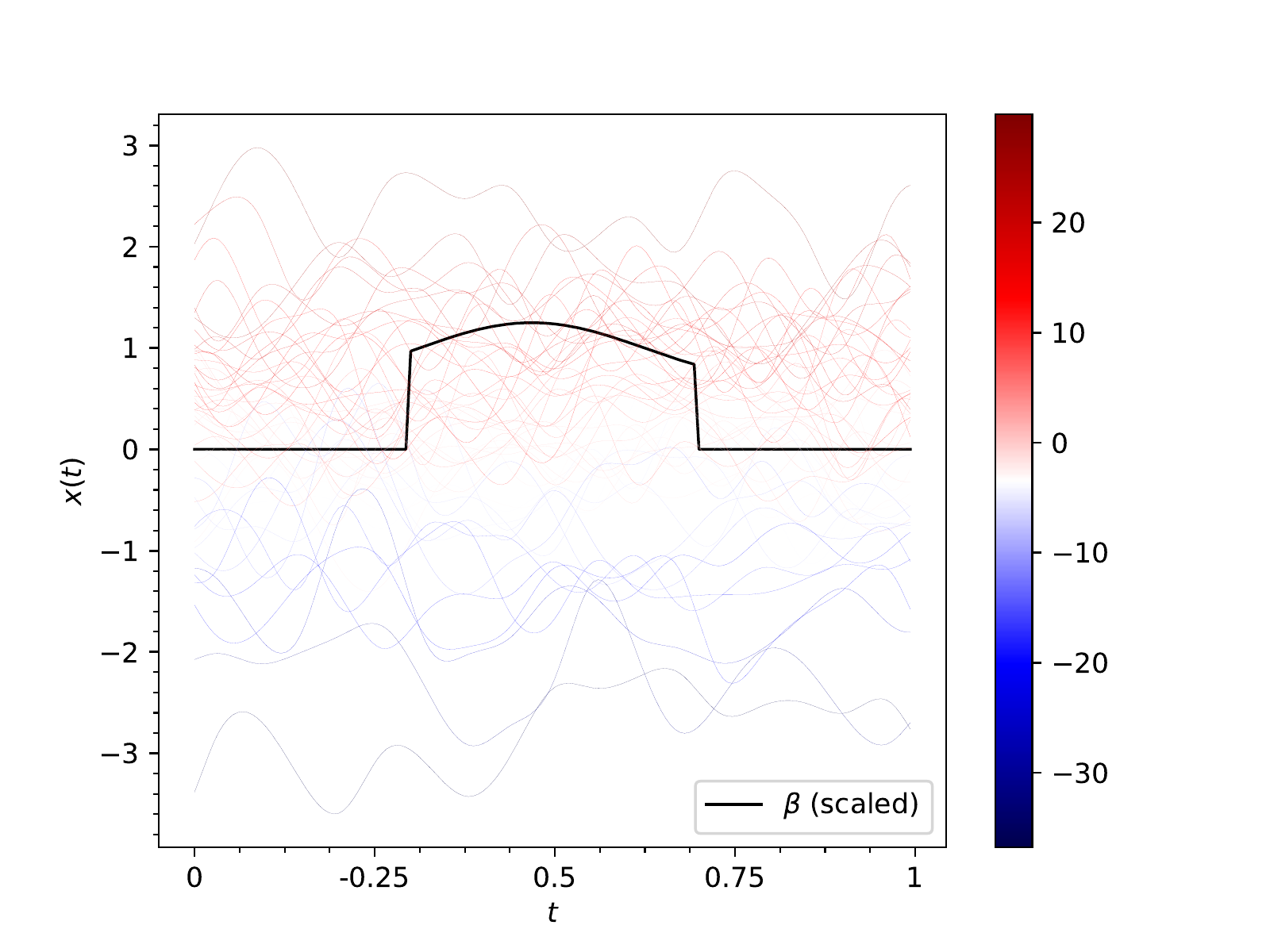}
      \caption{$x_{i}$ - dependent}
    \end{subfigure} 
    \begin{subfigure}[c]{0.32\textwidth}
      \includegraphics[width=\textwidth]{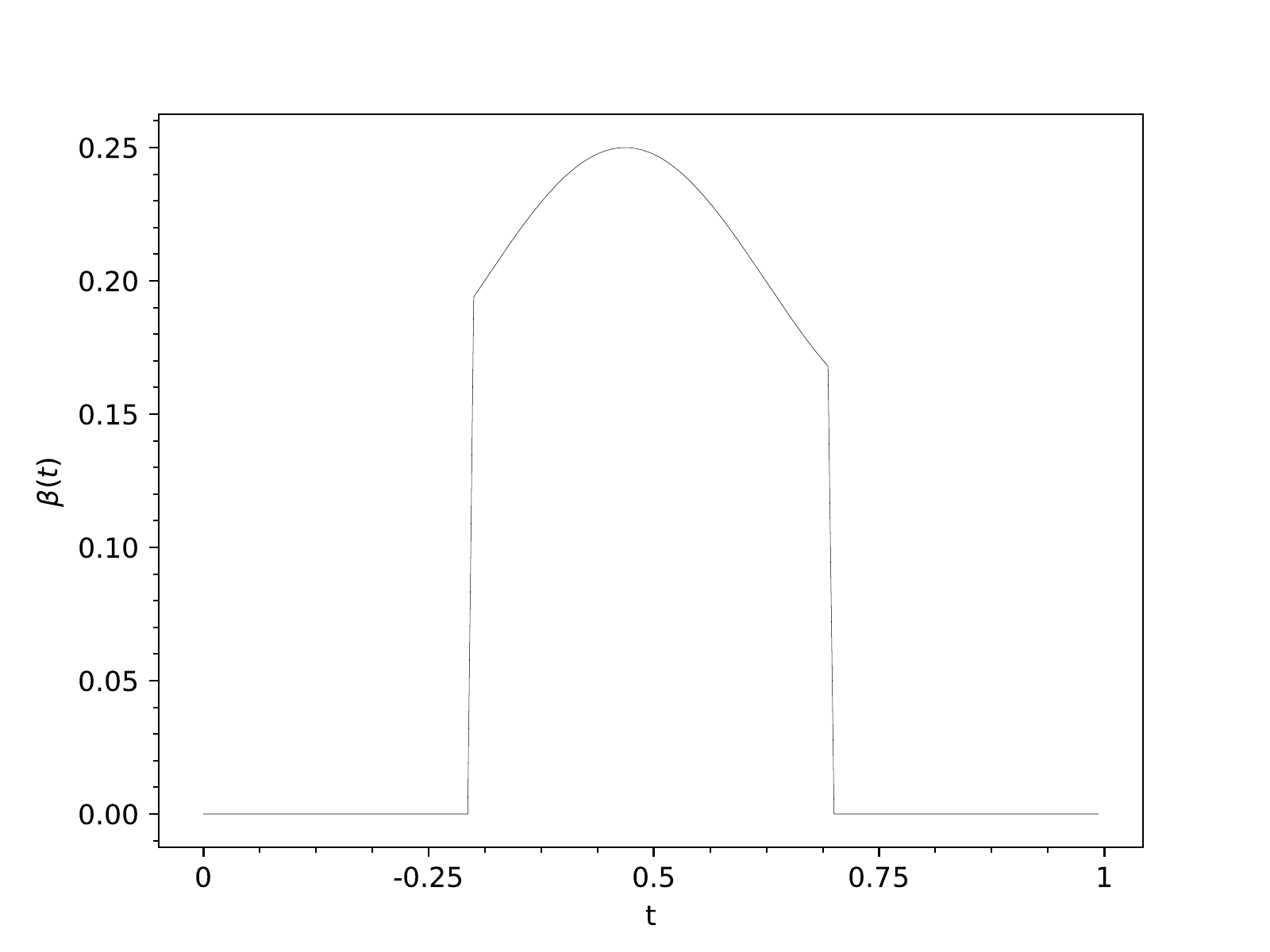}
      \caption{true $\beta$}
    \end{subfigure} 
  \caption{Simulated data with independent and highly dependent spline coefficients, true $\beta$}
  \label{fig:data_simulation}
\end{figure}

\begin{figure}[H]
  \centering
  	\begin{subfigure}[c]{0.22\textwidth}
      \includegraphics[width=\textwidth]{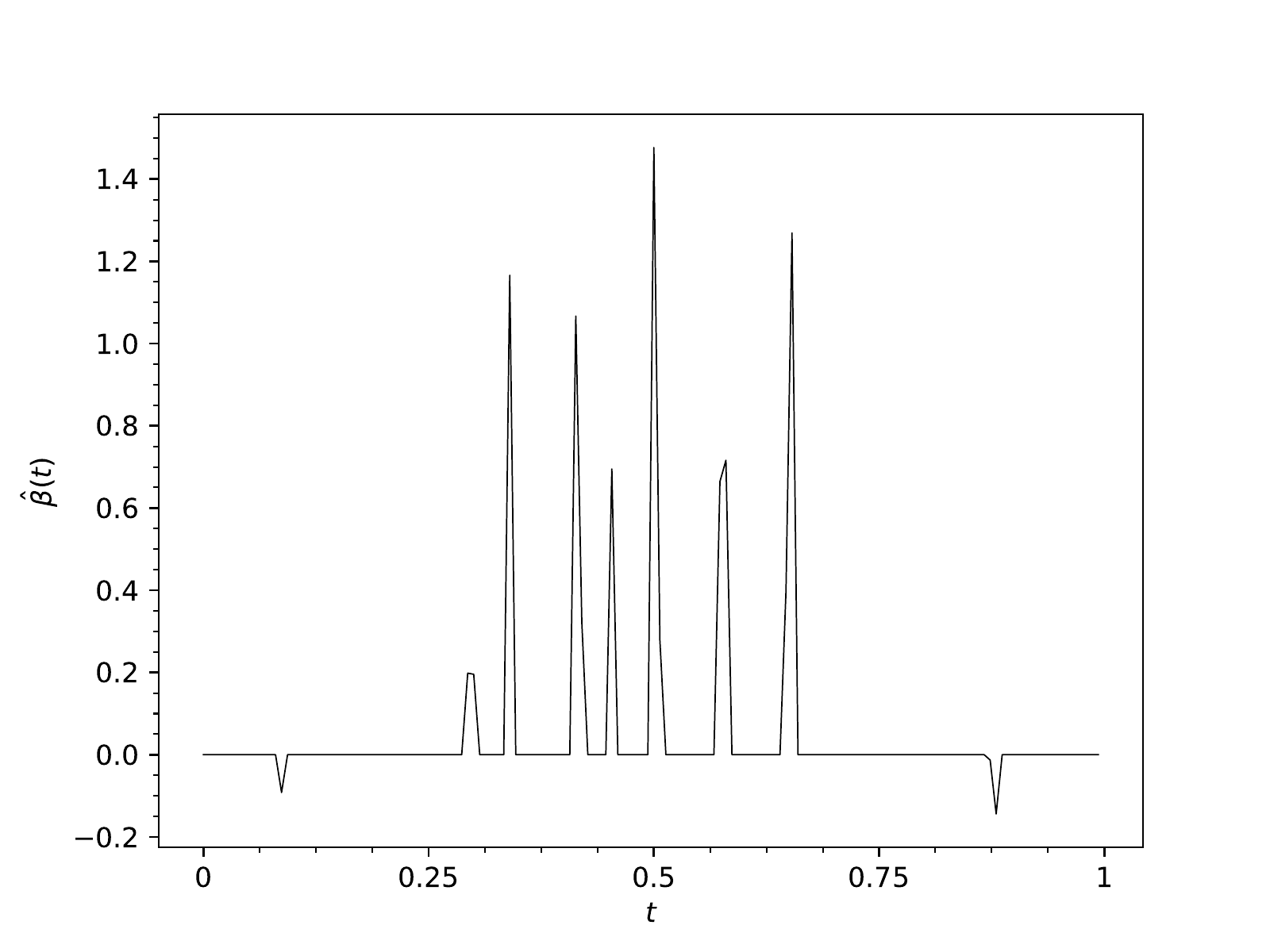}
      \caption{lasso}
    \end{subfigure}
    \begin{subfigure}[c]{0.22\textwidth}
      \includegraphics[width=\textwidth]{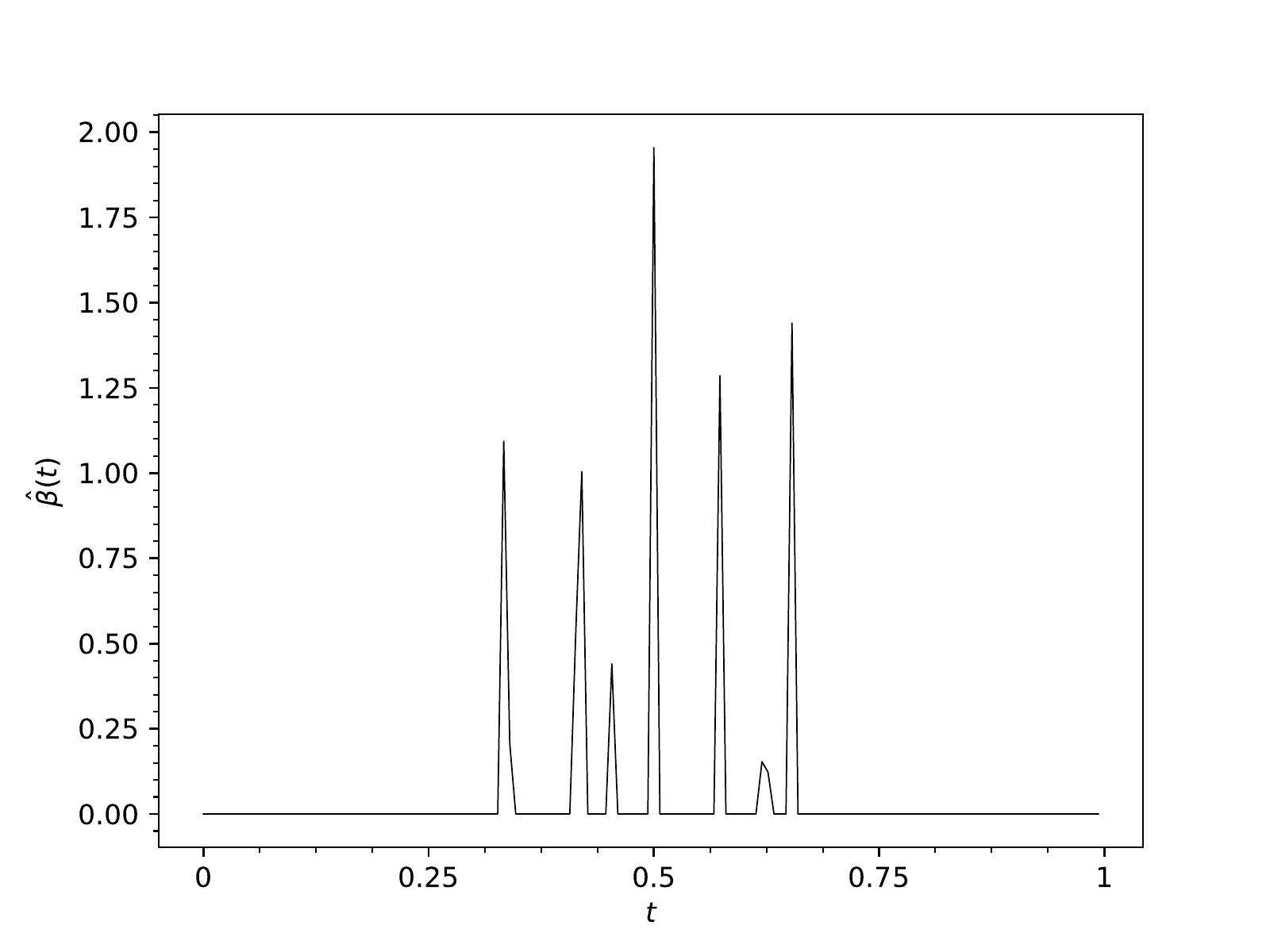}
      \caption{adaptive lasso}
    \end{subfigure}
    \begin{subfigure}[c]{0.22\textwidth}
      \includegraphics[width=\textwidth]{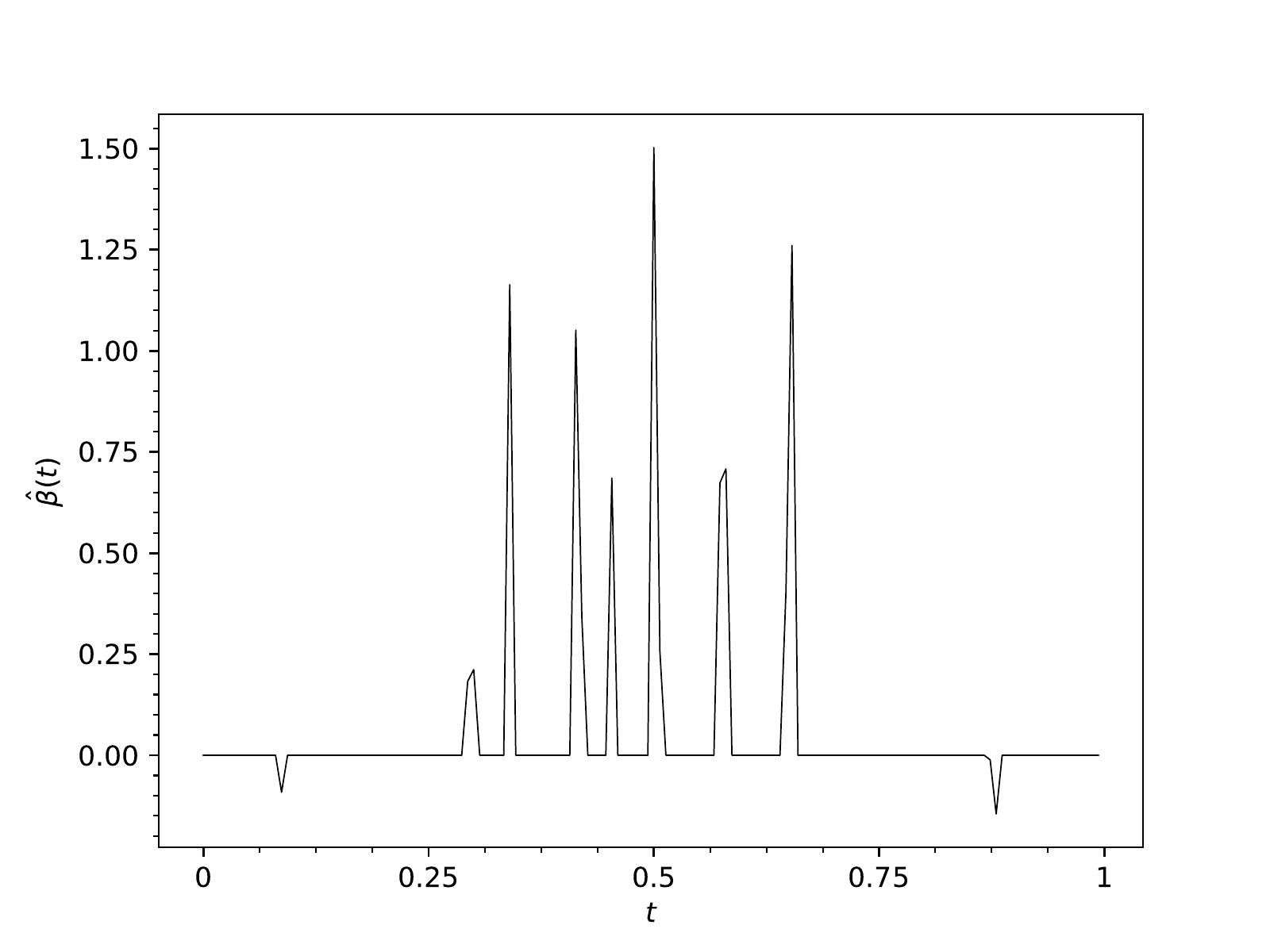}
      \caption{relaxed lasso}
    \end{subfigure}
    \begin{subfigure}[c]{0.22\textwidth}
      \includegraphics[width=\textwidth]{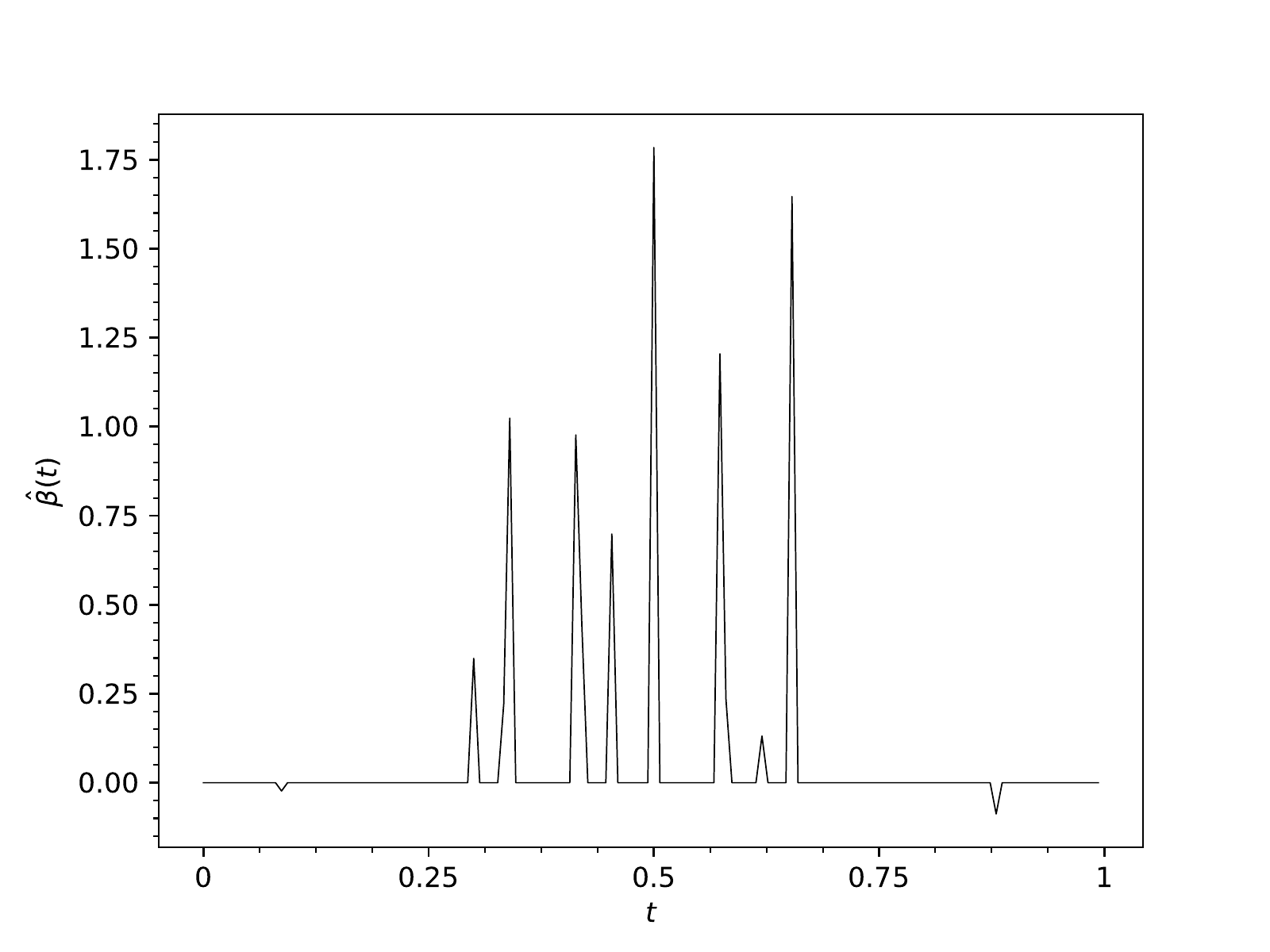}
      \caption{NNG}
    \end{subfigure}

  	\begin{subfigure}[c]{0.22\textwidth}
      \includegraphics[width=\textwidth]{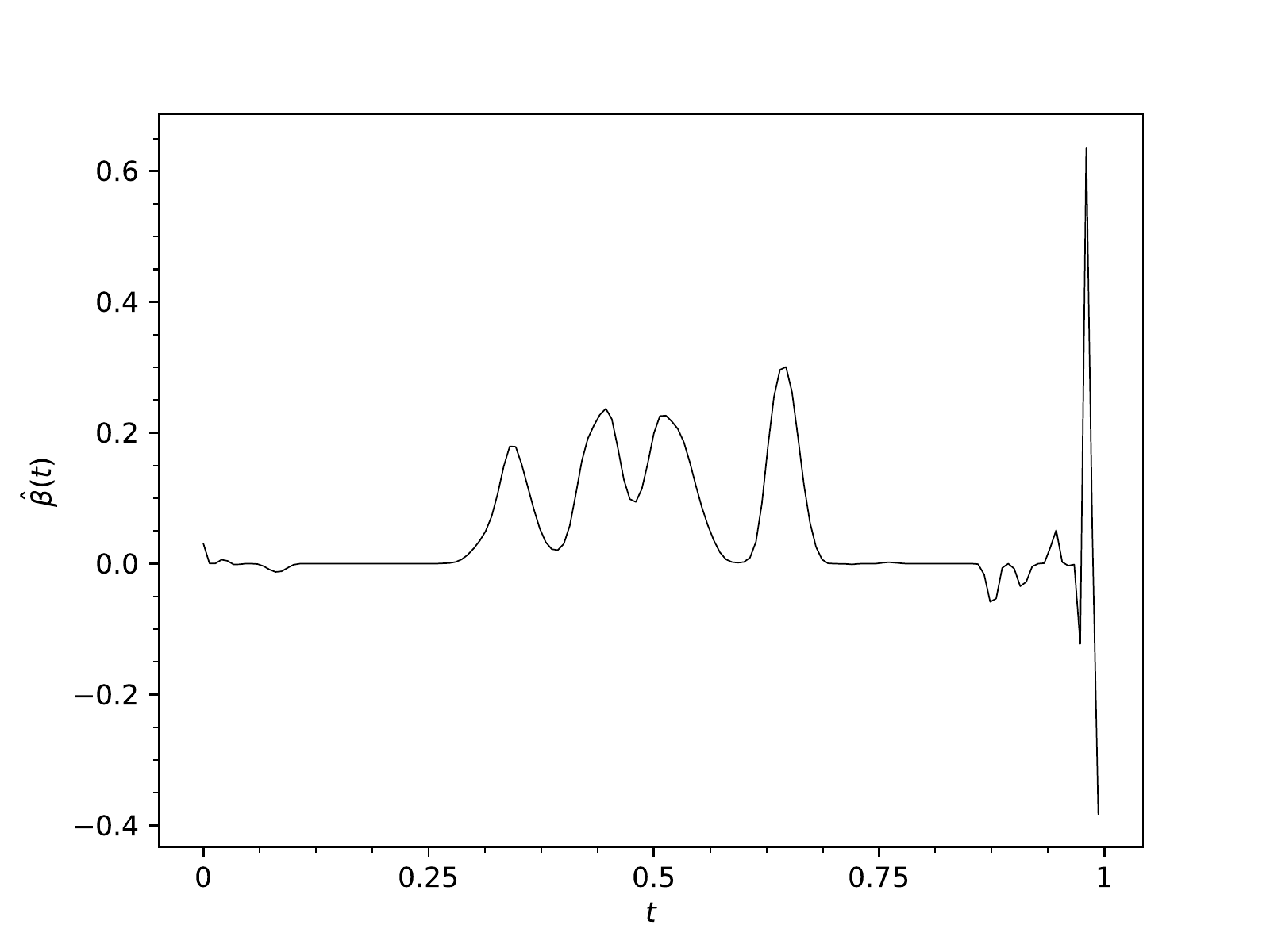}
      \caption{BAR}
    \end{subfigure}
    \begin{subfigure}[c]{0.22\textwidth}
      \includegraphics[width=\textwidth]{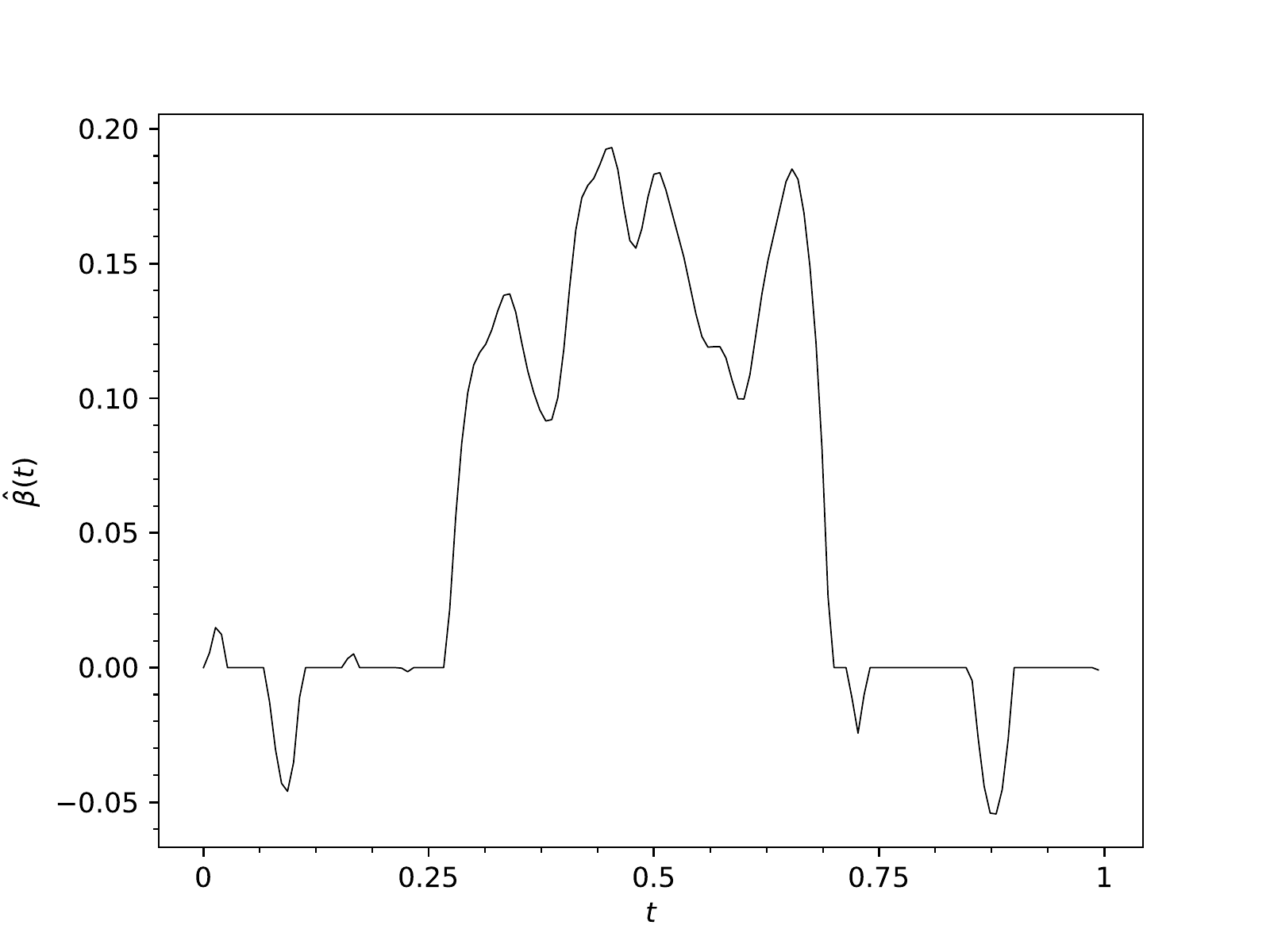}
      \caption{elastic net}
    \end{subfigure}
    \begin{subfigure}[c]{0.22\textwidth}
      \includegraphics[width=\textwidth]{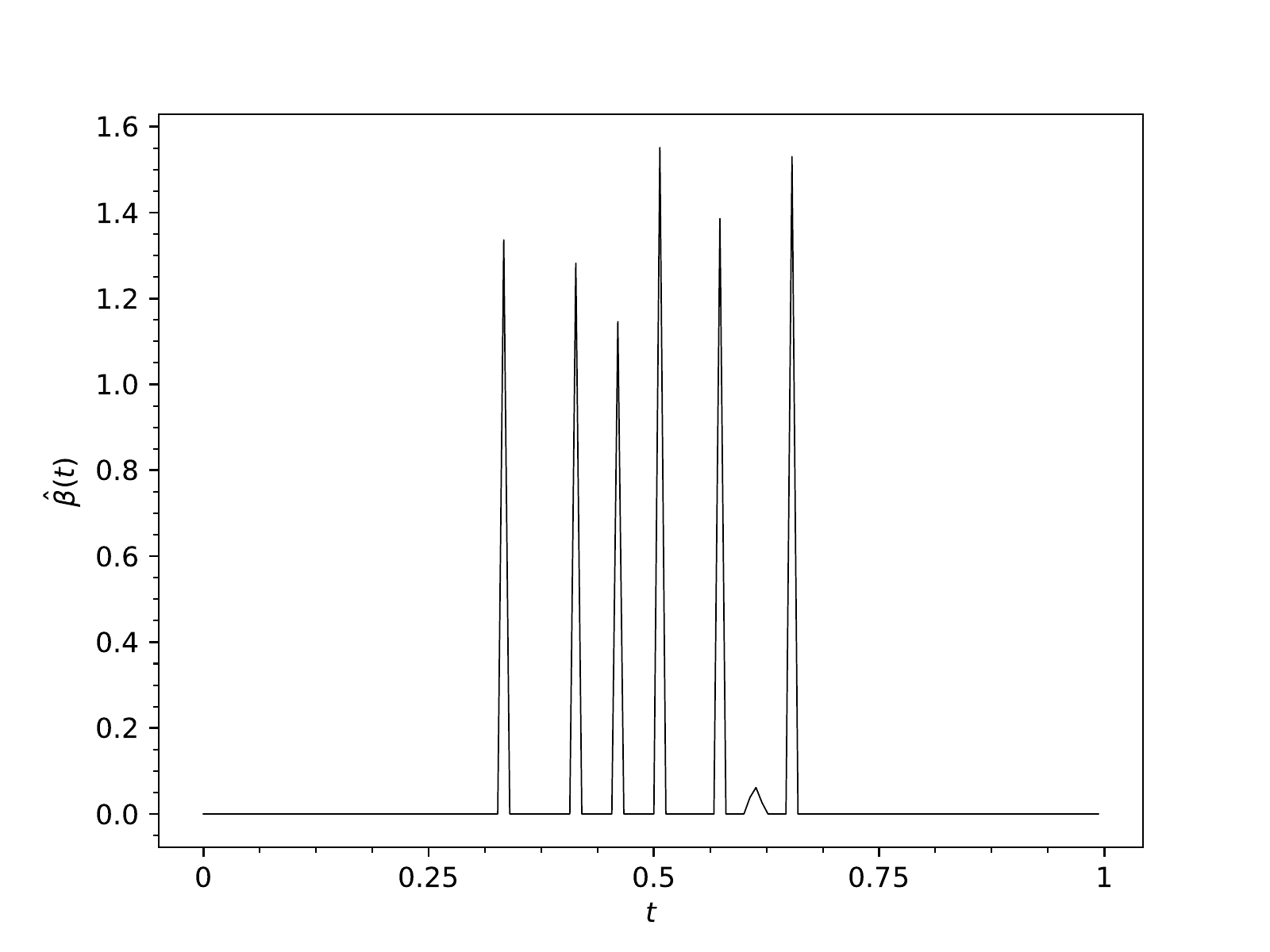}
      \caption{elastic SCAD}
    \end{subfigure}
    \begin{subfigure}[c]{0.22\textwidth}
      \includegraphics[width=\textwidth]{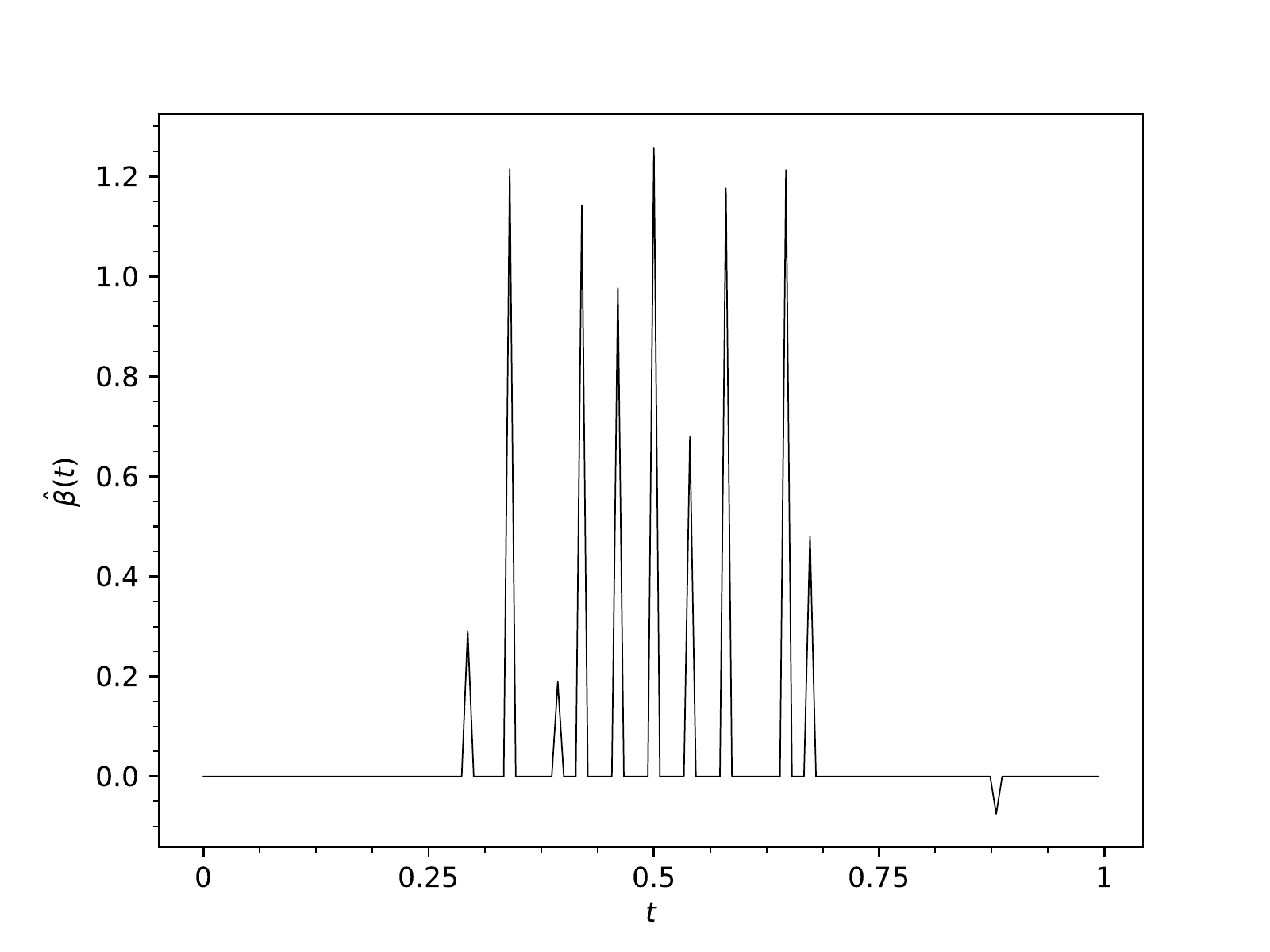}
      \caption{elastic MCP}
    \end{subfigure}

    \begin{subfigure}[c]{0.22\textwidth}
      \includegraphics[width=\textwidth]{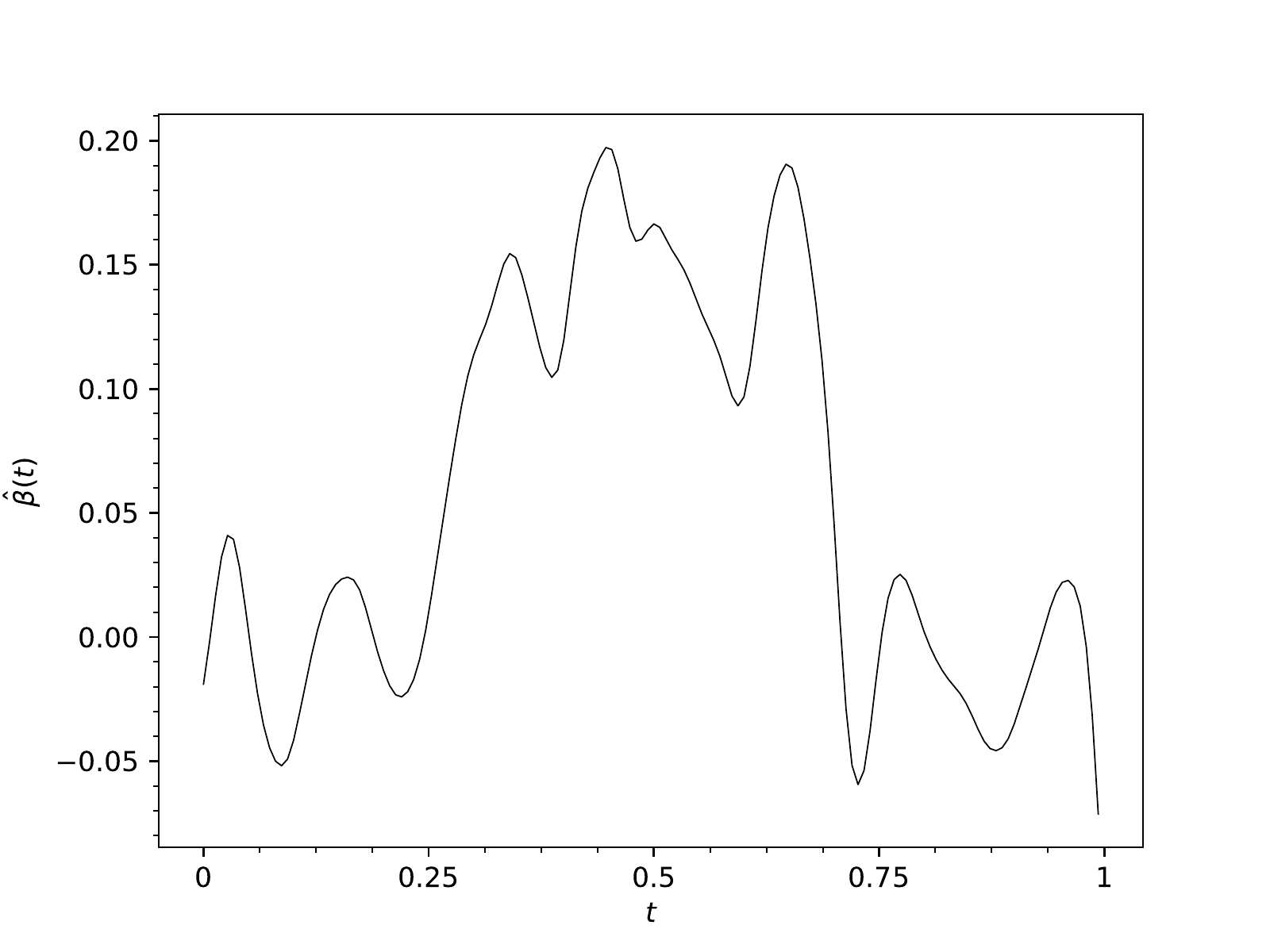}
      \caption{ridge}
    \end{subfigure}
    \begin{subfigure}[c]{0.22\textwidth}
      \includegraphics[width=\textwidth]{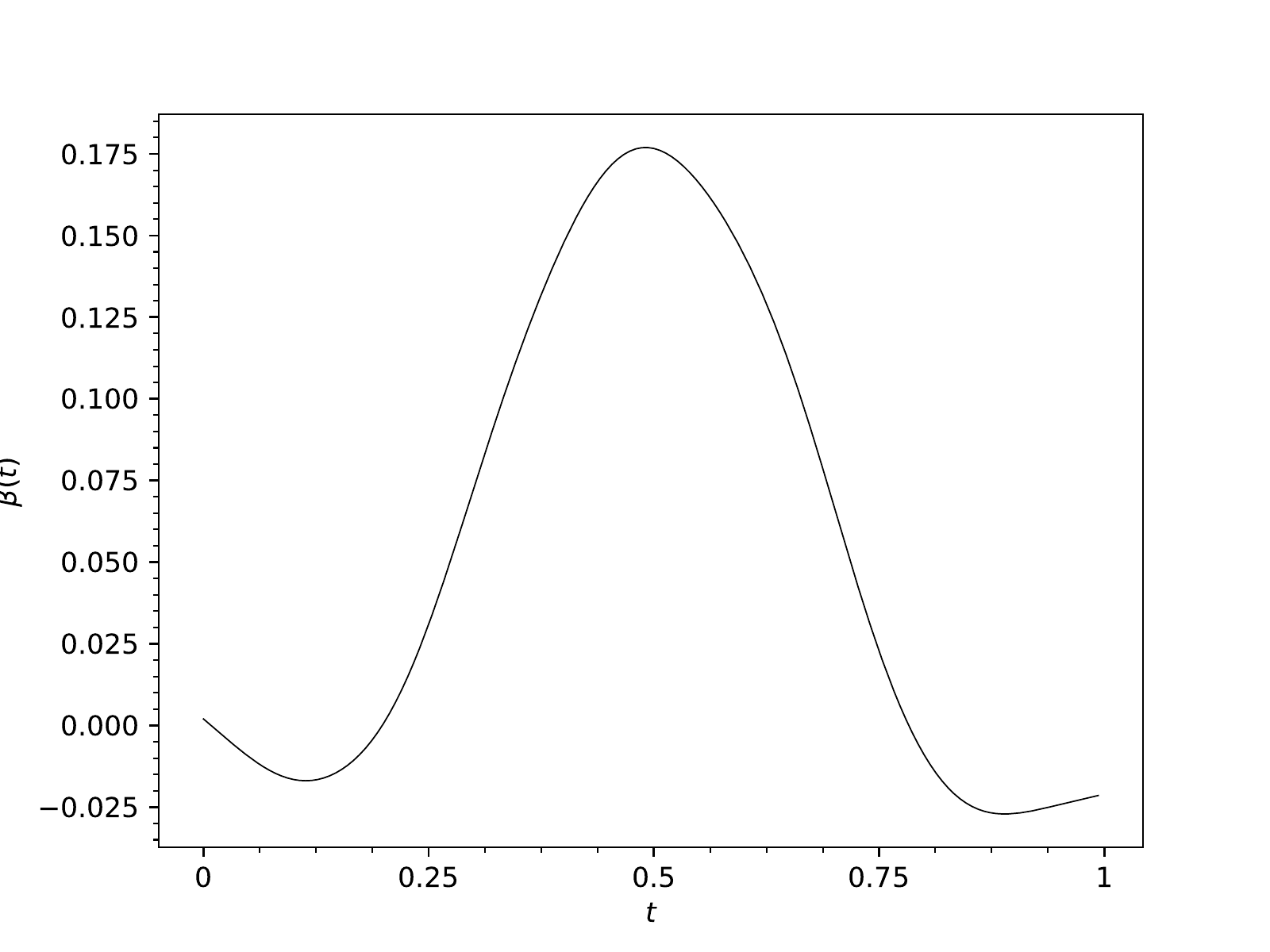}
      \caption{roughness}
    \end{subfigure}
    \begin{subfigure}[c]{0.22\textwidth}
      \includegraphics[width=\textwidth]{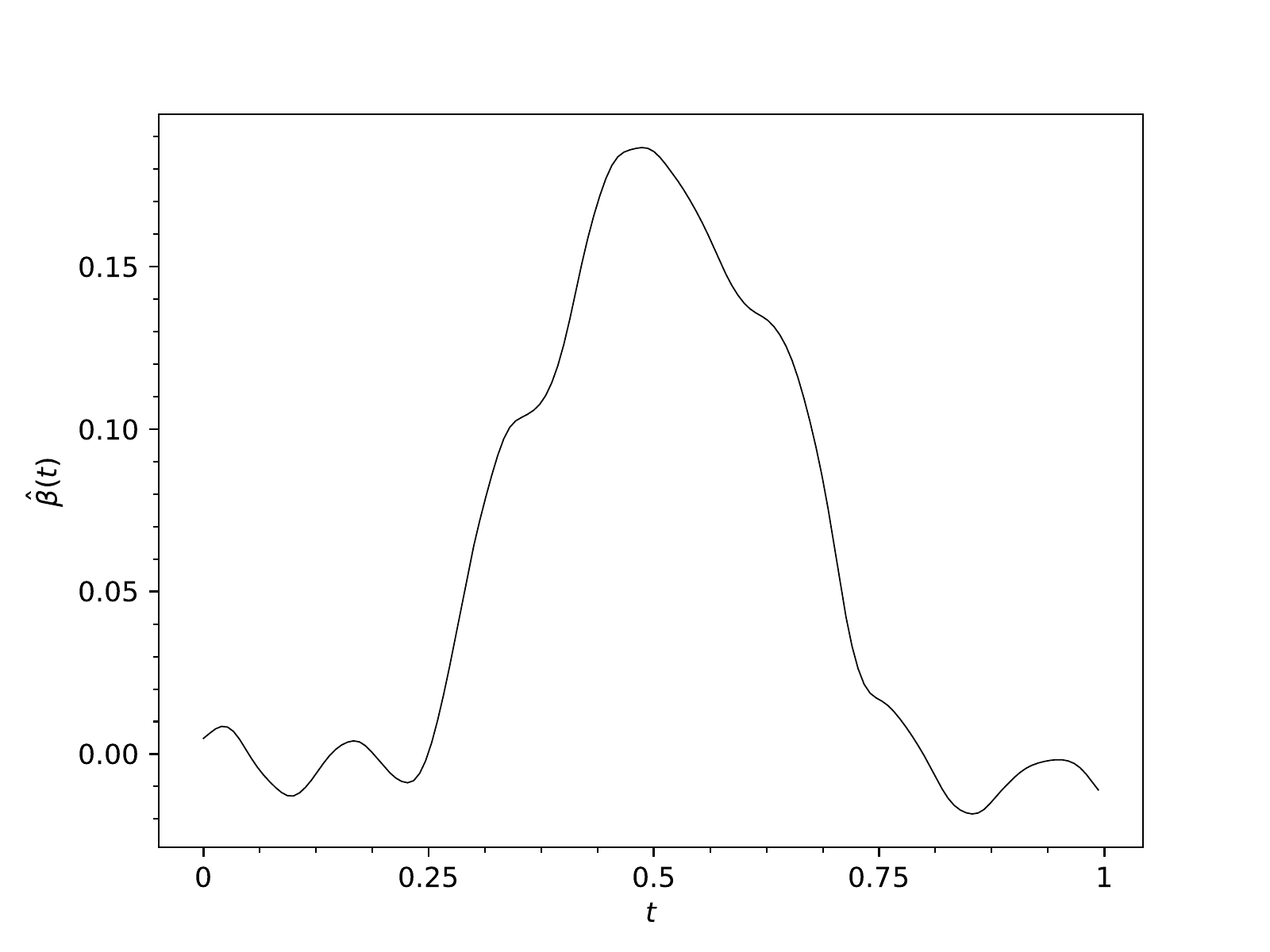}
      \caption{SACR}
    \end{subfigure}
    
  \caption{Independent spline coefficients: fitted coefficient functions $\hat{\beta}$ by penalty type}
  \label{fig:allbetas_ind}
\end{figure}

\begin{figure}[H]
  \centering
    \begin{subfigure}[c]{0.45\textwidth}
      \includegraphics[width=\textwidth]{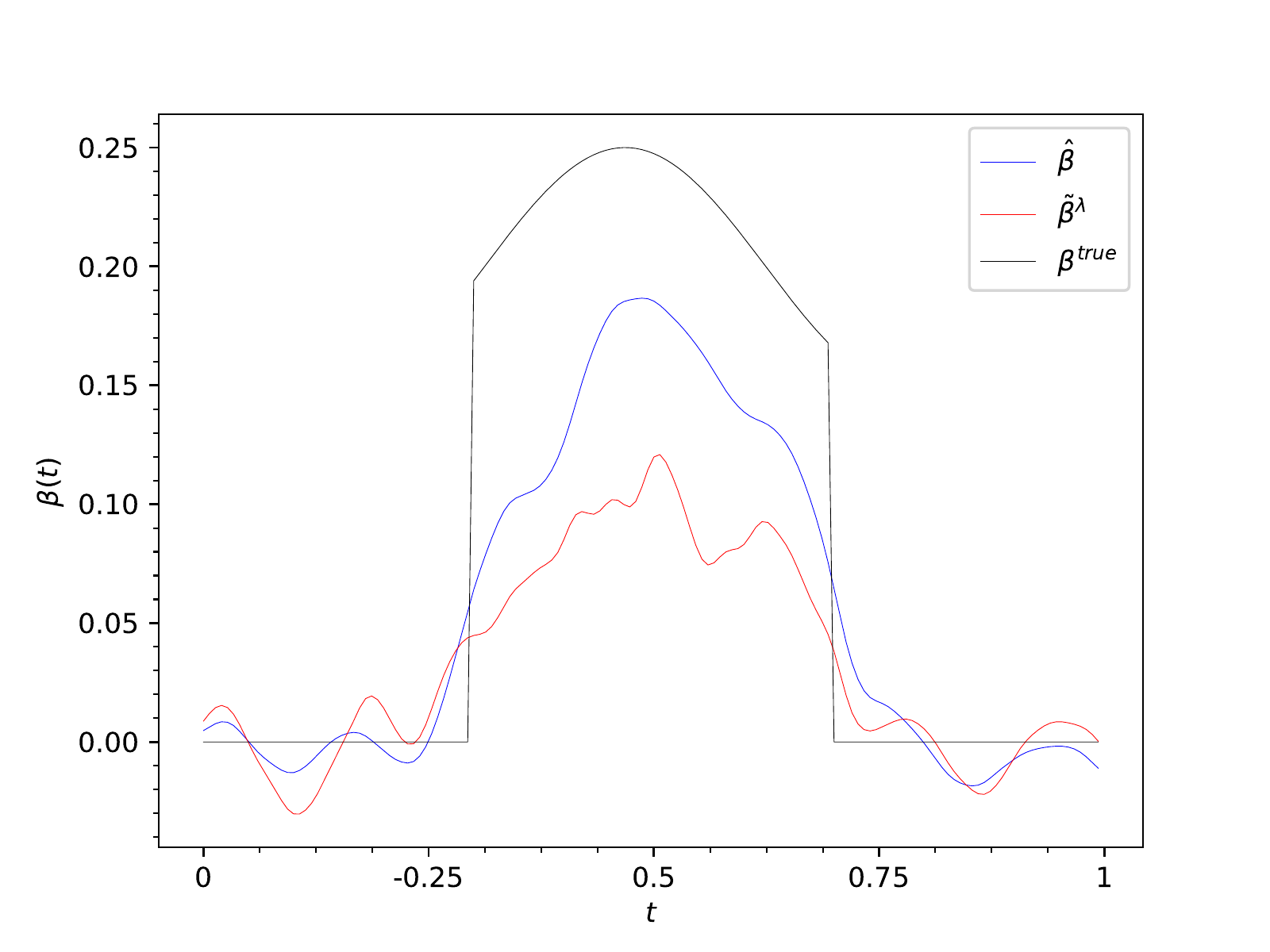}
      \caption{true $\beta$ - initial $\tilde{\beta}^{\lambda}$ - fitted $\hat{\beta}$}
    \end{subfigure}
    \begin{subfigure}[c]{0.45\textwidth}
      \includegraphics[width=\textwidth]{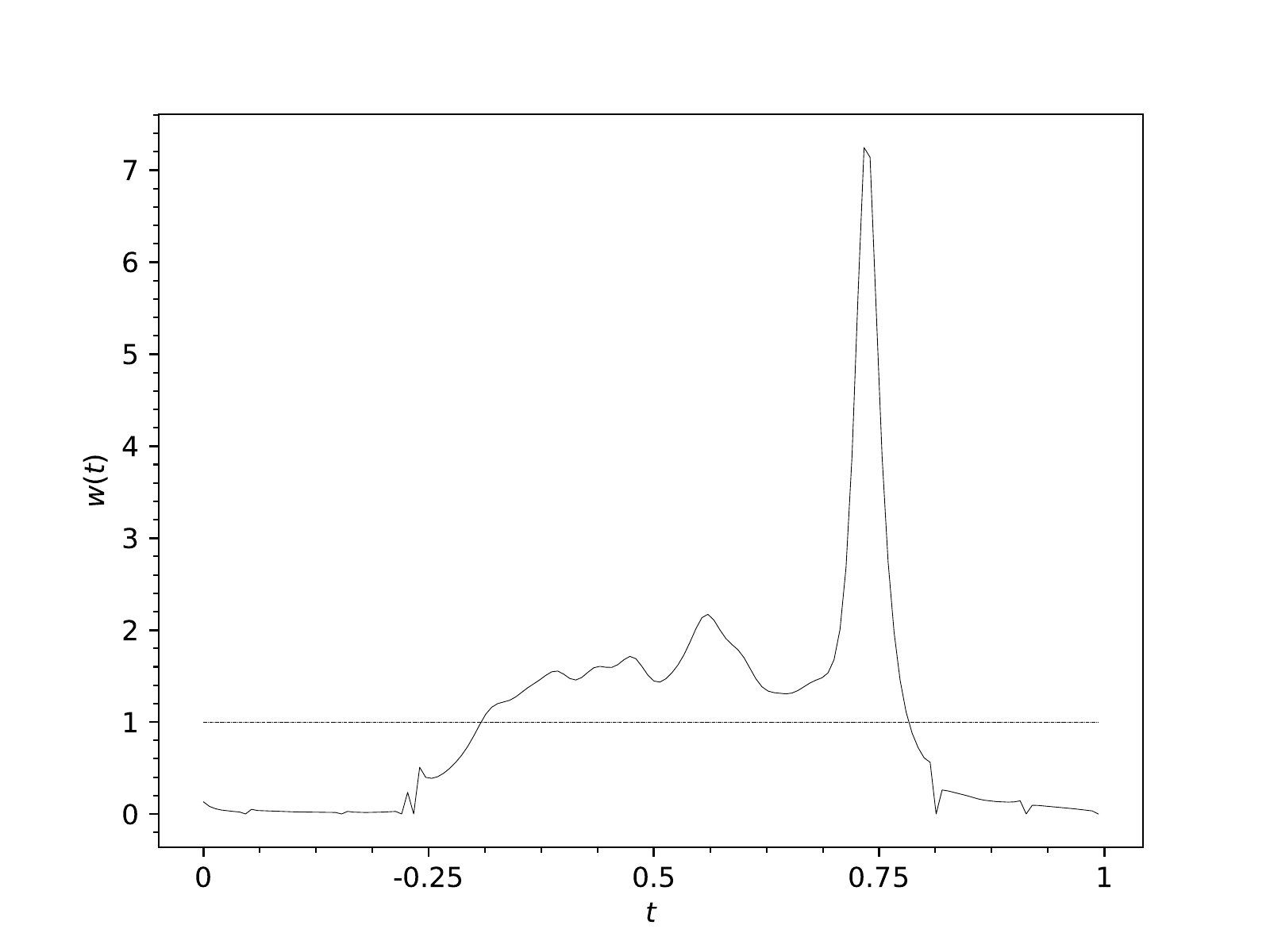}
      \caption{$w$}
    \end{subfigure} 
  \caption{Independent spline coefficients: comparison between true $\beta$, initial centerfunction $\tilde{\beta}^{\lambda}$ and fitted SACR estimator $\hat{\beta}$, with corresponding weight function $w$}
  \label{fig:sacrcomparison_ind}
\end{figure}

\begin{figure}[H]
  \centering
  	\begin{subfigure}[c]{0.22\textwidth}
      \includegraphics[width=\textwidth]{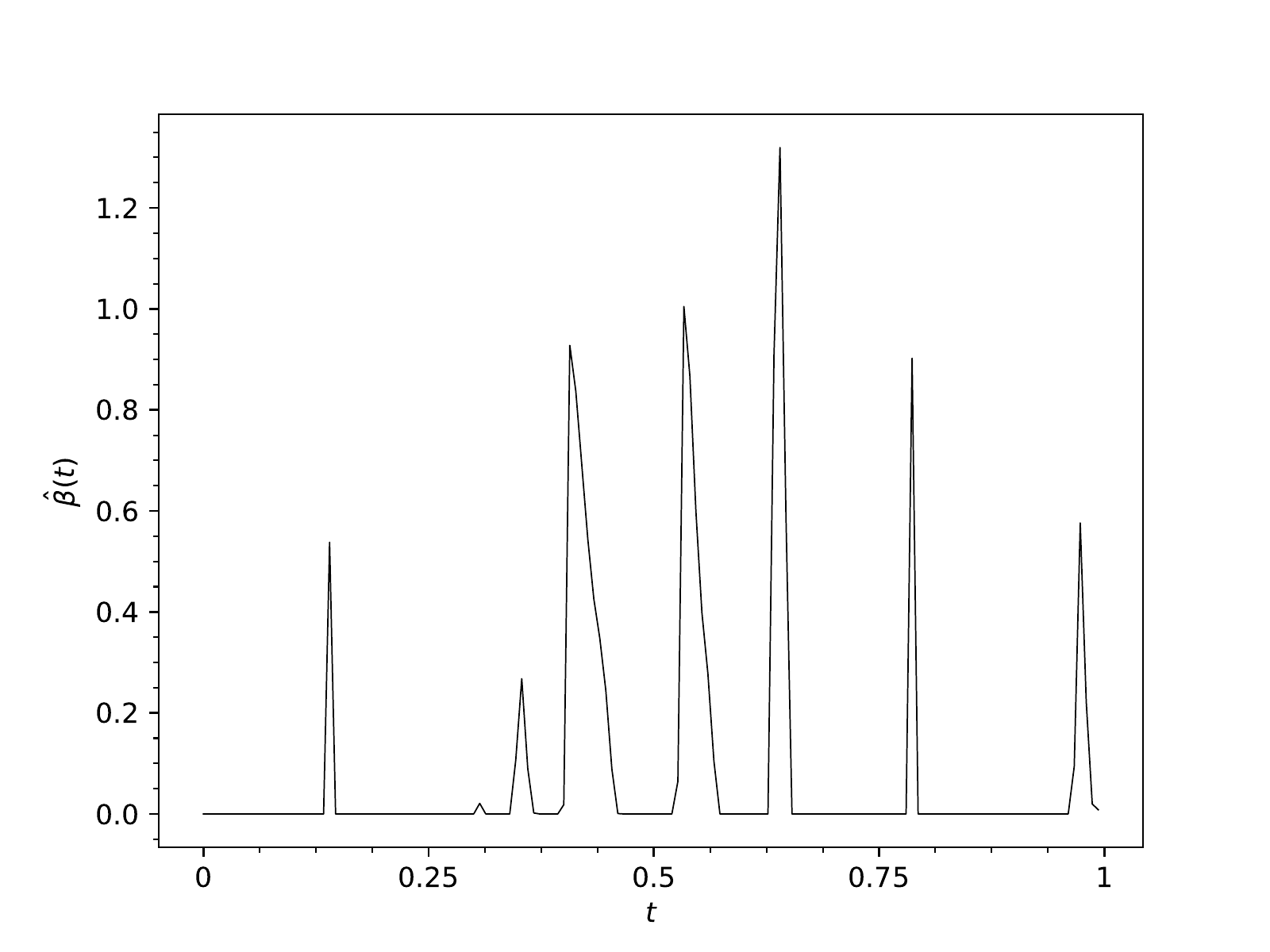}
      \caption{lasso}
    \end{subfigure}
    \begin{subfigure}[c]{0.22\textwidth}
      \includegraphics[width=\textwidth]{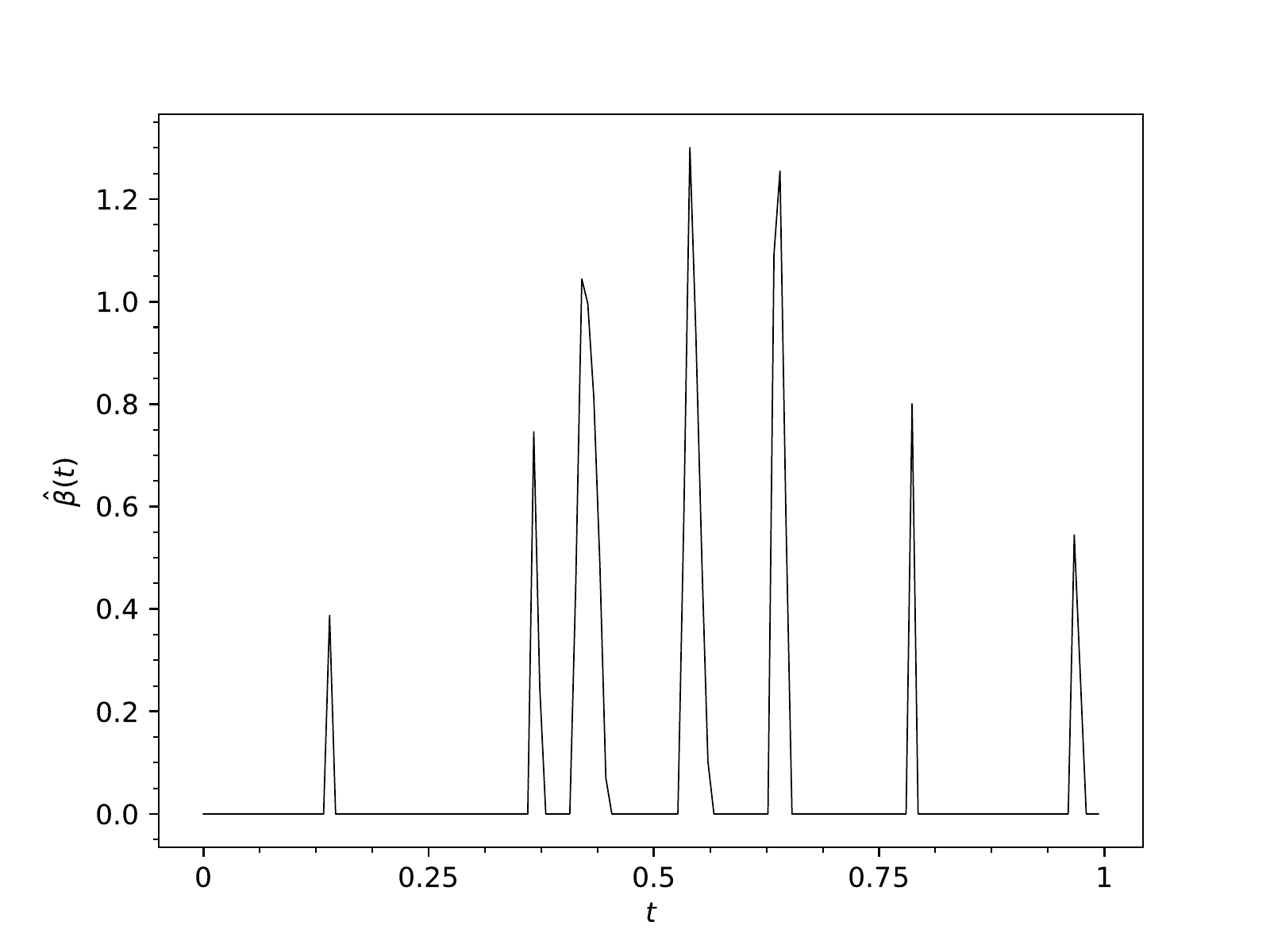}
      \caption{adaptive lasso}
    \end{subfigure}
    \begin{subfigure}[c]{0.22\textwidth}
      \includegraphics[width=\textwidth]{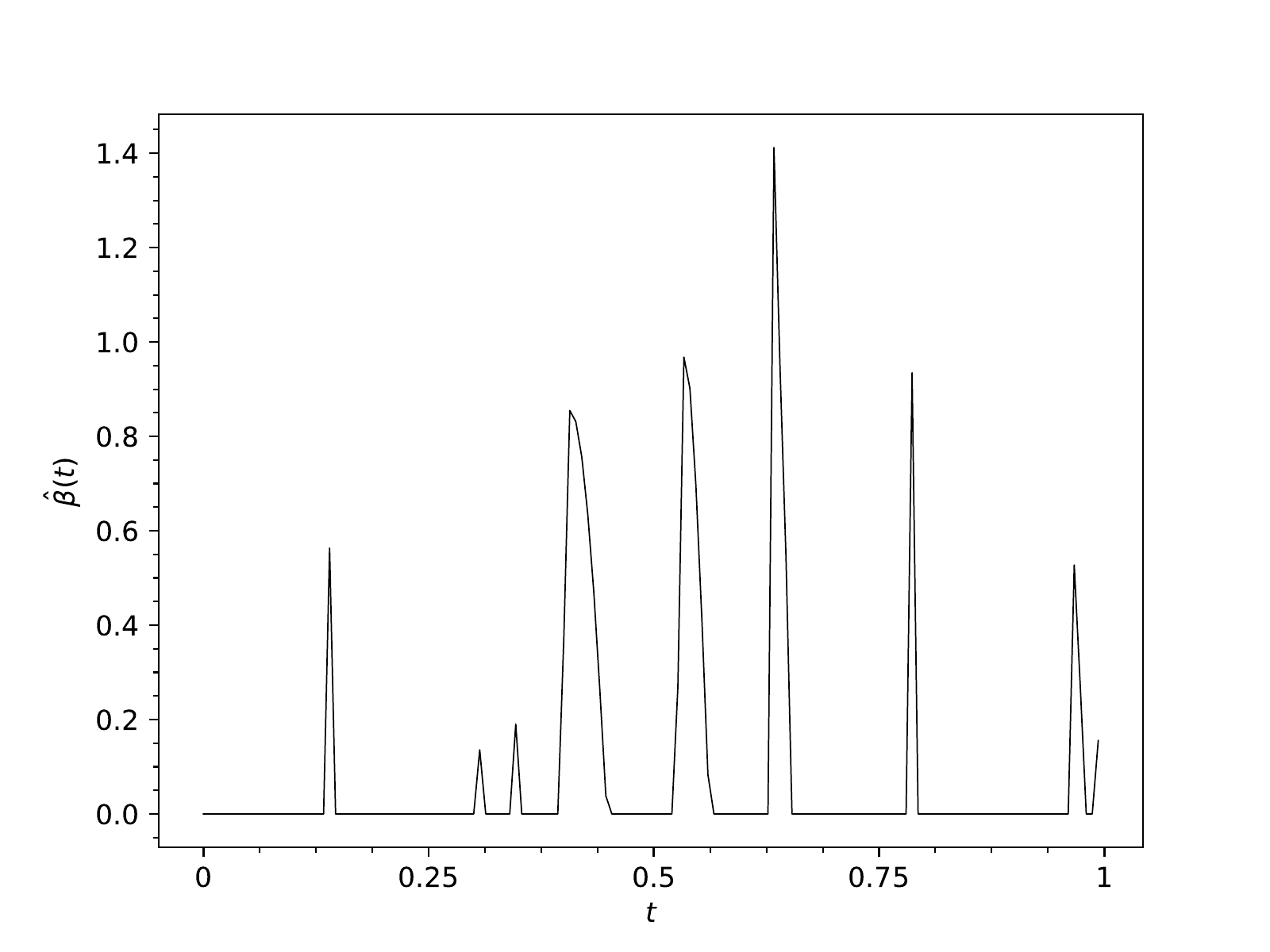}
      \caption{relaxed lasso}
    \end{subfigure}
    \begin{subfigure}[c]{0.22\textwidth}
      \includegraphics[width=\textwidth]{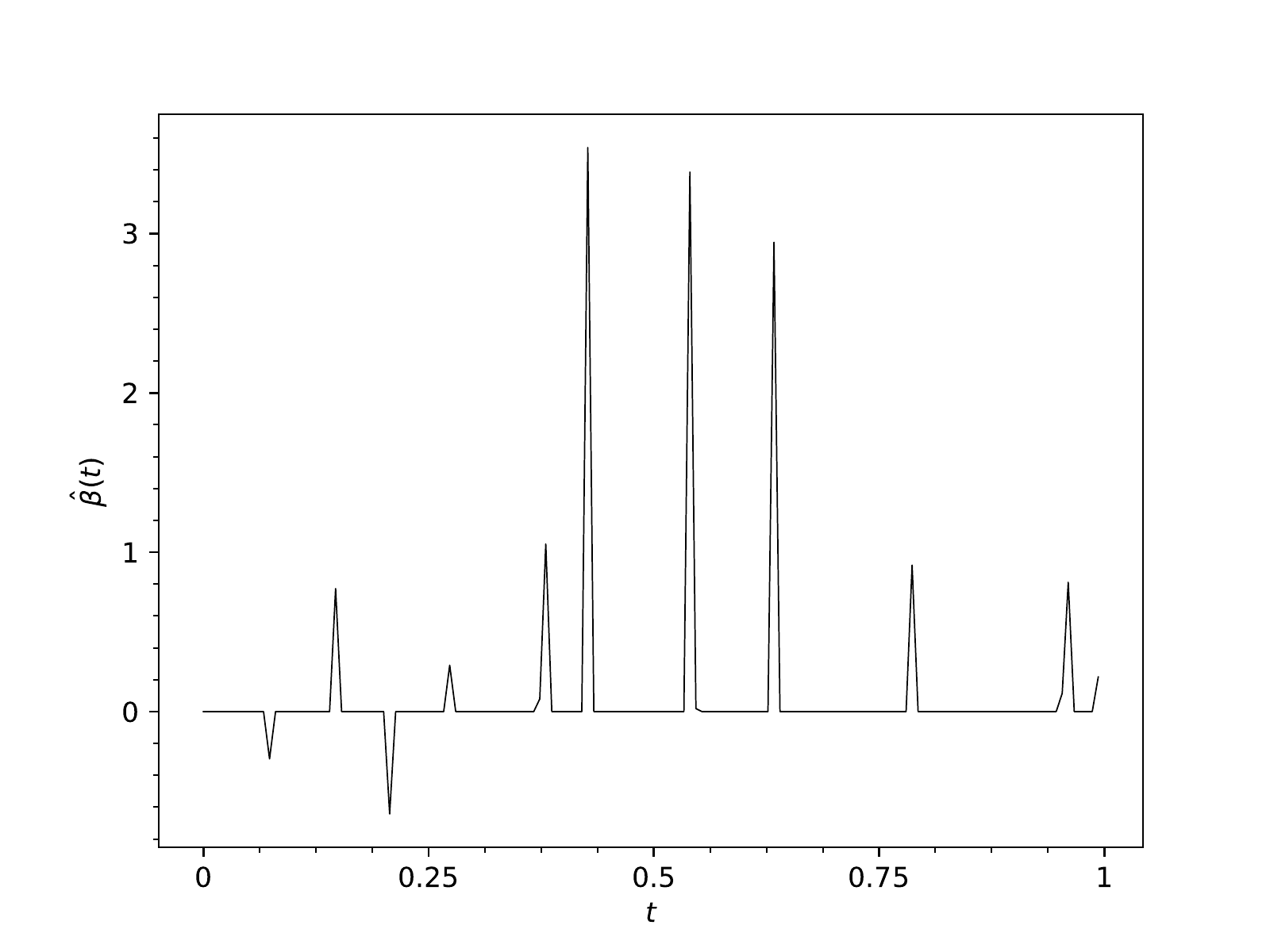}
      \caption{NNG}
    \end{subfigure}

  	\begin{subfigure}[c]{0.22\textwidth}
      \includegraphics[width=\textwidth]{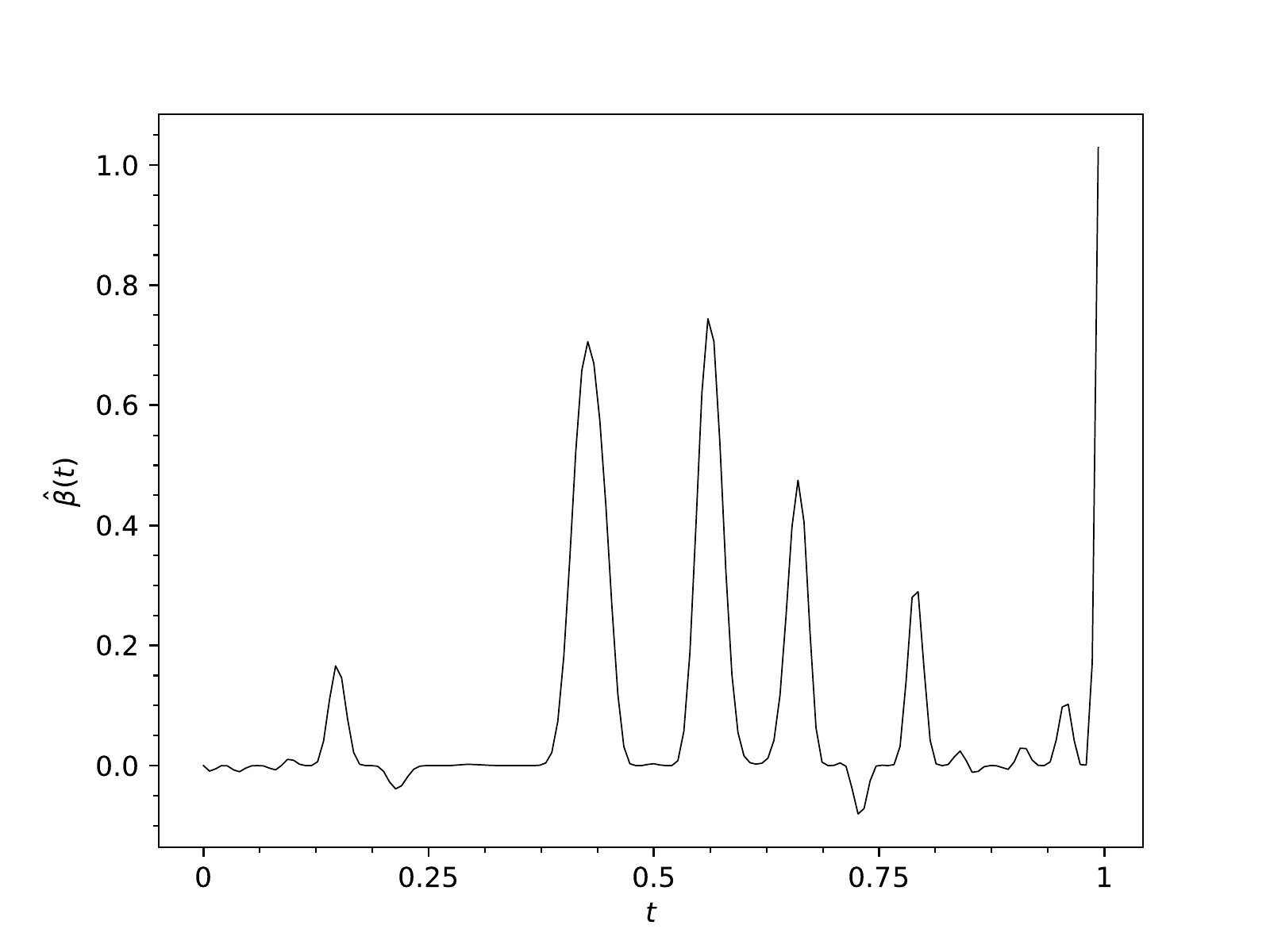}
      \caption{BAR}
    \end{subfigure}
    \begin{subfigure}[c]{0.22\textwidth}
      \includegraphics[width=\textwidth]{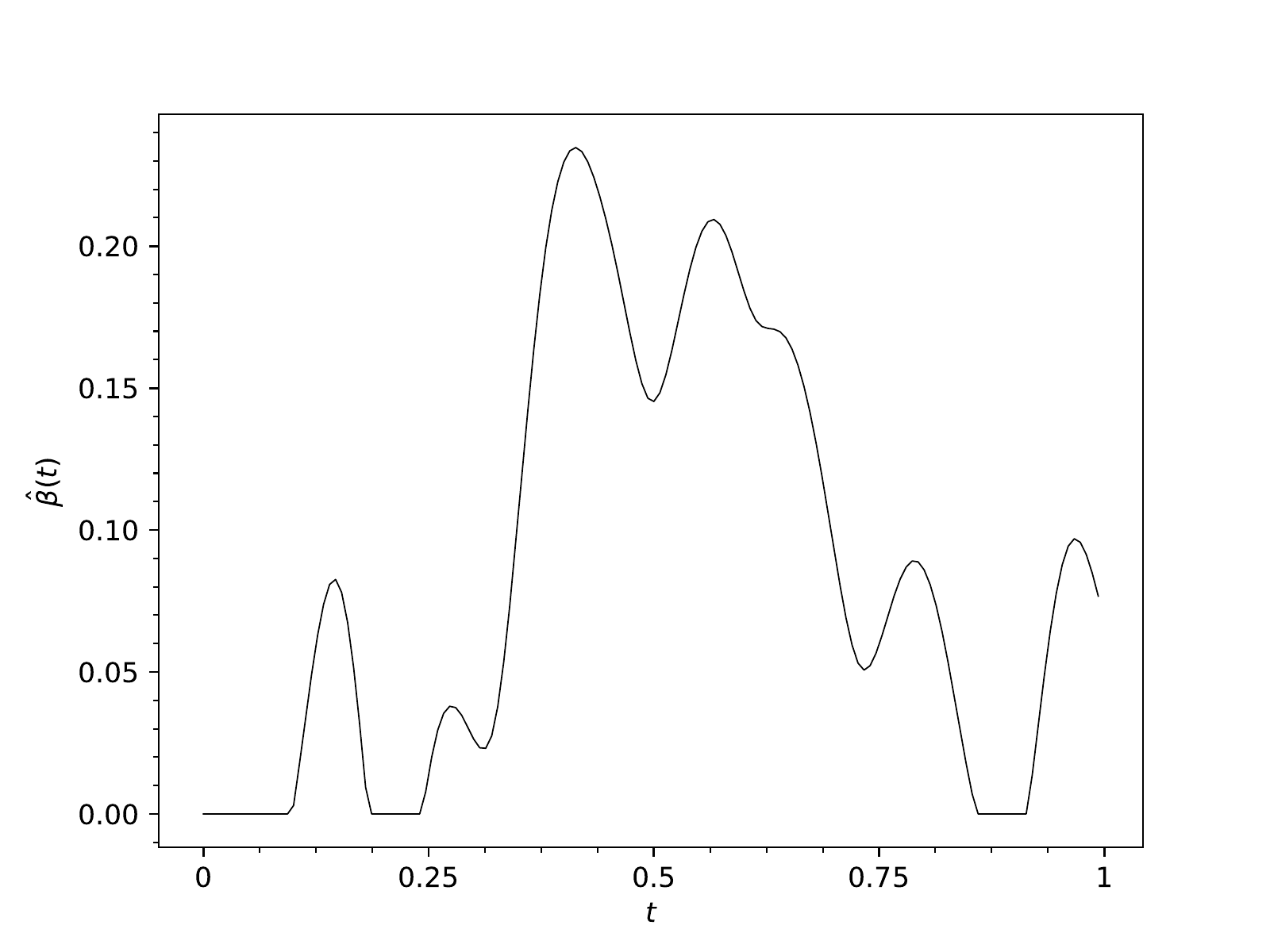}
      \caption{elastic net}
    \end{subfigure}
    \begin{subfigure}[c]{0.22\textwidth}
      \includegraphics[width=\textwidth]{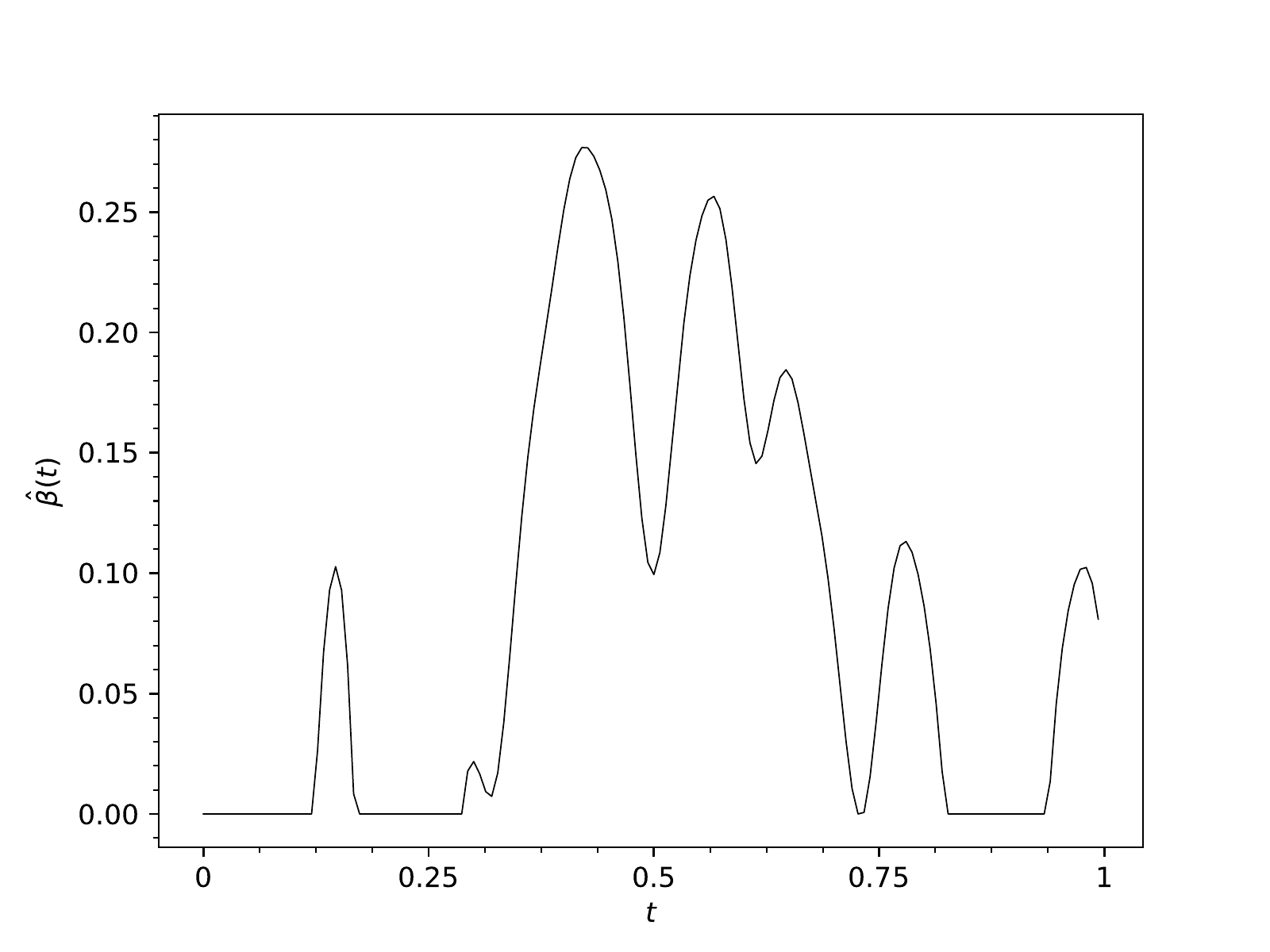}
      \caption{elastic SCAD}
    \end{subfigure}
    \begin{subfigure}[c]{0.22\textwidth}
      \includegraphics[width=\textwidth]{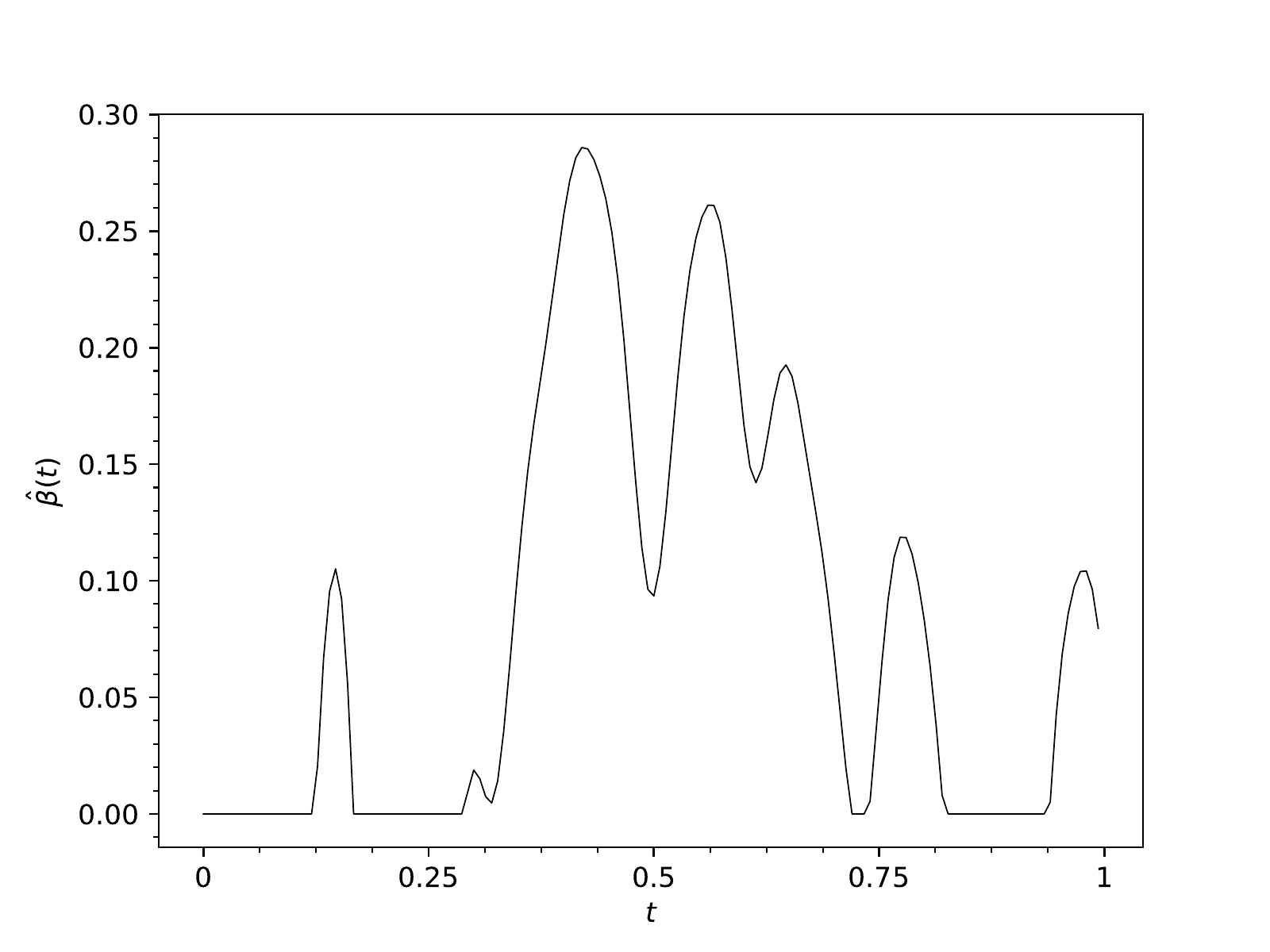}
      \caption{elastic MCP}
    \end{subfigure}

    \begin{subfigure}[c]{0.22\textwidth}
      \includegraphics[width=\textwidth]{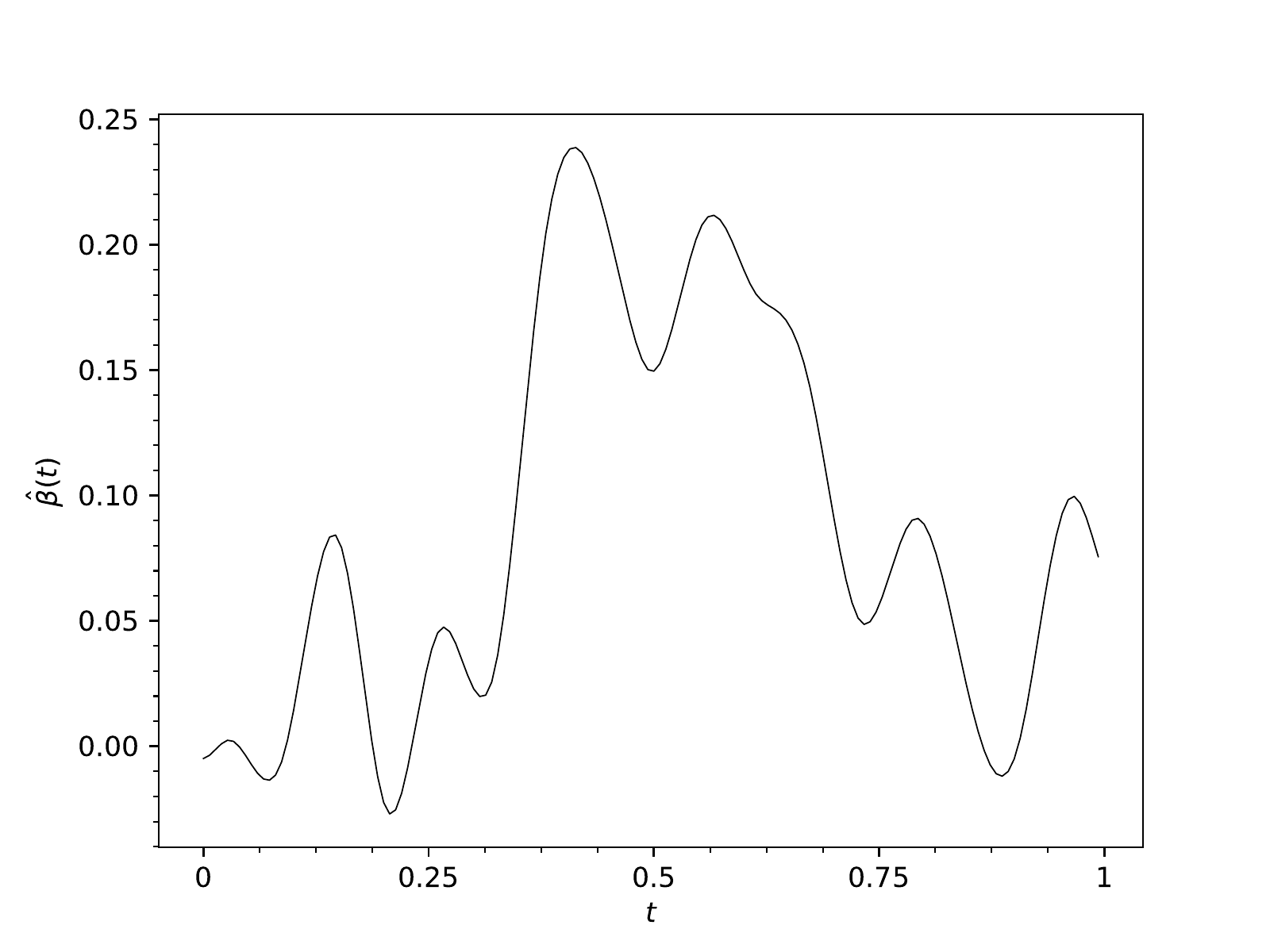}
      \caption{ridge}
    \end{subfigure}
    \begin{subfigure}[c]{0.22\textwidth}
      \includegraphics[width=\textwidth]{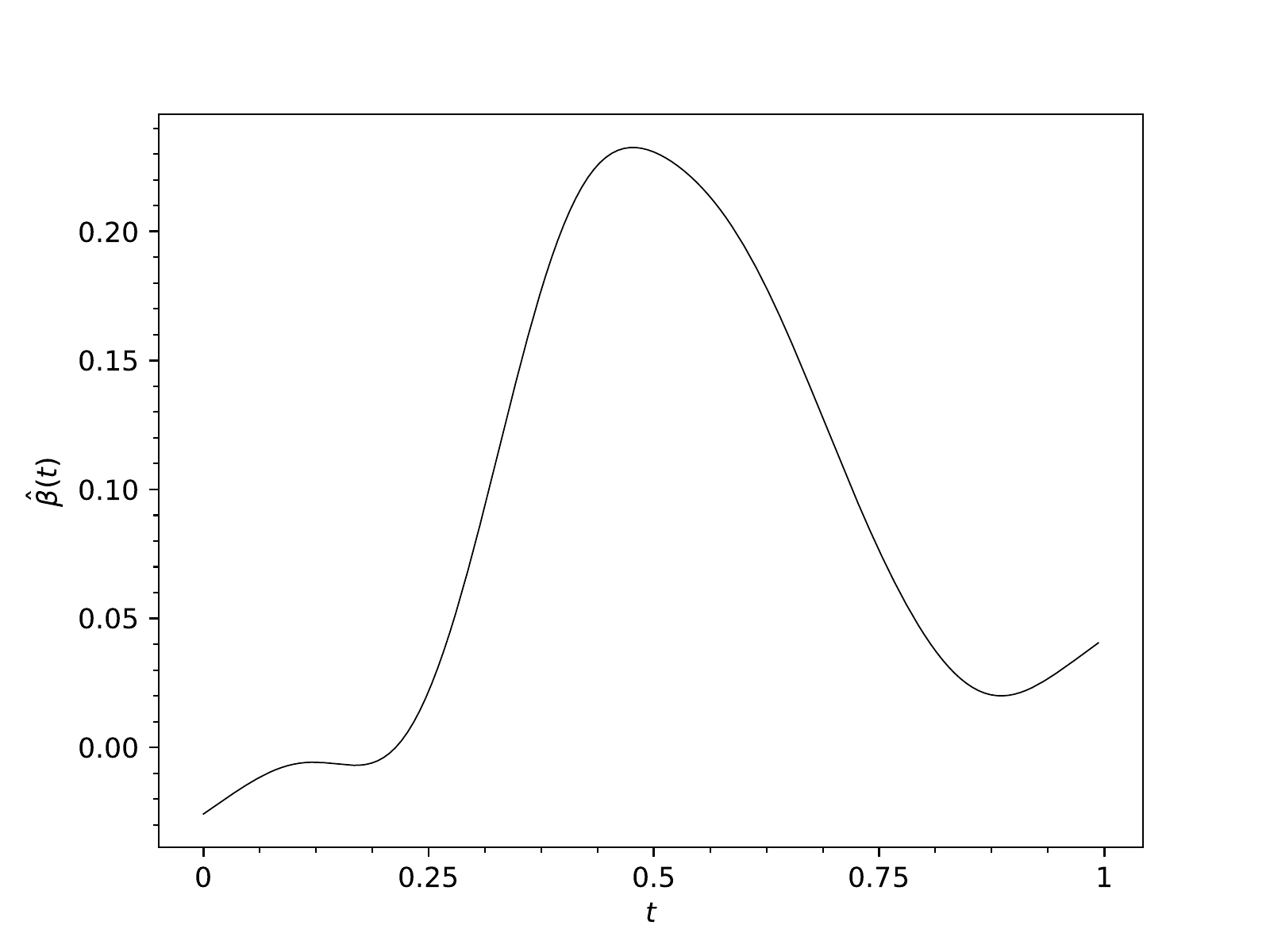}
      \caption{roughness}
    \end{subfigure}
    \begin{subfigure}[c]{0.22\textwidth}
      \includegraphics[width=\textwidth]{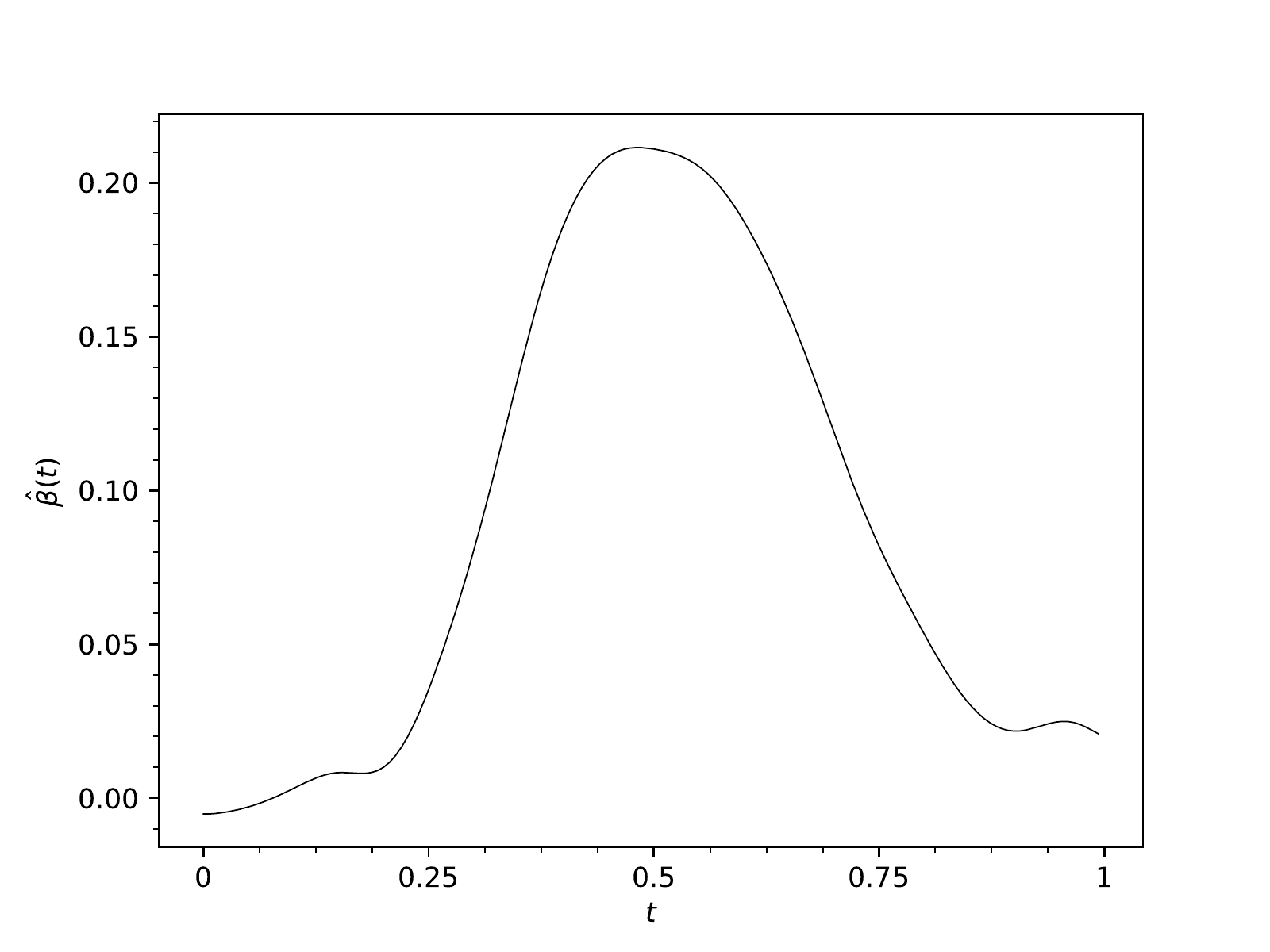}
      \caption{SACR}
    \end{subfigure}
    
  \caption{Dependent spline coefficients: fitted coefficient functions $\hat{\beta}$ by penalty type}
  \label{fig:allbetas_dep}
\end{figure}

\begin{figure}[H]
  \centering
    \begin{subfigure}[c]{0.45\textwidth}
      \includegraphics[width=\textwidth]{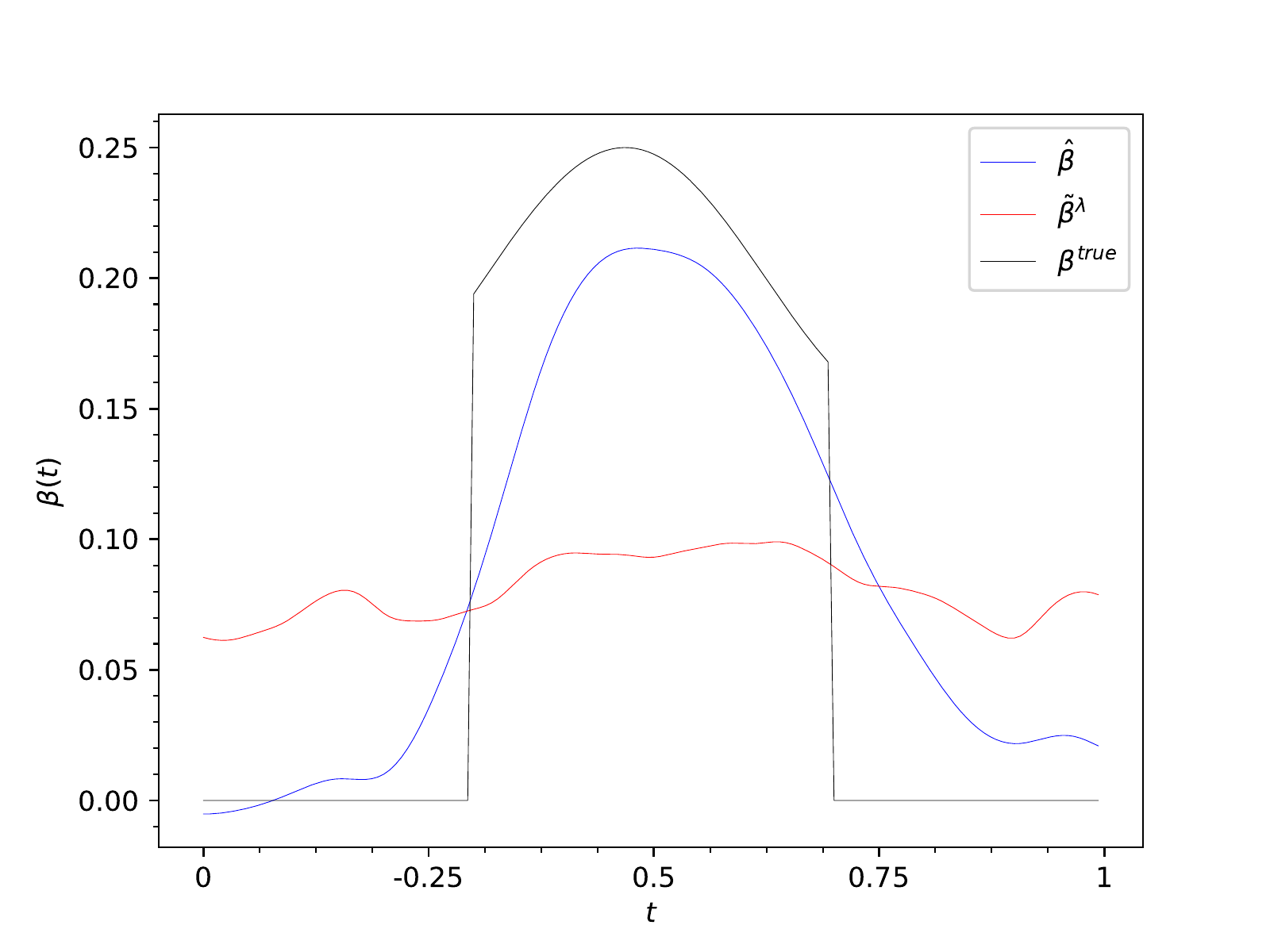}
      \caption{true $\beta$ - initial $\tilde{\beta}^{\lambda}$ - fitted $\hat{\beta}$}
    \end{subfigure}
    \begin{subfigure}[c]{0.45\textwidth}
      \includegraphics[width=\textwidth]{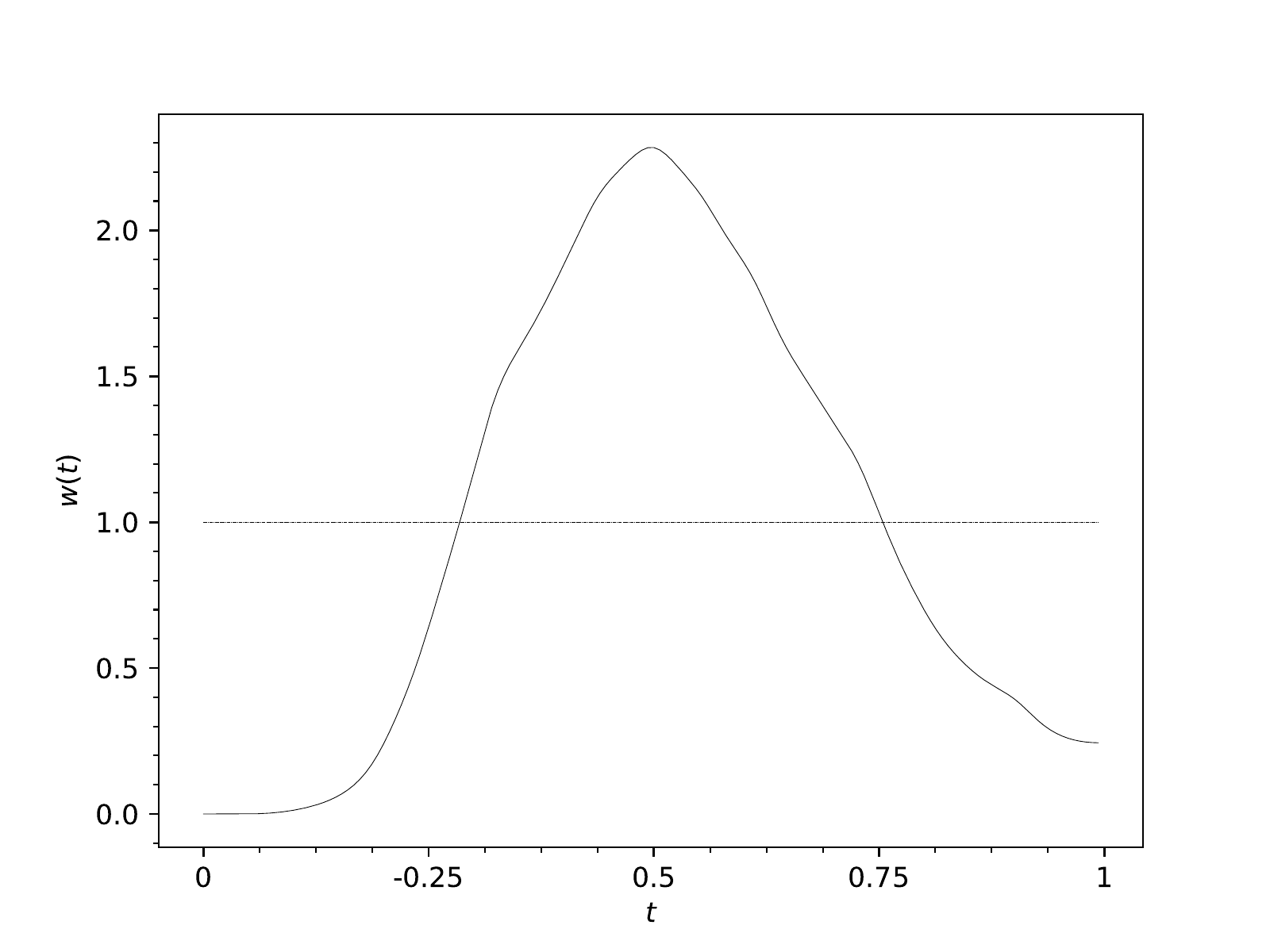}
      \caption{$w$}
    \end{subfigure} 
  \caption{Dependent spline coefficients: comparison between true $\beta$, initial centerfunction $\tilde{\beta}^{\lambda}$ and fitted SACR estimator $\hat{\beta}$, with corresponding weight function $w$}
  \label{fig:sacrcomparison_dep}
\end{figure}

This is a simulated regression problem where the true coefficient function $\beta$ is sparse and smooth, in order to show how the adaptive centering is able to jointly shrink towards a mixed target. In particular, we simulate the input curves with the same B-spline model but with two different configurations of dependency between the coefficients, resulting in two separate simulations. The shared base model is a cubic B-spline with inner knots equispaced between $[-0.5,1.5]$, while the spline coefficients are sampled from a multivariate normal for each of the $N=50$ observations, with either a diagonal covariance matrix and 35 inner knots as an edge case, or high positive correlation and 50 inner knots in order to simulate a standard FDA setting. The $N$ input functions $x_{i}$ are then evaluated on the same equispaced grid of length $p=150$ over $I=[0,1]$, and the responses are computed as $y_{i} = \int_{I} x_i(t)\beta(t)dt + \epsilon_{i}$ with $\epsilon_{i} \sim \mathcal{N}(0,1)$. Figure \ref{fig:data_simulation} shows the input curves for both simulations and the true coefficient function, while Figures \ref{fig:allbetas_ind} and \ref{fig:allbetas_dep} show the fitted coefficient functions for all methods tested. Regarding the strictly $L_{2}$-based methods, the ridge, the roughness penalty and the SACR provide similar solutions, with varying degrees of smoothness as expected. The SACR in fact reminds of a warped version of the ridge, with a much smooother behaviour in the highly collinear simulation, which is very close to the solution of the roughness penalized model. In Figures \ref{fig:sacrcomparison_ind} and \ref{fig:sacrcomparison_dep} we show the comparison between the initial center, the true coefficient function and the fitted SACR function, together with the corresponding fitted weight function. In particular, while our method is not able to recover exactly zero values of the coefficient function (given the $L_{2}$ norm), the overall sparsity pattern is arguably recognizable and is confirmed by the shape of the weight function, which is above one where the initial solution should be inflated, while tapering towards zero in the regions that should be sparse. Note that the fact of having a zero weight does not guarantee an exactly zero coefficient function $\beta$, since the weight acts on the center of the penalty and not on the coefficients themselves, with analogous considerations for very high values of the weight, as shown in the simulation with independent coefficients. The BAR estimator seems to have problems of instability, which could be a numerical issue of our implementation given its asymptotic definition, although the method is competitive in both the subsequent real world case studies. Regarding the pure sparsity inducing penalties like the lasso, the relaxed lasso, the adaptive lasso, and the NNG, there is no clear distinction between the two simulations, and in both cases the methods recover coefficient functions with the typical spikes on some of the variables, failing to recover the exact sparsity pattern and therefore excluding from the model many of the significant predictors. On the other hand, the thresholding effect of the $L_{1}$ penalty allows to set to zero the coefficients of the unwanted variables. Finally, the hybrid penalties show a clearly different behaviour in the two simulations, where all three methods visibly leverage the ridge part of the penalty in the highly collinear simulation, including in the model all the correct predictors and many unwanted ones, while in the independent simulation only the elastic net leverages the ridge part, and instead both the elastic SCAD and elastic MCP recover very sparse solutions. We report the regression results in Table \ref{tab:simulations}, obtained by 5-fold cross-validation with 3-fold cross-validation for grid-search hyperparameter selection. It is worth noting that in the simulation with independent coefficients, the adaptive lasso, the elastic SCAD and elastic MCP have lower mean-square error than the SACR, despite the fact that they leave out of the model many of the relevant predictors. 

\begin{table}[H]
\centering
\caption{\label{tab:simulations} Regression results for both simulations: mean-square error}
\scalebox{1}{
\begin{tabular}{l D{,}{\, \pm \,}{-1} D{,}{\, \pm \,}{-1}}
\toprule
\midrule
           
 & \multicolumn{1}{c}{independent} & \multicolumn{1}{c}{dependent} \\
\midrule

lasso             & 1.807,1.03 & 2.011,1.2   \\

adaptive lasso    & 1.479,.067 & 2.078,1.4   \\

relaxed lasso     & 1.807,1.03 & 1.982,1.2   \\

NNG               & 1.571,1.01 & 2.171,1.3  \\

BAR               & 5.669,2.60 & 2.538,1.1   \\
  
elastic net       & 1.708,.548 & 1.949,1.2   \\

elastic SCAD      & 1.263,.347 & 2.414,1.2  \\

elastic MCP       & 1.291,.407 & 2.407,1.1   \\
  
ridge             & 2.408,.458 & 1.948,1.2   \\

roughness         & 1.540,.391 & 1.894,1.3   \\

SACR              & 1.521,.308 & 1.795,1.1   \\

\midrule
\bottomrule  
\end{tabular}}
\end{table}

\subsection{IDRC 2018}
\begin{figure}[H]
  \centering
  	\begin{subfigure}[c]{0.32\textwidth}
      \includegraphics[width=\textwidth]{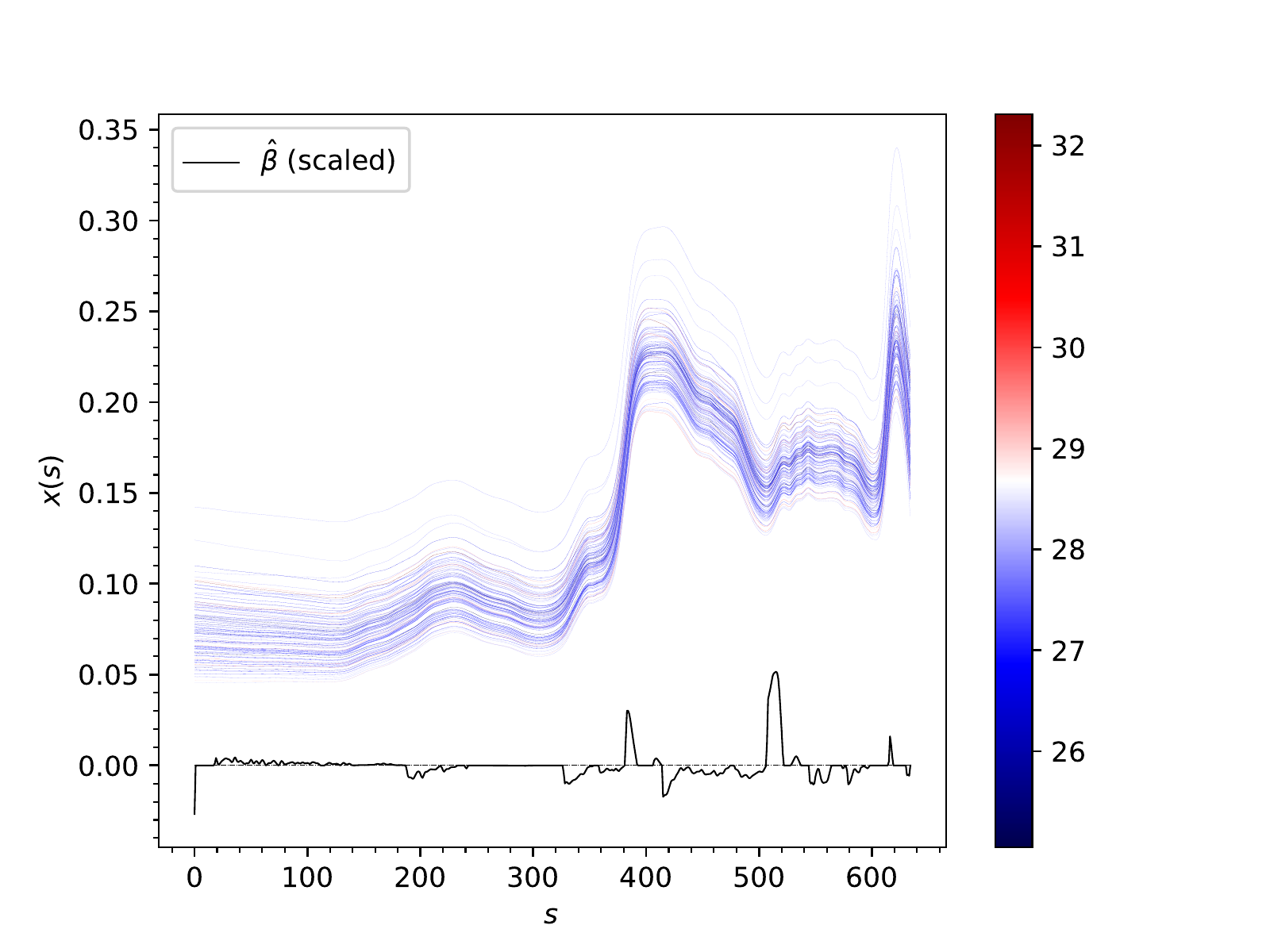}
      \caption{lasso}
    \end{subfigure}
    \begin{subfigure}[c]{0.32\textwidth}
      \includegraphics[width=\textwidth]{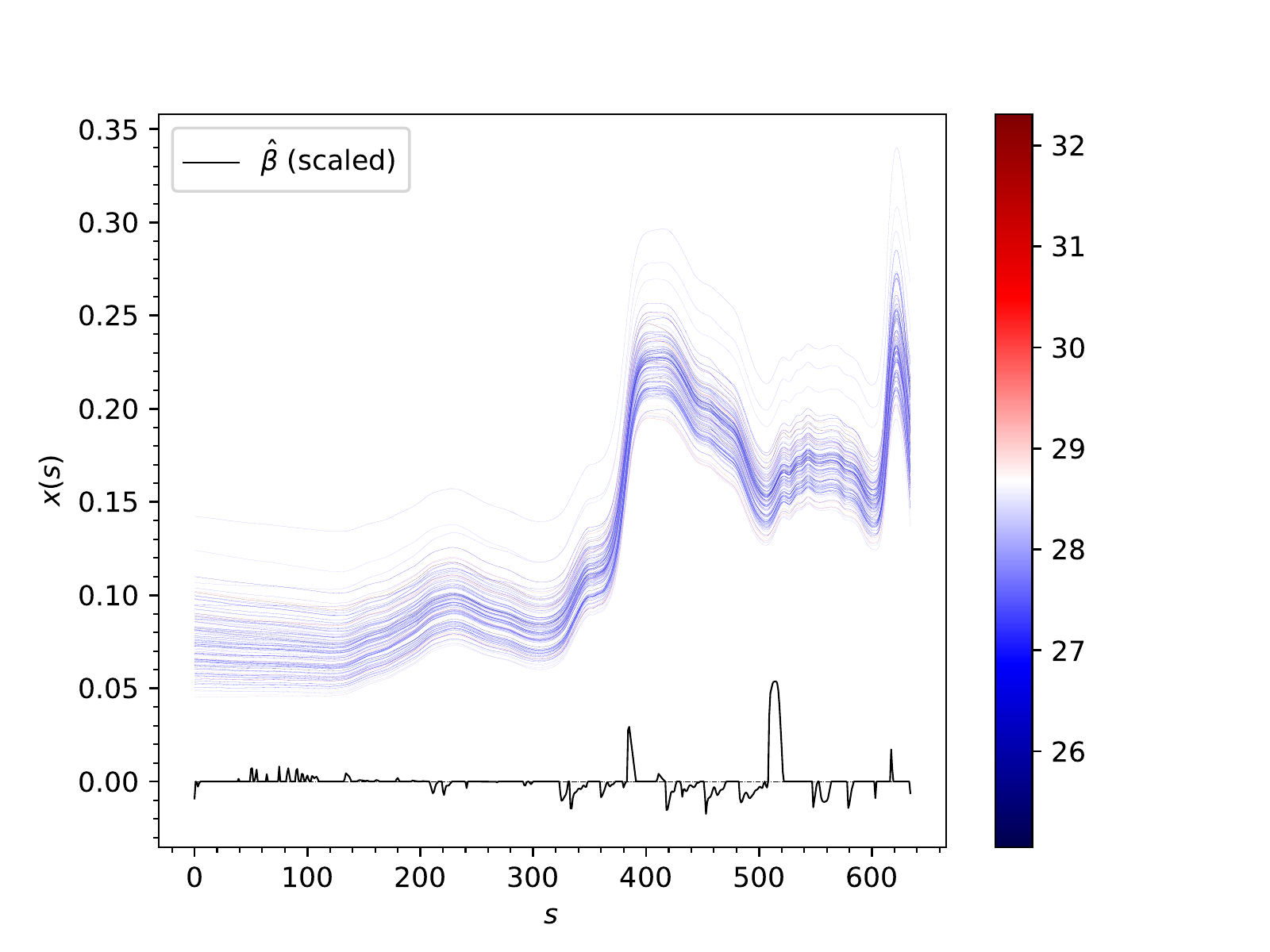}
      \caption{adaptive lasso}
    \end{subfigure}
    \begin{subfigure}[c]{0.32\textwidth}
      \includegraphics[width=\textwidth]{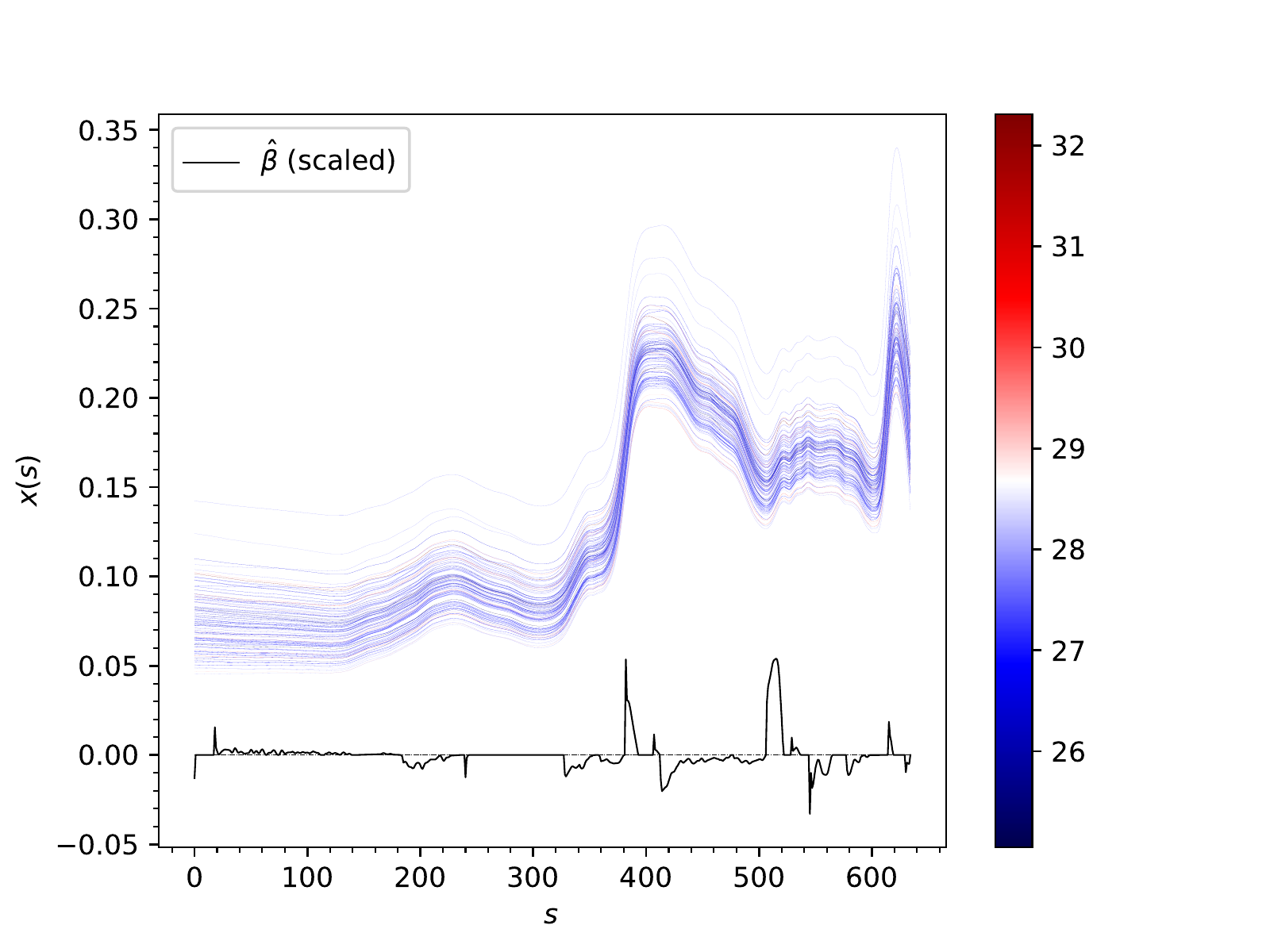}
      \caption{relaxed lasso}
    \end{subfigure}
    
    \begin{subfigure}[c]{0.32\textwidth}
      \includegraphics[width=\textwidth]{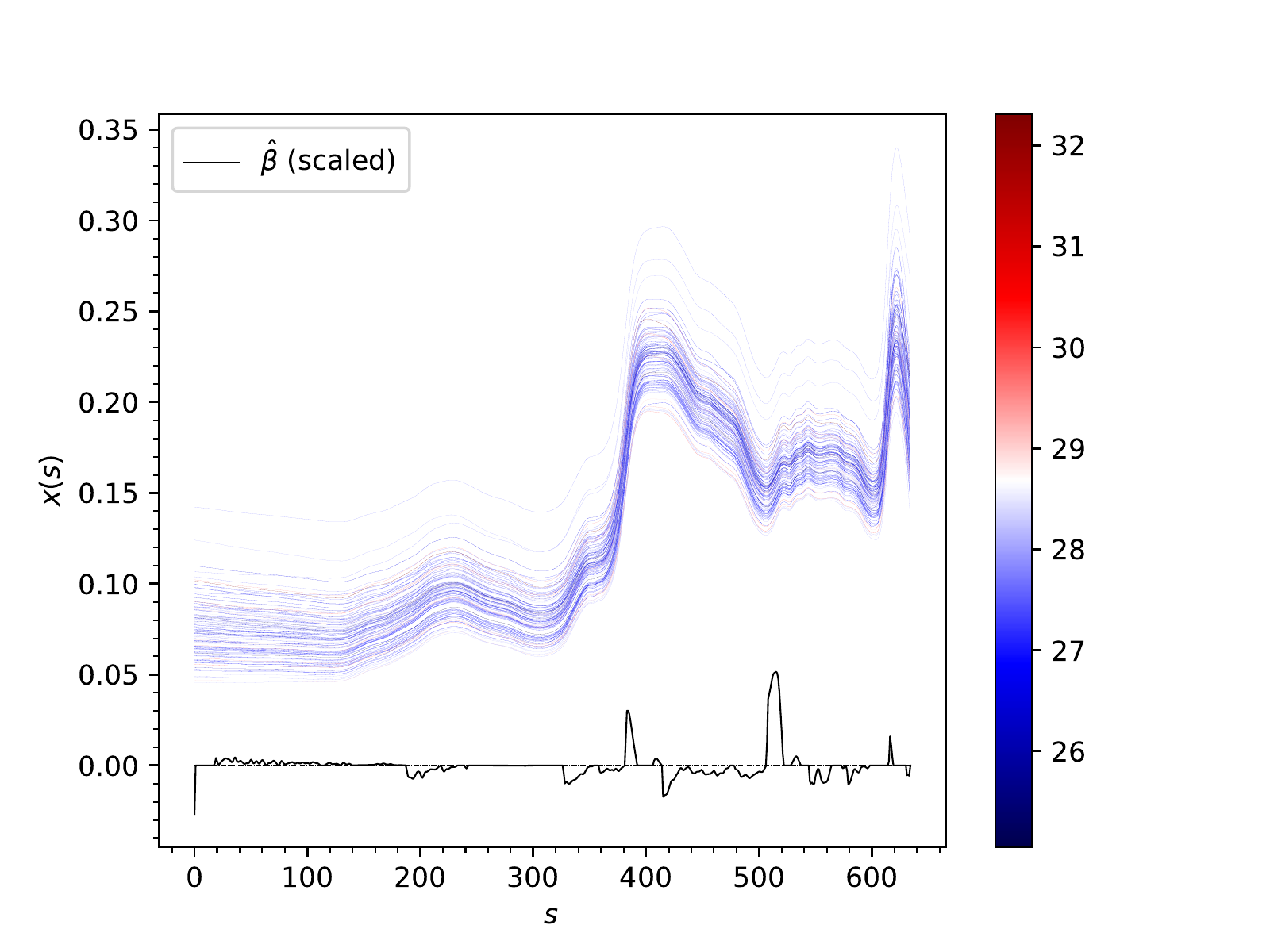}
      \caption{elastic net}
    \end{subfigure}
    \begin{subfigure}[c]{0.32\textwidth}
      \includegraphics[width=\textwidth]{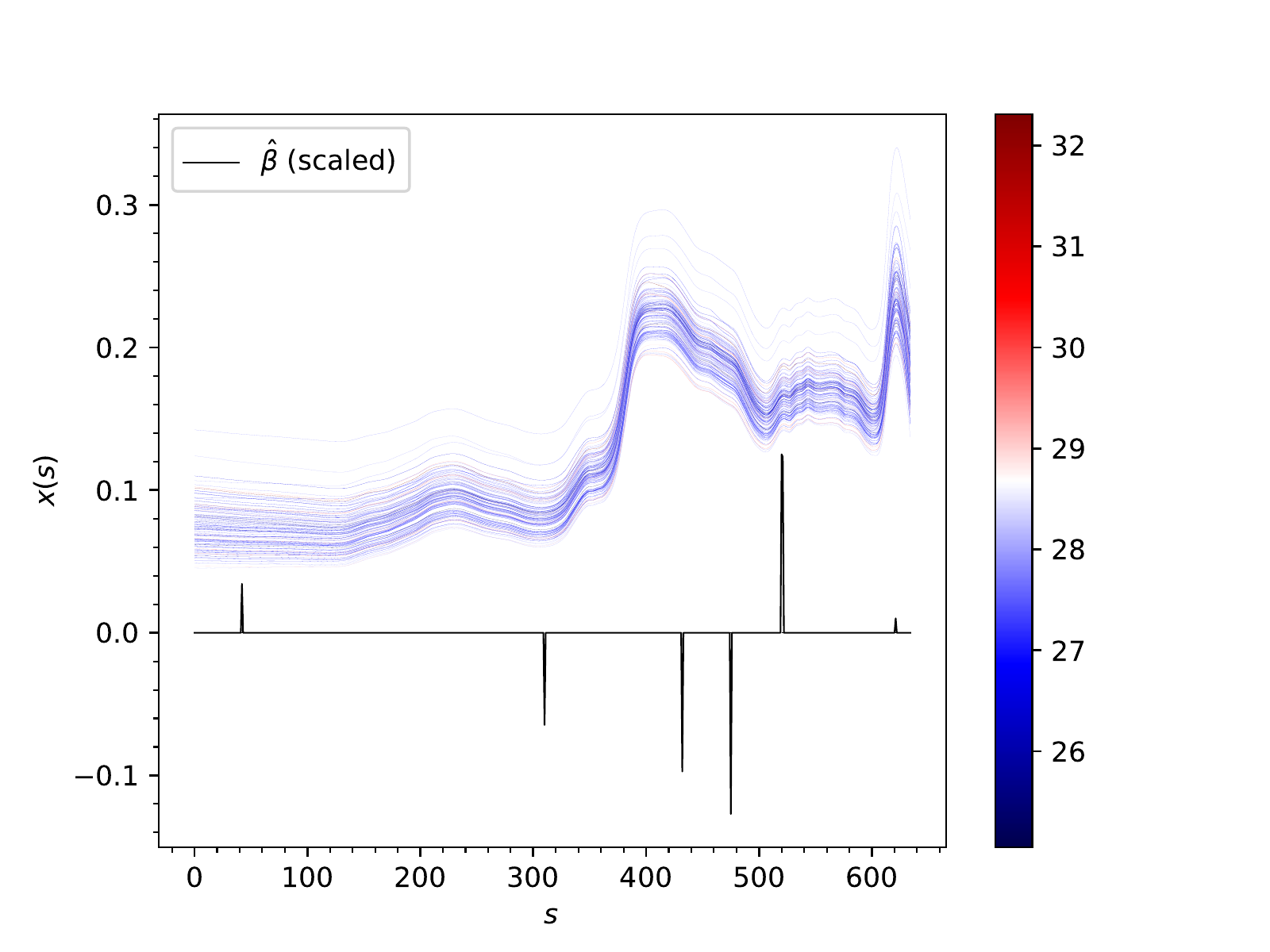}
      \caption{elastic SCAD}
    \end{subfigure}
    \begin{subfigure}[c]{0.32\textwidth}
      \includegraphics[width=\textwidth]{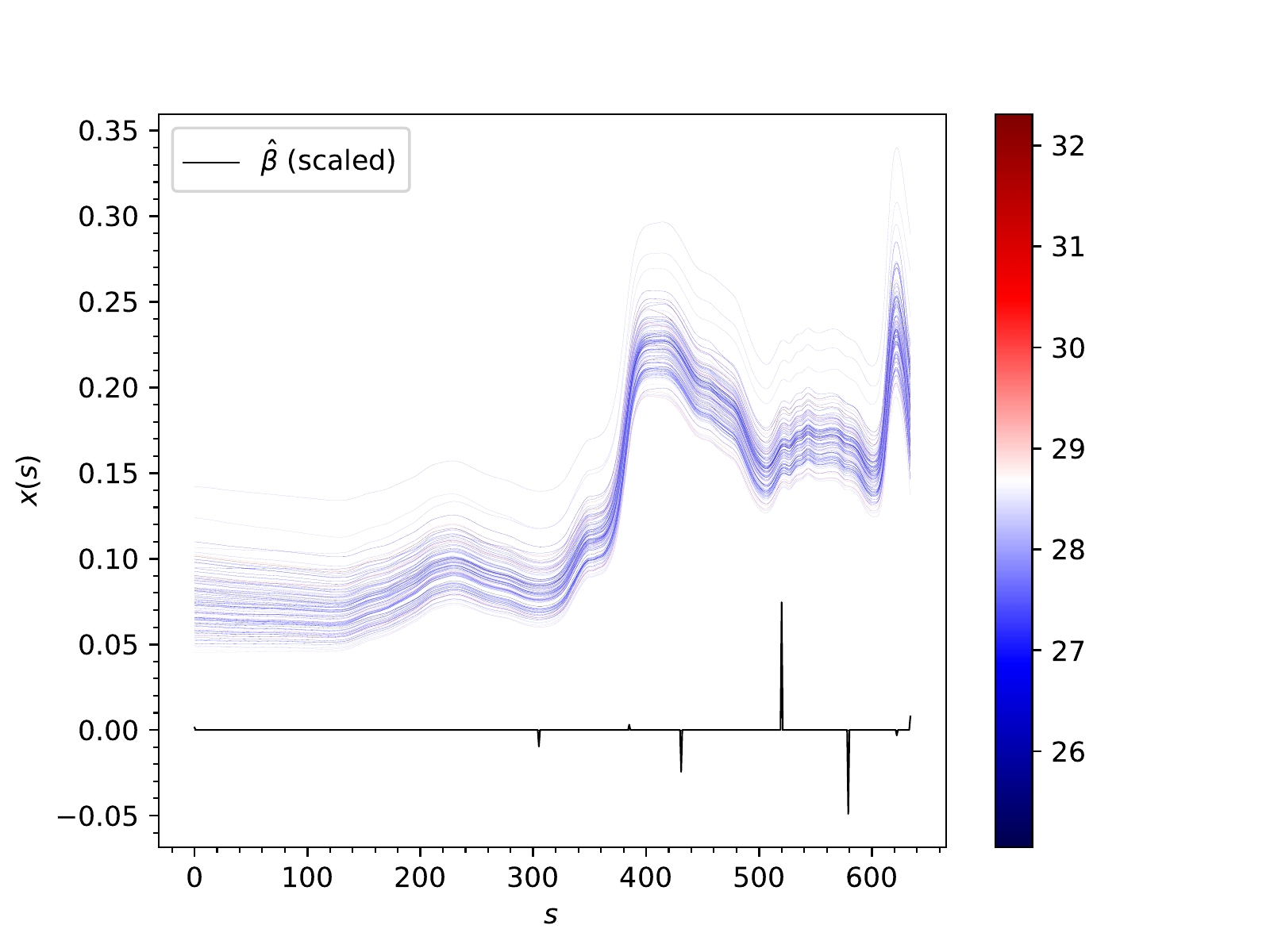}
      \caption{elastic MCP}
    \end{subfigure}
    
    \begin{subfigure}[c]{0.32\textwidth}
      \includegraphics[width=\textwidth]{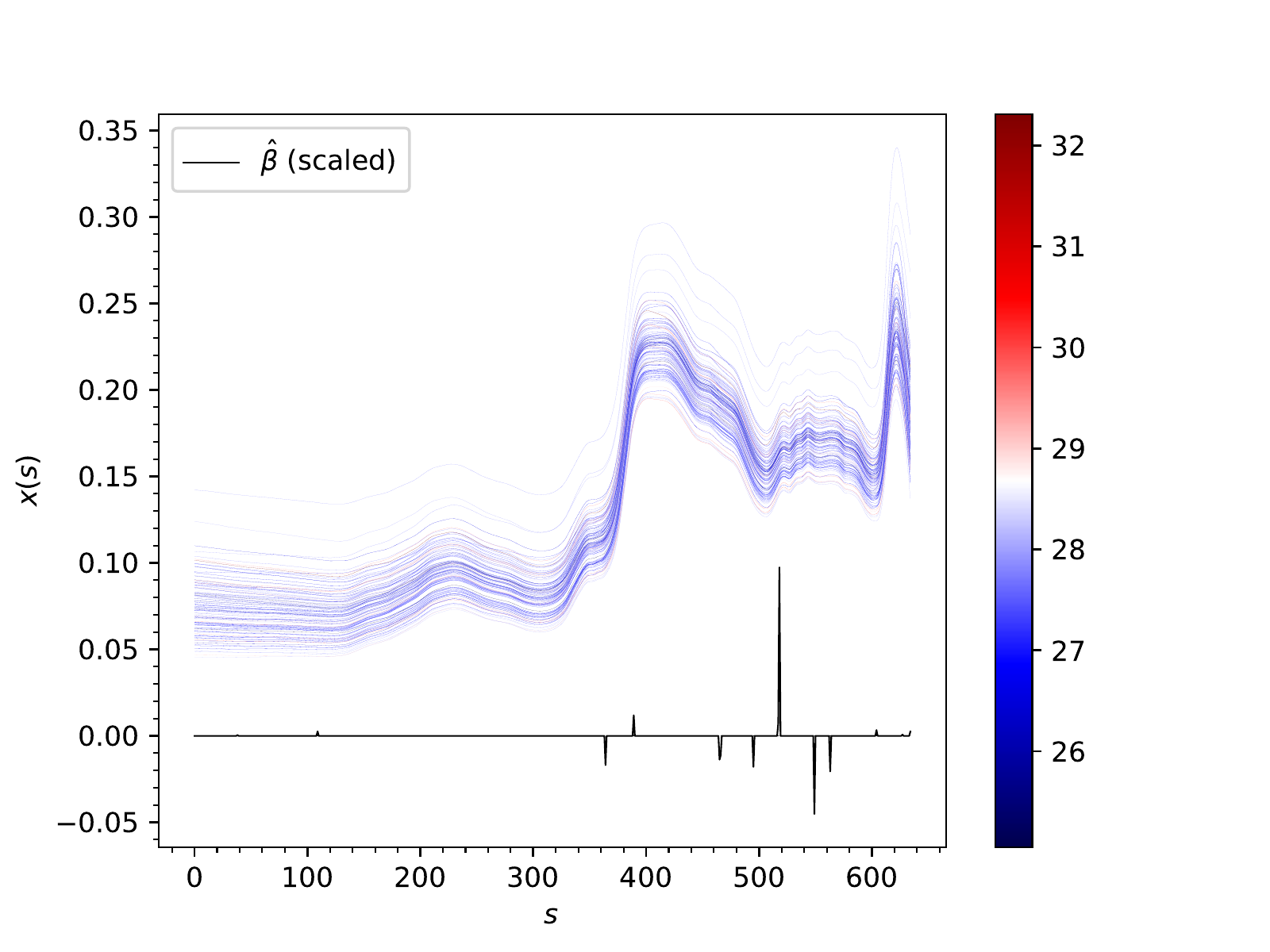}
      \caption{NNG}
    \end{subfigure}
  	\begin{subfigure}[c]{0.32\textwidth}
      \includegraphics[width=\textwidth]{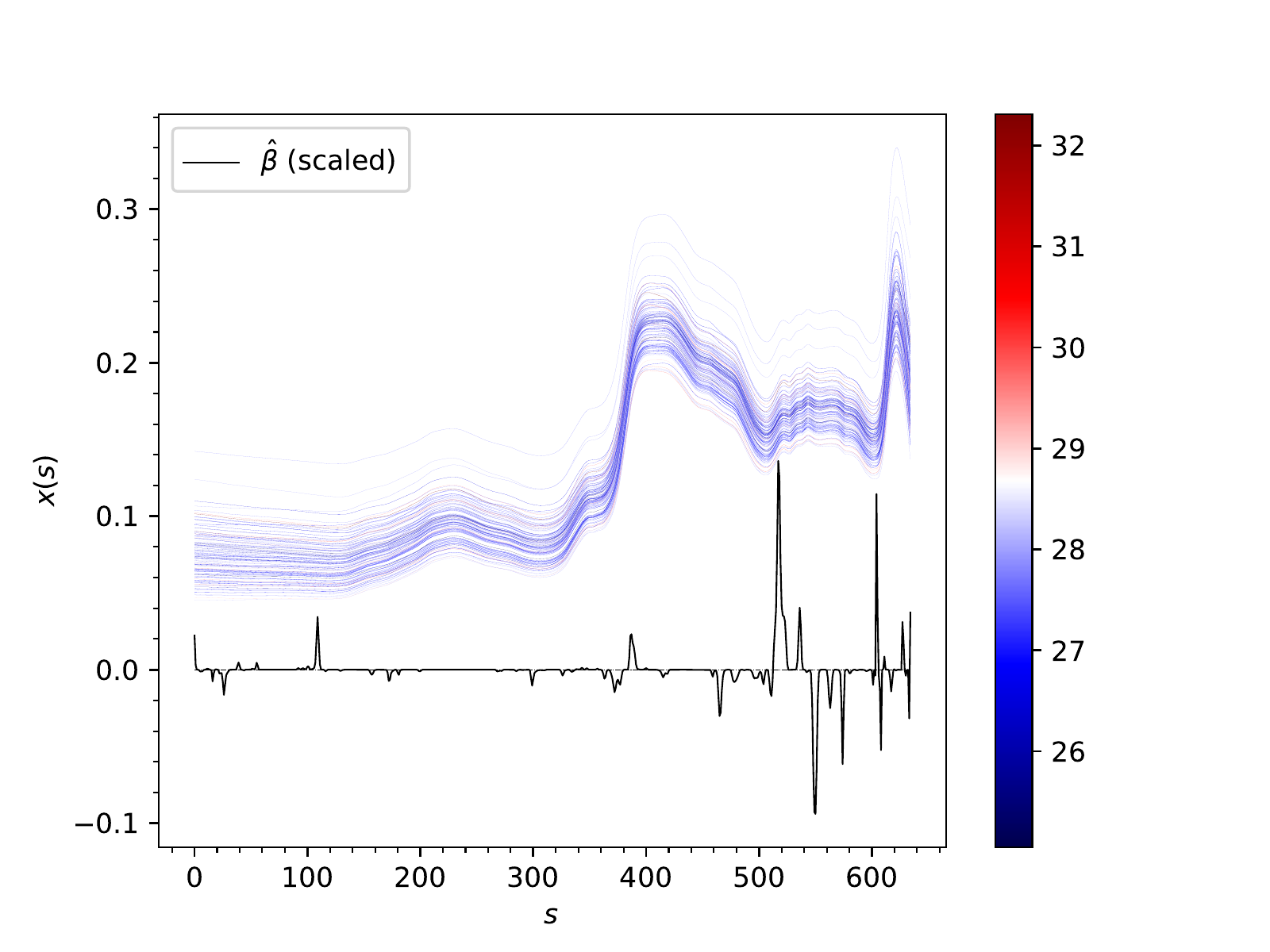}
      \caption{BAR}
    \end{subfigure}   
    \begin{subfigure}[c]{0.32\textwidth}
      \includegraphics[width=\textwidth]{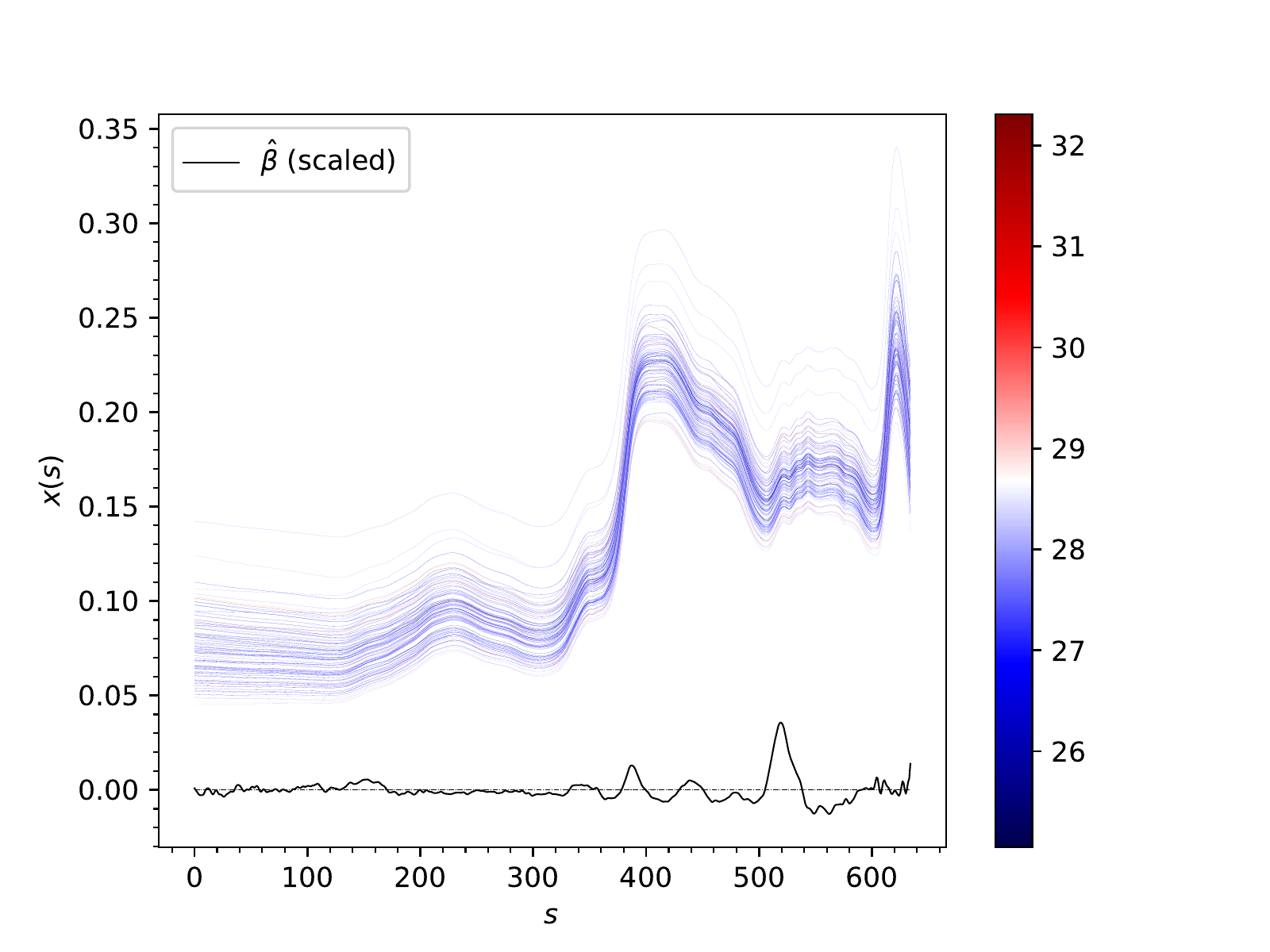}
      \caption{ridge}
    \end{subfigure}
    
    \begin{subfigure}[c]{0.32\textwidth}
      \includegraphics[width=\textwidth]{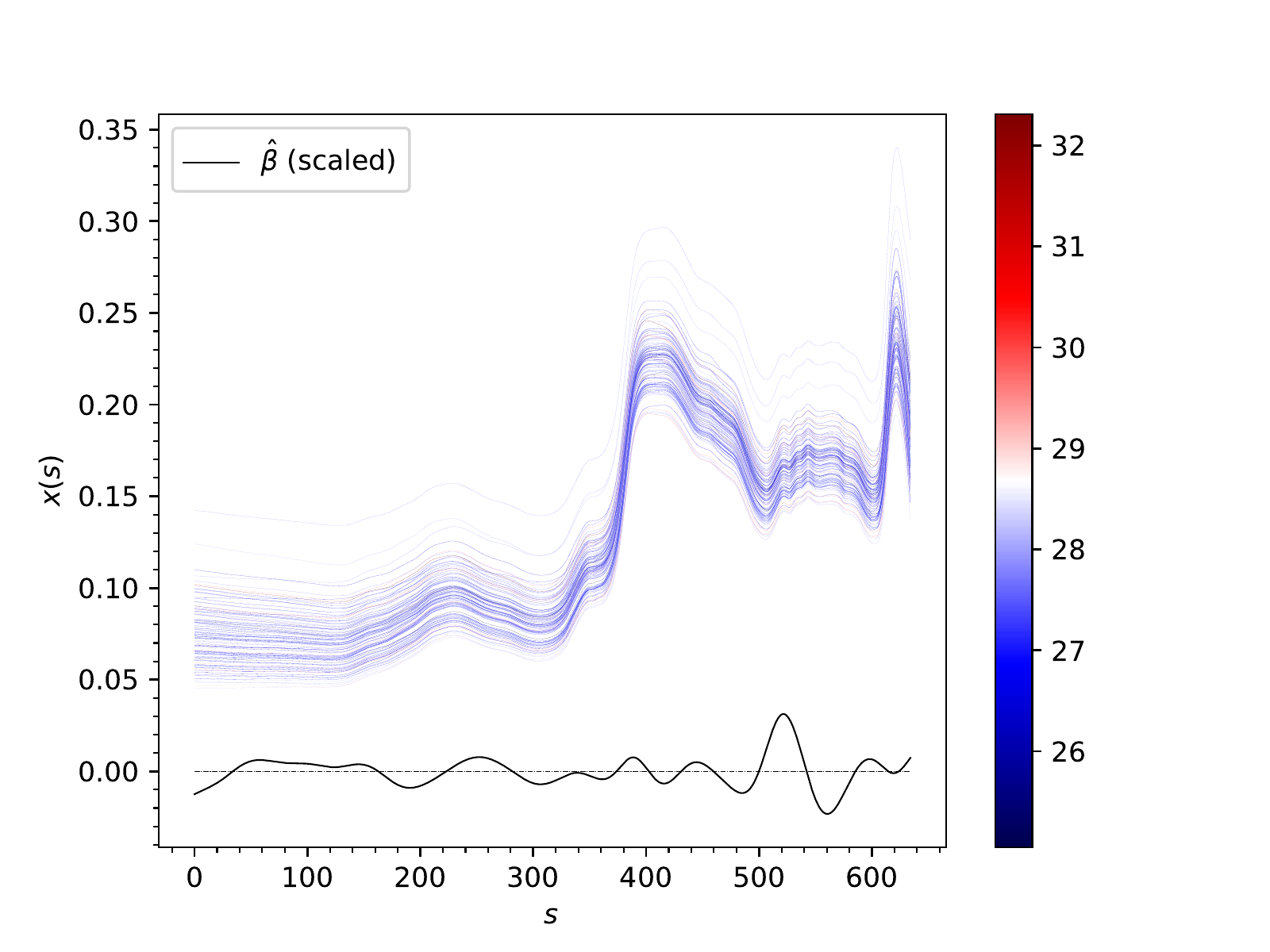}
      \caption{roughness}
    \end{subfigure}
    \begin{subfigure}[c]{0.32\textwidth}
      \includegraphics[width=\textwidth]{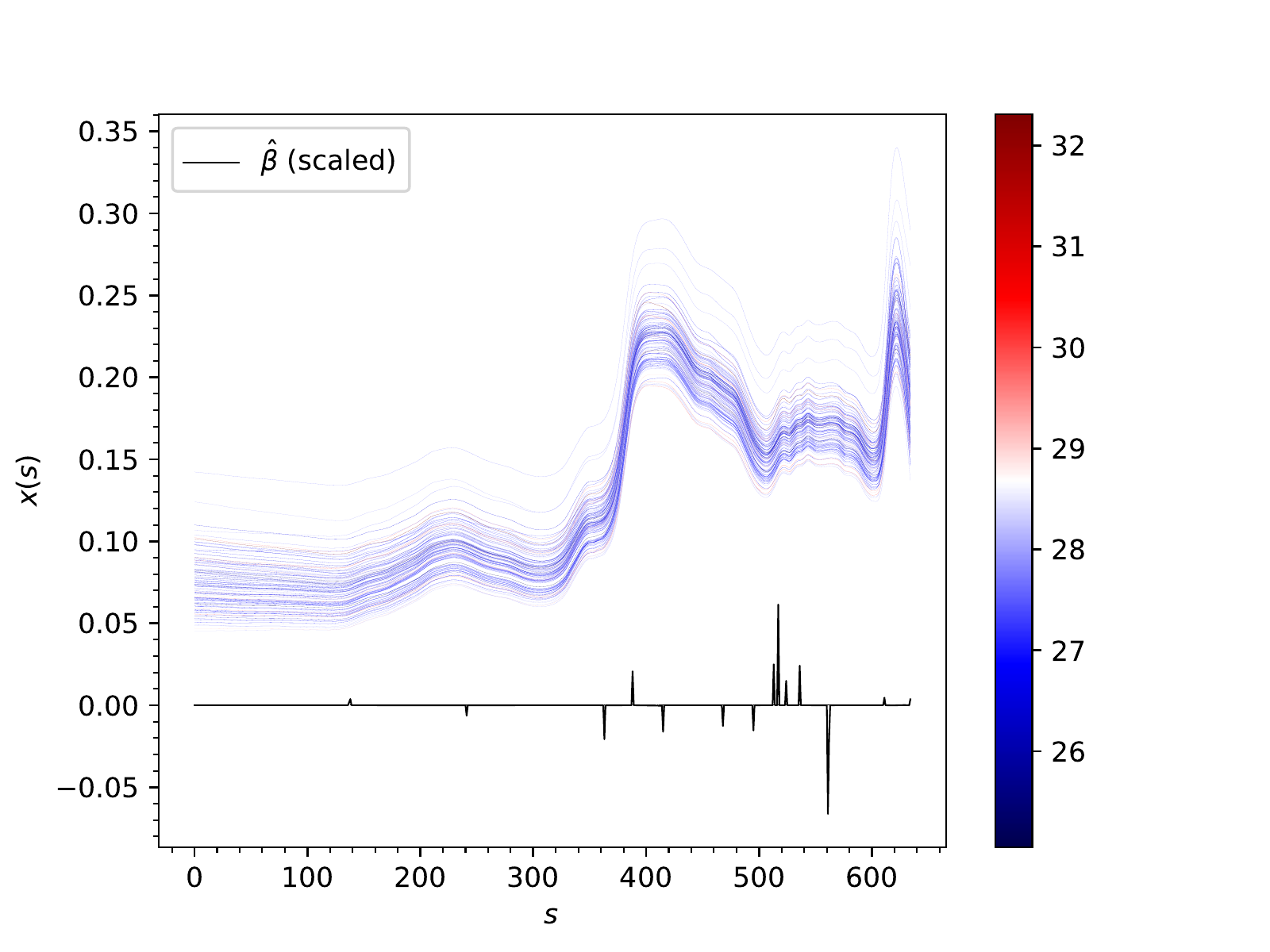}
      \caption{SACR}
    \end{subfigure}
  \caption{IDRC 2018: comparison of the fitted coefficient functions $\hat{\beta}$ scaled with respect to the spectra, the black dashed line represents the zero level for $\hat{\beta}$}
  \label{fig:idrc2018}
\end{figure}

For regression we present a spectroscopy application that was originally proposed for the on-site competition of the 2018 International Diffuse Reflectance Conference. The data is already smooth and is available at \url{https://www.cnirs.org/content.aspx?page_id=22&club_id=409746&module_id=276203}, with $N=150$ and $p=635$. The response variable has values over the whole dataset of 27.7 $\pm$ 1 ($\mu\pm\sigma$), with no information about the nature of the data. Figure \ref{fig:idrc2018} shows the input curves with the corresponding (rescaled) coefficient function for each of the methods that we tested, where the dashed line indicates the zero of the coefficient function. The results are reported in Table \ref{tab:idrc2018} and are obtained by three random repetitions of 5-fold cross-validation, with 3-fold cross-validation for grid search hyperparameter selection. In this application, the SACR and the ridge both achieve comparable mean-square error (mse) scores, but with a clear difference in the fitted coefficient functions. In fact, while the ridge recovers a noisy solution, the SACR is able to recover a sparse coefficient function which is easier to interpret. Overall, most methods tend to include the same variables in the model, with the lasso variants and the elastic net that show a noisy behaviour similar to the ridge, which may be given by the high collinearity. The roughness penalty instead recovers a smooth but oscillating coefficient function, despite achieving a similar score to the BAR estimator, which is sparse and slightly noisy. Finally, the NNG, the elastic SCAD and elastic MCP all recover very sparse solutions as expected.

\begin{table}
\caption{IDRC 2018: regression results, mean-square error}
\label{tab:idrc2018}
\begin {center}
\begin{tabular}{l D{,}{\, \pm \,}{-1}}
\toprule
\midrule
lasso             & .1029,.019  \\

adaptive lasso    & .0926,.018  \\

relaxed lasso     & .0951,.017  \\

NNG               & .0750,.015  \\

BAR               & .0717,.015  \\
  
elastic net       & .1048,.019  \\

elastic SCAD      & .1559,.061  \\

elastic MCP       & .1415,.072  \\
  
ridge             & .0695,.013  \\

roughness         & .0711,.016  \\

SACR              & .0691,.014  \\
\midrule
\bottomrule
\end{tabular}
\end {center}
\end{table}

\newpage 

\subsection{Wine}
\begin{figure}[H]
  \centering
  	\begin{subfigure}[c]{0.32\textwidth}
      \includegraphics[width=\textwidth]{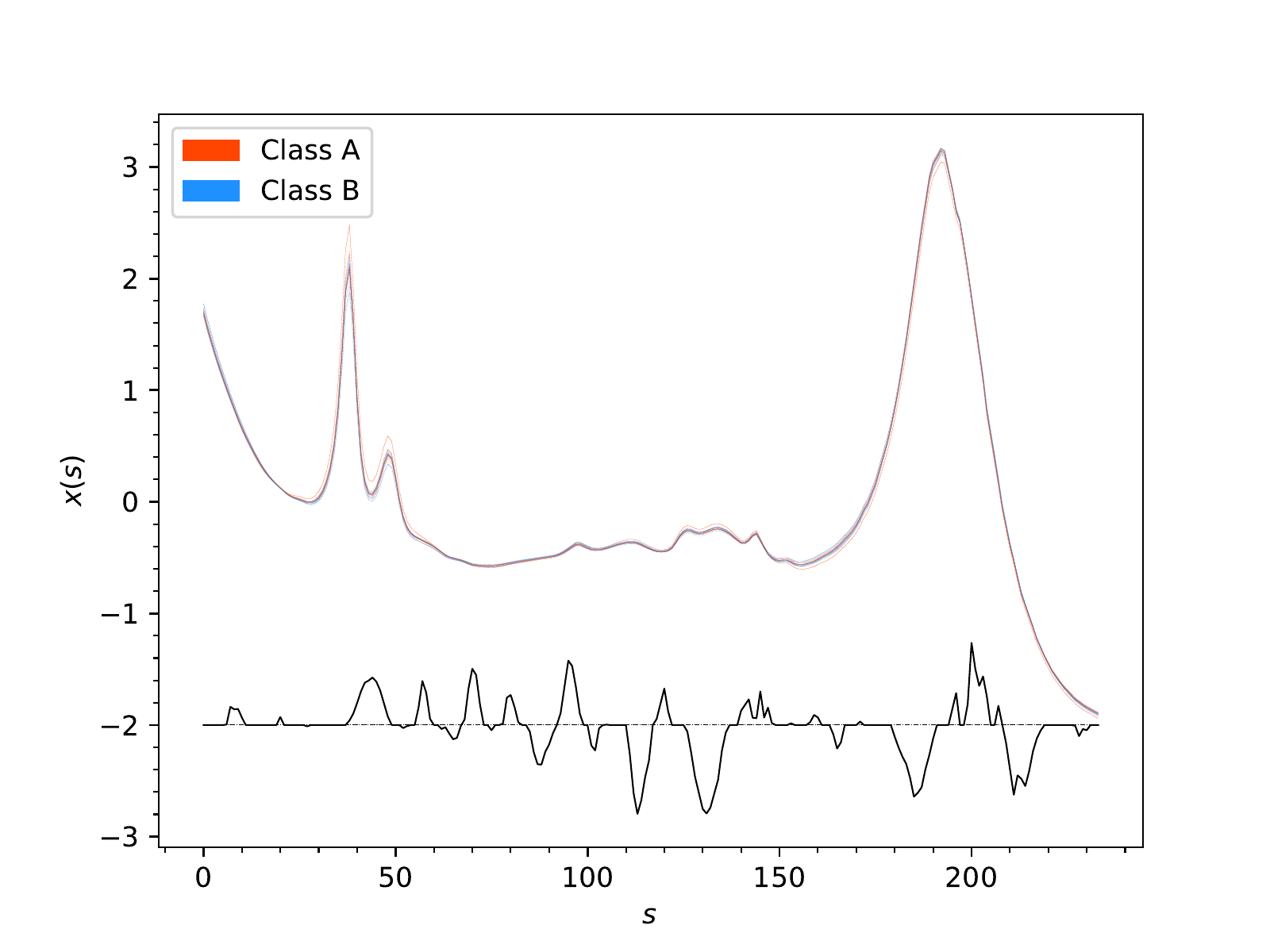}
      \caption{lasso}
    \end{subfigure}
    \begin{subfigure}[c]{0.32\textwidth}
      \includegraphics[width=\textwidth]{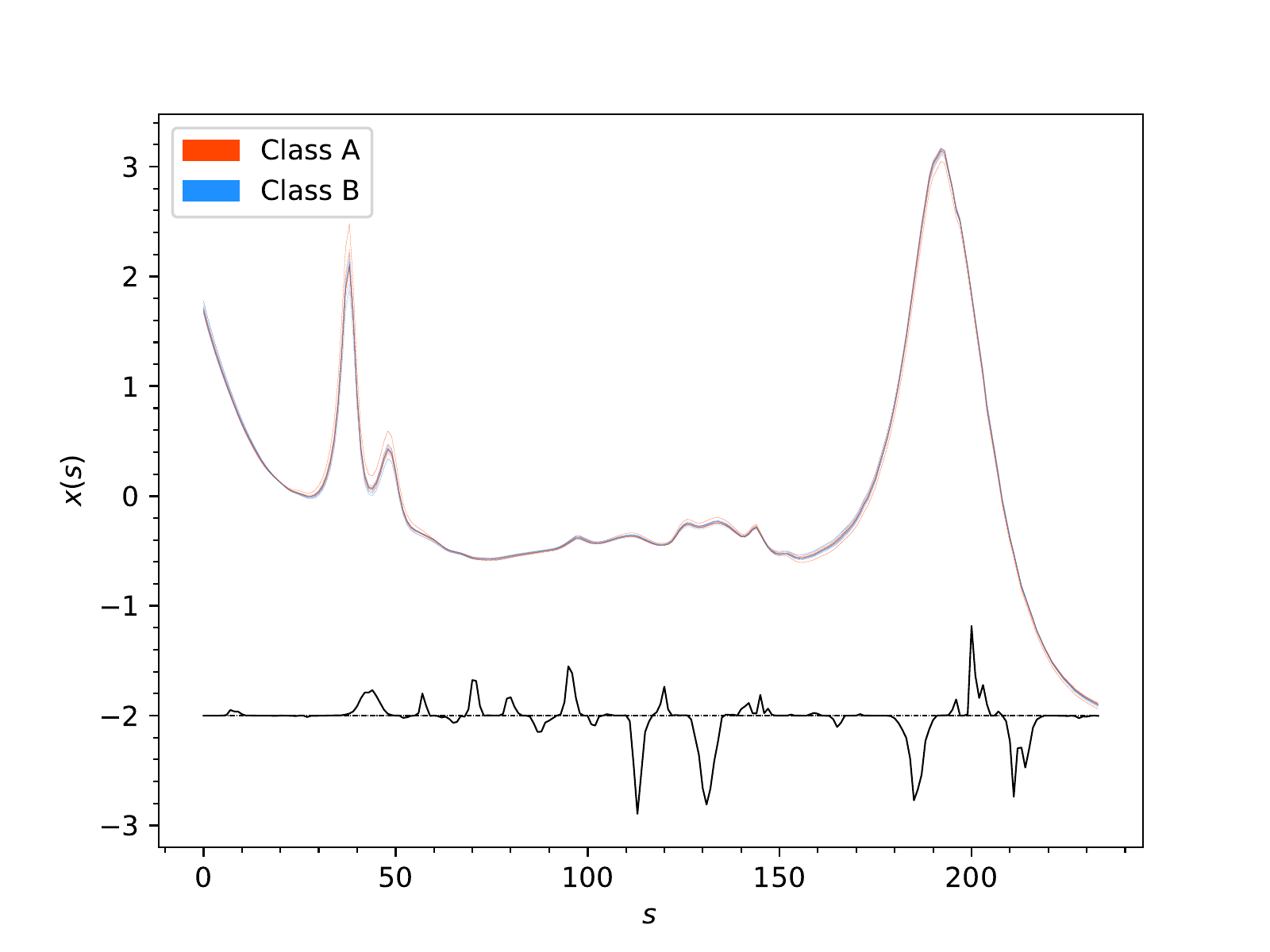}
      \caption{adaptive lasso}
    \end{subfigure}
    \begin{subfigure}[c]{0.32\textwidth}
      \includegraphics[width=\textwidth]{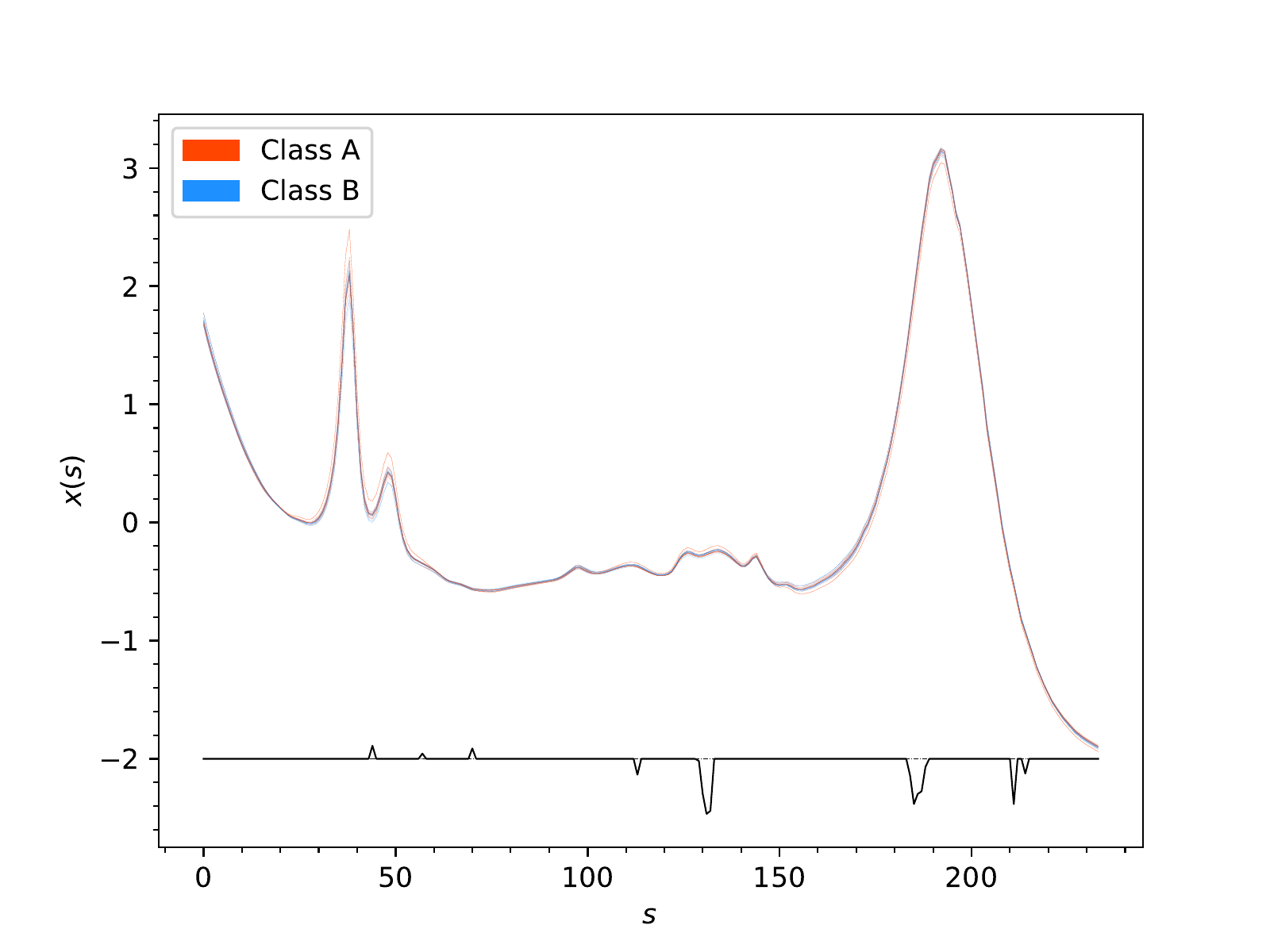}
      \caption{relaxed lasso}
    \end{subfigure}
    
    \begin{subfigure}[c]{0.32\textwidth}
      \includegraphics[width=\textwidth]{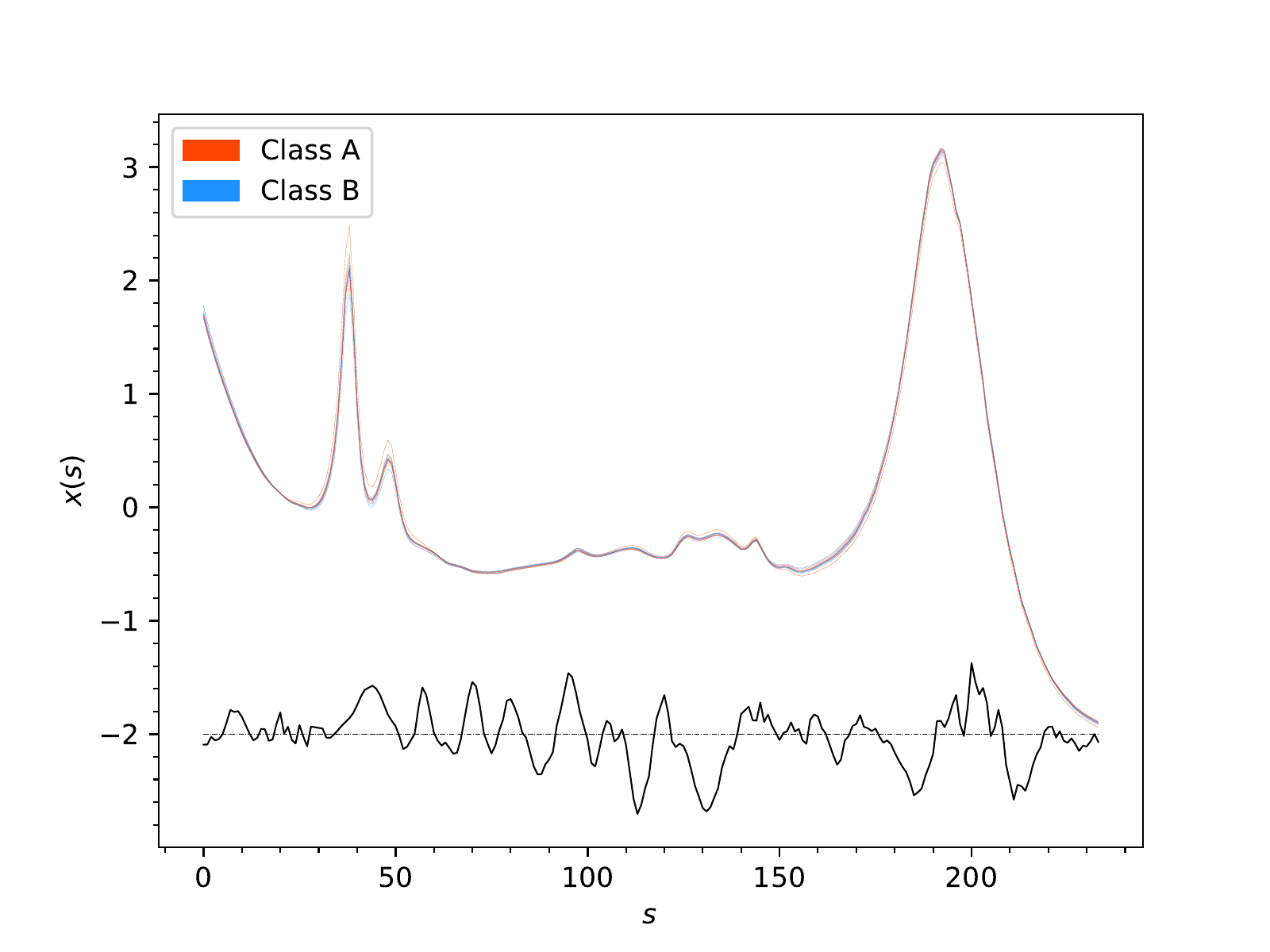}
      \caption{elastic net}
    \end{subfigure}
    \begin{subfigure}[c]{0.32\textwidth}
      \includegraphics[width=\textwidth]{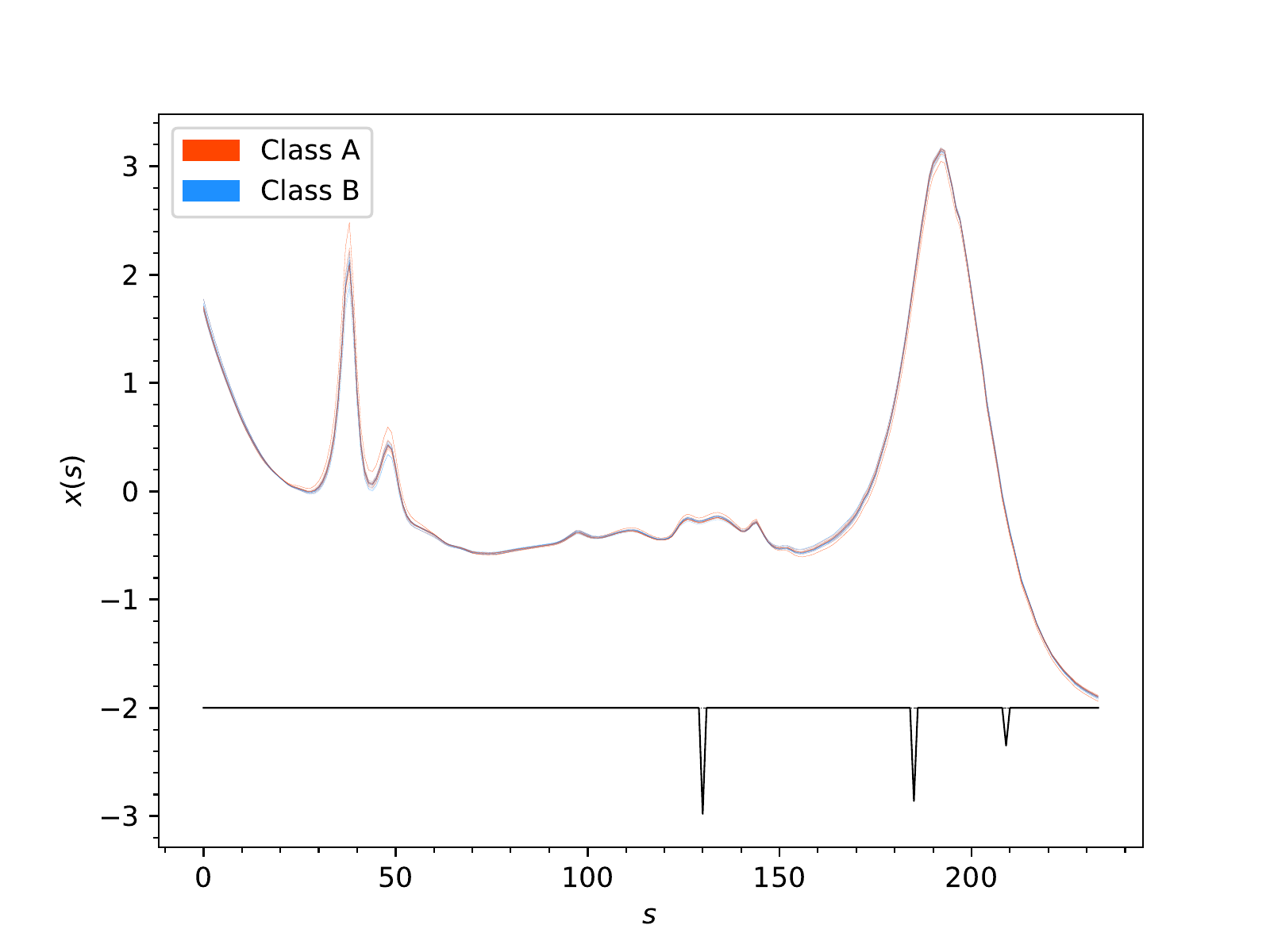}
      \caption{elastic SCAD}
    \end{subfigure}
    \begin{subfigure}[c]{0.32\textwidth}
      \includegraphics[width=\textwidth]{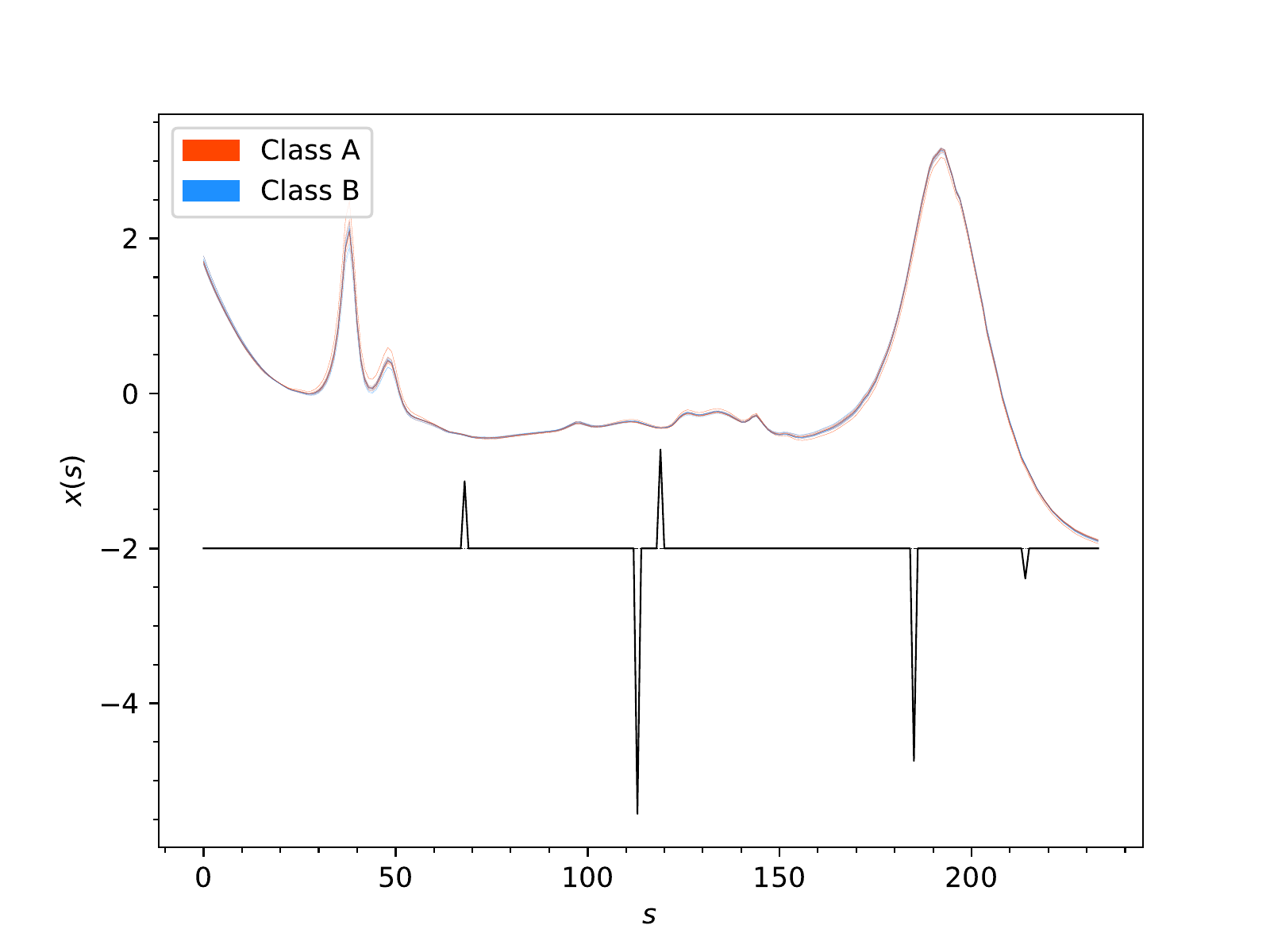}
      \caption{elastic MCP}
    \end{subfigure}
    
    \begin{subfigure}[c]{0.32\textwidth}
      \includegraphics[width=\textwidth]{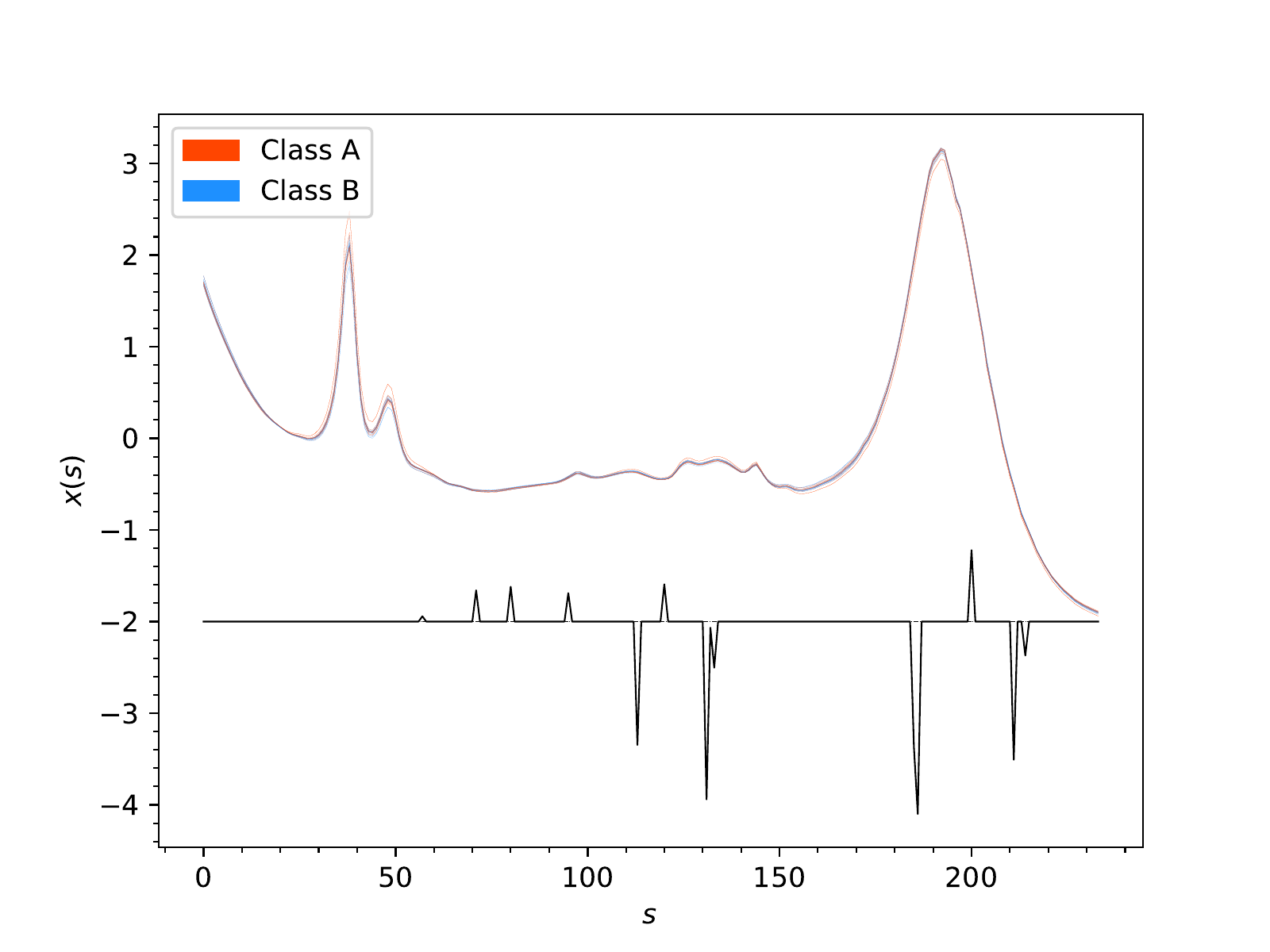}
      \caption{NNG}
    \end{subfigure}
  	\begin{subfigure}[c]{0.32\textwidth}
      \includegraphics[width=\textwidth]{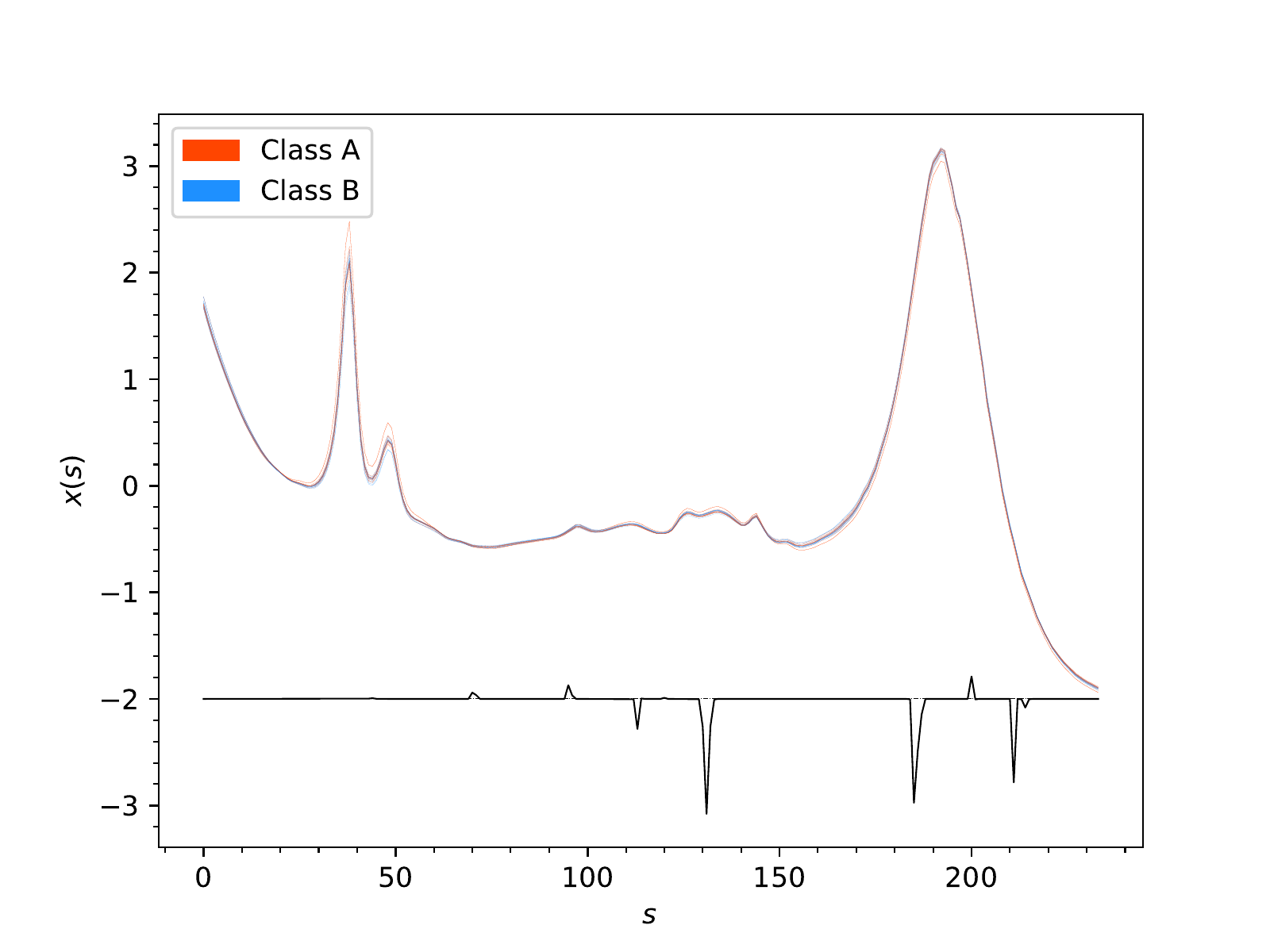}
      \caption{BAR}
    \end{subfigure}   
    \begin{subfigure}[c]{0.32\textwidth}
      \includegraphics[width=\textwidth]{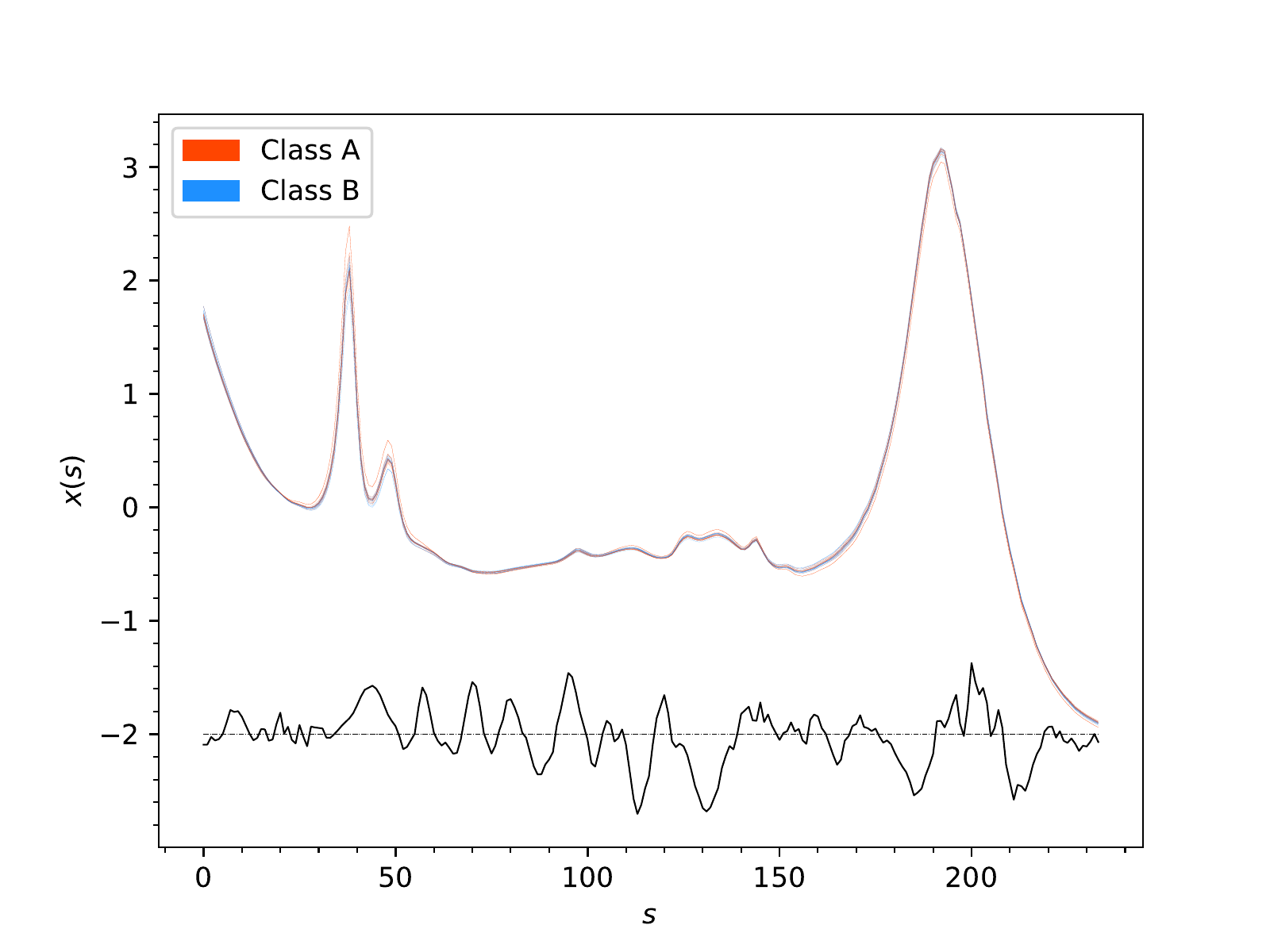}
      \caption{ridge}
    \end{subfigure}
    
    \begin{subfigure}[c]{0.32\textwidth}
      \includegraphics[width=\textwidth]{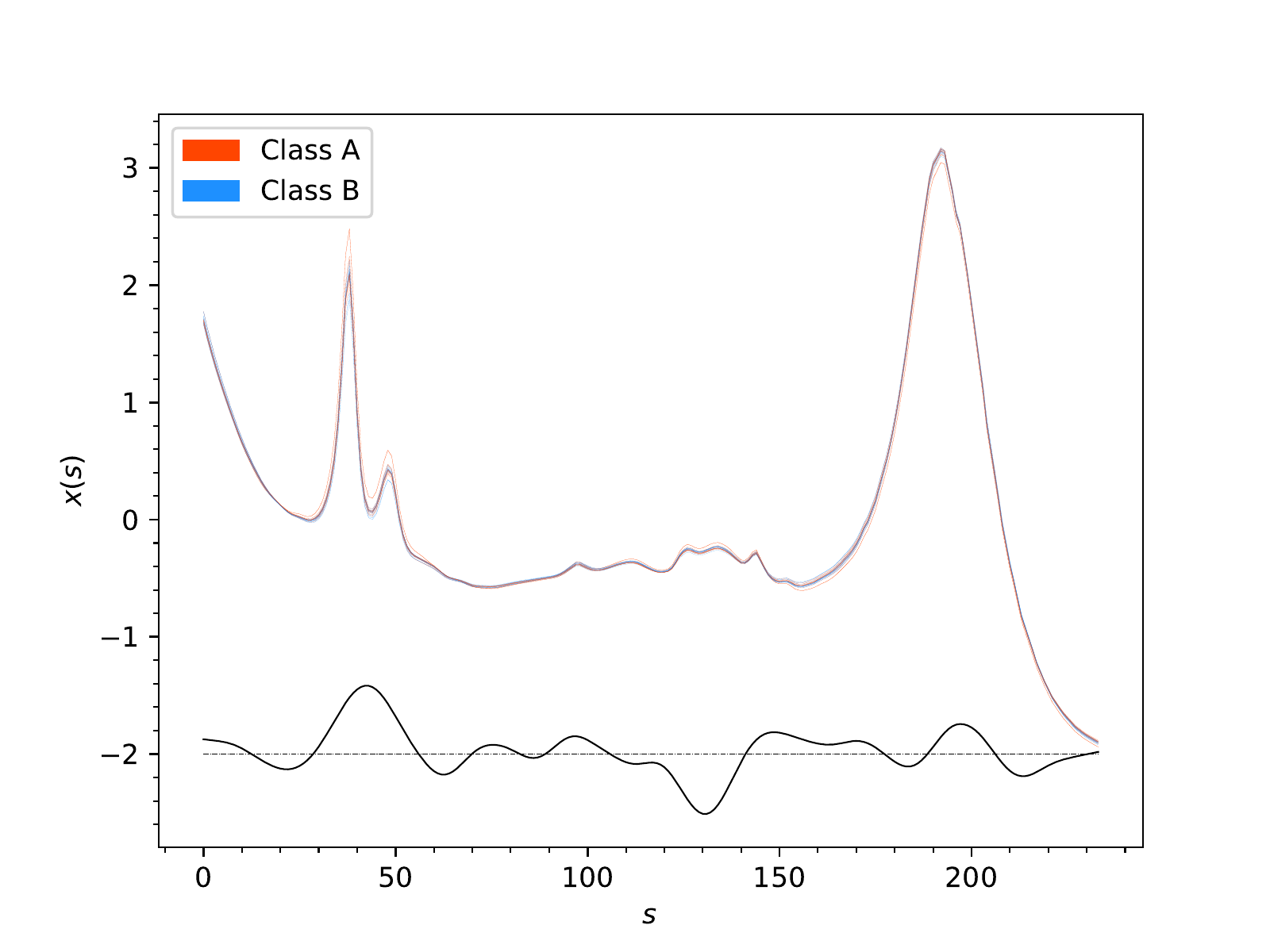}
      \caption{roughness}
    \end{subfigure}
    \begin{subfigure}[c]{0.32\textwidth}
      \includegraphics[width=\textwidth]{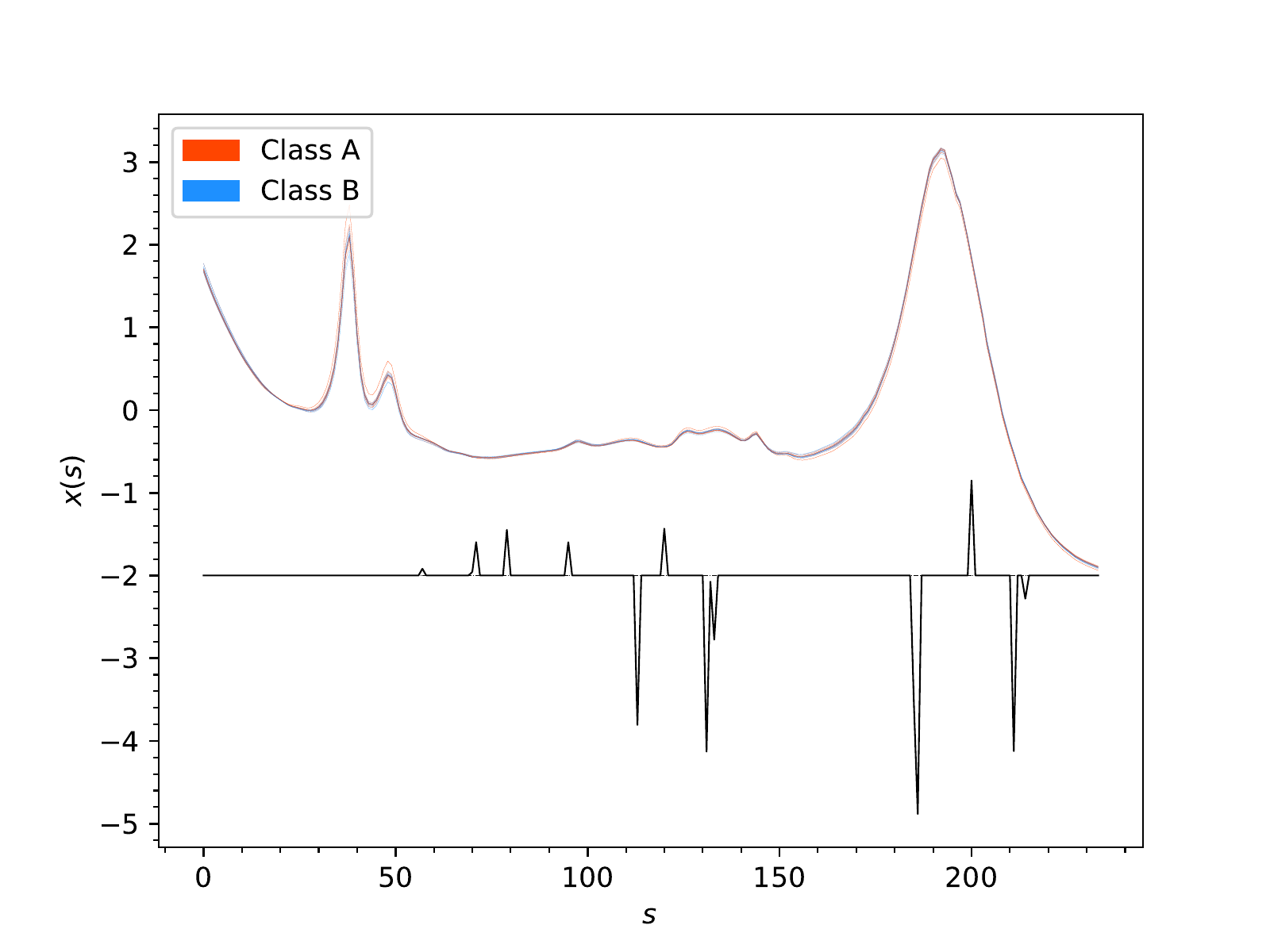}
      \caption{SACR}
    \end{subfigure}
  \caption{Wine: comparison of the fitted coefficient functions $\hat{\beta}$ scaled with respect to the spectra, the black dashed line represents the zero level for $\hat{\beta}$}
  \label{fig:wine}
\end{figure}

This spectroscopy application is a binary classification problem in which we want to discriminate between two different wine types. The data is available at \url{http://www.timeseriesclassification.com/description.php?Dataset=Wine} and is already smoothed with $N=111$ and $p=234$, although there is no additional information about the acquisition process. The results are reported in Table \ref{tab:wine} and are obtained by three random repetitions of 5-fold cross-validation, with additional 3-fold cross-validation for grid search hyperparameter selection. While there is no clear visual distinction between the spectra of the two classes, this problem is not exceptionally hard and it is well suited for a linear model. In particular, the roughness penalty has the second lowest accuracy, suggesting that a very smooth coefficient function is not appropriate. In fact, the SACR provides the highest accuracy without leveraging the smoothing term of the penalty, as the resulting coefficient function is sparse and with multiple spikes, similar to what is usually obtained with $L_1$-based methods, as shown in Figure \ref{fig:wine}. Given the relatively high sample size, the lasso is also able to include many variables in the model, with the adaptive lasso and the relaxed lasso that gradually produce sparser solutions. It is interesting to note that the elastic net instead yields a coefficient function that is almost identical to the one obtained with the ridge, while the elastic SCAD and elastic MCP do not seem to leverage the ridge part of the penalty. Despite that, both approaches include different variables in the model and their resulting accuracy is lower than the other sparse methods, which may be related to the known difficulties of optimizing nonconcave penalties. The NNG has the second highest accuracy and the fitted coefficient function is in fact very similar to the one resulting from the SACR, further suggesting that a sparse solution is indeed adequate for this application, which is also confirmed by the BAR estimator, that for the most part recovers the same variables.

\newpage

\begin{table}
\caption{Wine: classification results, accuracy (\%)}
\label{tab:wine}
\begin {center}
\begin{tabular}{l D{,}{\, \pm \,}{-1}}
\toprule
\midrule
lasso             & 94.64,5.9  \\

adaptive lasso    & 96.77,4.6  \\

relaxed lasso     & 95.88,5.0  \\

NNG               & 97.95,3.1  \\

BAR               & 96.82,3.3  \\
  
elastic net       & 94.34,5.9  \\

elastic SCAD      & 93.35,3.4  \\

elastic MCP       & 89.49,6.7  \\
  
ridge             & 95.25,5.5  \\

roughness         & 90.28,8.4  \\

SACR              & 99.44,1.1  \\
\midrule
\bottomrule
\end{tabular}
\end {center}
\end{table}

\newpage

\section{Conclusions}
In the context of high dimensional linear models, the ordinary ridge penalty is widely known to shrink the coefficients uniformly towards zero, resulting in stable solutions at the price of intruducing some bias. In order to reduce unwanted shrinkage on a subset of the coefficients, generalized and adaptive ridge estimators introduce coefficient-wise penalty parameters that allow for a non-uniform regularization effect, with the downside that tuning such parameters is a nonconvex problem. The nonzero centered ridge instead allows for a convex formulation that uniformly shrinks the coefficients towards a specific target, which in turn has to be specified by the user. In this work we have provided a convex formulation that leverages the nonzero centered ridge and allows for variable shrinkage of the coefficient function along its domain, mitigating the downside of uniform shrinkage towards zero, without the need to specify a center for the penalty, as it is learned from the data in a supervised way. In particular, we introduced a constrained weight function that is jointly estimated while fitting the model and acts as a scaling transformation on the initial centerfunction, which is the ordinary ridge solution. We referred to our method as smoothly adaptively centered ridge (SACR), since the centerfunction is adaptively scaled with respect to the loss and is further penalized for its roughness, as it is common in the functional data setting. Regarding the computational aspect, our approach doubles the number of variables to be estimated but not the ones introduced in the model, and for the numerical optimization we resorted to known primal-dual interior point methods with line search. Finally, we provided some empirical evidence with a simulation study, and two real world spectroscopy applications for both  classification and regression.

\section*{Acknowledgments}
Edoardo Belli was financially supported by the ABB-Politecnico di Milano Joint Research Center through the PhD scholarship \textit{"Development and prototyping of distributed control systems for electric networks based on advanced statistical models for the analysis of complex data"}.

\clearpage

\bibliographystyle{elsarticle-harv} 
\bibliography{mypaper}

\end{document}